\documentclass[11pt]{article}
\usepackage{jheppub}

\usepackage{bbm}

\usepackage{mathrsfs,amssymb}

\newcommand{\be}{\begin{equation}}
\newcommand{\ee}{\end{equation}}
\newcommand{\bea}{\begin{eqnarray}}
\newcommand{\eea}{\end{eqnarray}}

\def\cO{{\mathcal{O}}}
\def\cC{{\mathcal{C}}}
\def\cE{{\mathcal{E}}}
\def\cG{{\mathcal{G}}}
\def\cA{{\mathcal{A}}}
\def\cG{{\mathcal{G}}}
\def\cB{{\mathcal{B}}}
\def\cN{{\mathcal{N}}}

\def\cJ{{\mathcal{J}}}
\def\cI{{\mathcal{I}}}
\def\cH{{\mathcal{H}}}
\def\cK{{\mathcal{K}}}
\def\cQ{{\mathcal{Q}}}
\def\cL{{\mathcal{L}}}

\def\cD{{\mathcal{D}}}
\def\cR{{\mathcal{R}}}
\def\cS{{\mathcal{S}}}
\def\cM{{\mathcal{M}}}
\def\cX{{\mathcal{X}}}
\def\cW{{\mathcal{W}}}
\def\cT{{\mathcal{T}}}

\def\al{\alpha}
\def\ad{\dot{\alpha}}
\def\la{\lambda}
\def\tla{{\tilde\lambda}}
\def\b{\beta}
\def\bd{\dot{\beta}}

\def\te{\theta}
\def\teb{\bar{\theta}}
\def\tx{\tilde{\mathrm{x}}}
\def\tl{\tilde}

\def\mx{{\mathrm{x}}}
\def\Xbf{{\mathbf{X}}}
\def\ub{{\mathbf{u}}}
\def\wb{{\mathbf{w}}}
\def\zb{{\mathbf{z}}}
\def\twb{{\tilde{\mathbf{w}}}}
\def\tzb{{\tilde{\mathbf{z}}}}

\def\de{\delta}

\def\pd{\partial}

\def\nn{\nonumber}

\def\<{\langle}
\def\>{\rangle}
\def\Sa{{S_{-}^{\ell}}}
\def\Sb{{S_{+}\, S_{-}^{\ell-1}}}
\def\Sc{{S_{+}^2\, S_{-}^{\ell-2}}}
\def\Ta{{T_{-}^{\ell}}}
\def\Tb{{T_{+}\, T_{-}^{\ell-1}}}
\def\Tc{{T_{+}^2\, T_{-}^{\ell-2}}}
\def\Dh{\overleftrightarrow{\mathcal{D}}}

\title{\boldmath Superconformal Partial Waves for Stress-tensor Multiplet Correlator in $4D$ $\cN=2$ SCFTs}

\author[]{Zhijin Li}

\affiliation[]{Centro de F\'\i sica do Porto,
Departamento de F\'\i sica e Astronomia\\
Faculdade de Ci\^encias da Universidade do Porto\\
Rua do Campo Alegre 687,
4169--007 Porto, Portugal}
\affiliation[]{CAS Key Laboratory of Theoretical Physics, Institute of Theoretical Physics\\
 Chinese Academy of Sciences, Beijing 100190, China}

\emailAdd{lizhijin18@gmail.com}

\abstract{We compute the superconformal partial waves of the four-point correlator $\langle JJJJ\rangle$, in which the external operator $J$ is the  superconformal primary of the $4D$ $\cN=2$ stress-tensor multiplet $\cJ$. We develop the superembedding formalism for the superconformal field theories (SCFTs) with extended supersymmetry. In $\cN=2$ SCFTs, the
three-point functions $\langle \cJ\cJ\cO\rangle$ with general multiplet $\cO$ contain two independent nilpotent superconformal invariants and new superconformal tensor structures, which can be nicely constructed from variables in superembedding space, and the three-point functions can be solved in compact forms. We compute the superconformal partial waves corresponding to the exchange of long multiplets using supershadow approach. The results are consistent with the non-trivial constraints by decomposing the $\cN=2$ superconformal blocks into $\cN=1$ superconformal blocks. Our results provide the necessary ingredient to study the fascinating $4D$ $\cN=2$ SCFTs using conformal bootstrap. }

\begin{document}
\maketitle
\flushbottom


\section{Introduction}
The conformal bootstrap program \cite{Polyakov:1974gs,Ferrara:1973yt,Mack:1975jr} has made remarkable progress to provide strong constraints on the conformal field theories (CFTs) in higher dimensions $D>2$ \cite{Rattazzi:2008pe}. The constraints are obtained from general consistency conditions of the theories, including crossing symmetry of four-point correlator and unitarity while no information on the Lagrangian is needed. Further developments show that the critical $3D$ Ising model  can be numerically solved using conformal bootstrap with reasonable assumptions \cite{ElShowk:2012ht, El-Showk:2014dwa, Kos:2014bka}.\footnote{A comprehensive review on these developments is provided in \cite{Poland:2018epd}} It is tempting to ask how the conformal bootstrap can improve our understanding on more complex CFTs.

$4D$ superconformal field theories (SCFTs) with $\cN=2$ supersymmetry provide a fascinating laboratory to study the dynamics of quantum field theories. There are abundant $\cN=2$ SCFTs that can be constructed from different ways. In the conventional Lagrangian approach, these theories can be built from gauge theories with proper matter representations which lead to vanishing beta function of all the gauge couplings. A large set of SCFTs can be obtained from the class $\cS$ constructions \cite{Gaiotto:2009we, Gaiotto:2009hg}. They correspond to a web of dualities, most of which do not admit Lagrangian description. Hopefully these SCFTs with or without Lagrangian description could be classified in a more fundamental frame, such as conformal bootstrap.

The program to study general $4D$ $\cN=2$ SCFTs using conformal bootstrap has been initiated in \cite{Beem:2014zpa}, in which the four-point correlators of the moment map and chiral operators have been studied. These correlators have been explored further in \cite{Lemos:2015awa} and \cite{Cornagliotto:2017snu} with emphasis on the simplest Argyres-Douglas fixed point. The stress-tensor four-point correlator\footnote{In $4D$ $\cN=2$ SCFTs, the stress-tensor stays in the supercurrent multiplet $\cJ$ in which the superconformal primary $J$ is a scalar. In this work, the stress-tensor four-point correlator means the four-point correlator of the superconformal primary operator.} is expected to be quite important to carve out the space of $\cN=2$ SCFTs. The stress-tensor operator is universal in any local CFT so it is a natural candidate to bootstrap.
The $3D$ stress-tensor four-point correlator without supersymmetry has been bootstrapped in \cite{Dymarsky:2017yzx}.
Remarkably the conformal collider bound on the conformal anomaly coefficients $a/c$ \cite{Hofman:2008ar} automatically appears without extra assumptions besides the
general consistency conditions.\footnote{The conformal collider bound also appears in the $3D$ conserved current bootstrap \cite{Dymarsky:2017xzb}.}
The supersymmetric stress-tensor correlator has been bootstrapped in \cite{Beem:2013qxa, Beem:2016wfs} for $4D$ $\cN=4$ SCFT and in \cite{Beem:2015aoa} for $6D$ $(2, 0)$ SCFTs. The results are more restrictive due to the extra constraints from supersymmetry.
Nevertheless, for $4D$ $\cN=2$ theories, the superconformal partial wave expansion, or the superconformal blocks of the stress-tensor four-point correlator was not unknown before. The three-point function of the stress-tensor, or the selection rule of the $\cJ\times \cJ$ OPE  has been studied in \cite{Liendo:2015ofa, Ramirez:2016lyk}. In this work, we compute the superconformal partial wave expansion of the stress-tensor four-point correlator.

Conformal partial wave (or conformal block differing by kinematic factors) describes the contributions on the four-point correlator from exchange of a primary operator and its conformal descendants. The unitarity and crossing symmetry of four-point correlator, as consistent conditions employed in conformal bootstrap, are imposed based on the conformal partial wave expansions of the four-point correlator. The conformal blocks for external scalars in general spacetime dimensions have been solved in \cite{Dolan:2000ut, Dolan:2003hv, Dolan:2011dv} as series expansion of the two conformal invariant cross ratios. The expressions are further simplified into compact forms based on Hypergeometric functions in even dimensions.
The superconformal block consists of finite many conformal blocks and its explicit form depends on the specific superconformal algebra and the representations, which in general is an involved technical problem. The major challenge is to fix the coefficients for each conformal block  appearing in the decomposition of superconformal block. For the theories with $\cN=1$ supersymmetry, the superconformal block can be obtained from straightforward approach, i.e., decomposing the exchange supermultiplet into several conformal primary families and fixing the coefficients from two and three-point functions \cite{Poland:2010wg, Fortin:2011nq, Li:2017ddj}. However, this approach turns into extremely cumbersome and out of control for theories with extended supersymmetry.
For external operators which are superconformal primaries in $\frac{1}{2}$-BPS multiplet, the superconformal blocks can be solved from the superconformal ward identities \cite{Eden:2000qp, Dolan:2001tt, Dolan:2004mu}. This method has been used to compute the superconformal blocks of shortened operators in SCFTs with extended supersymmetry \cite{Chester:2014fya, Chester:2014mea, Beem:2014zpa, Liendo:2015cgi, Liendo:2016ymz, Lemos:2016xke, Beem:2016wfs}.  The superconformal Casimir approach can also be used to compute the superconformal blocks of shortened operators as solutions of the superconformal Casimir eigen-equation \cite{Bobev:2015jxa, Fitzpatrick:2014oza, Bobev:2017jhk}. While for more general operators, the superconformal blocks involve complex structures and above methods are not quite helpful. Specifically, the $4D$ $\cN=2$ stress-tensor multiplet has no shortening at the first level. The shortening appears at level 2 while it does not lead to the BPS-like condition for the four-point function.

We employ the superembedding formalism to compute the superconformal partial waves of the $4D$ $\cN=2$ stress-tensor four-point correlator. The superembedding formalism has been developed in \cite{Fitzpatrick:2014oza, Khandker:2014mpa} for $4D$ $\cN=1$ superconformal blocks and it has been applied to obtain the superconformal blocks for general external scalars \cite{Li:2016chh}. It is the supersymmetric generalization of the embedding formalism in which the conformal transformations are realized linearly and the conformal correlators can be simplified drastically \cite{Dirac:1936fq,Mack:1969rr,Ferrara:1973eg,Weinberg:2010fx,Costa:2011mg,Costa:2011dw,SimmonsDuffin:2012uy, Elkhidir:2014woa, Costa:2014rya, Echeverri:2015rwa, Rejon-Barrera:2015bpa, Iliesiu:2015akf, Echeverri:2016dun, Costa:2016xah, Costa:2016hju}. The superembedding space and its relation with the superspace has been studied in \cite{Goldberger:2011yp, Goldberger:2012xb, Siegel:2012di, Maio:2012tx, Kuzenko:2012tb}.
For $4D$ $N=1$ superconformal theories, the superembedding formalism can nicely reproduce the superconformal invariants and tensor structures obtained in superspace \cite{Park:1997bq, Osborn:1998qu, Park:1999pd}, which are the fundamental elements of the superconformal correlators.

The $4D$ $\cN=2$ superconformal correlators have been studied in superspace \cite{Park:1999pd, Kuzenko:1999pi, Arutyunov:2001qw, Liendo:2015ofa}.
There are two independent nilpotent superconformal invariants that can be built from three points \cite{Park:1999pd}, and they are involved in the correlator $\langle J J\cO\rangle$ where $\cO$ is a general long multiplet \cite{Liendo:2015ofa}. In particular, it has been shown from the analysis in superspace that the three-point correlators $\langle J J\cO\rangle$ contain new tensor structures that are completely different from these known for $\cN=1$ theories \cite{Liendo:2015ofa}. One of the challenges to study the $\cN=2$ stress-tensor four-point correlator is to construct the new independent nilpotent invariant and tensor structures in superembedding space.
In this work, we construct the new nilpotent superconformal invariant and tensor structures in $\cN=2$ superembedding space.
The constructions follow a systematical correspondence between the superconformal invariants (tensor structures) in superembedding space and superspace. We also use the supershadow approach \cite{Fitzpatrick:2014oza, Khandker:2014mpa} to
project the four-point correlator into the contributions from a specific exchanged multiplet.

This paper is organized as follows. In the next section, we briefly review the representations of $4D$ $\cN=2$ superconformal algebra that are relevant to the stress-tensor OPE, and also the selection rules for the $\cJ\times\cJ$ OPE. In section 3 we introduce the superembedding space with $\cN=2$ supersymmetry and construct the two independent superconformal invariants and tensor structures. We show their relations with the results obtained from $4D$ $\cN=2$ superspace analysis. In section 4 we solve the three-point functions $\langle J J\cO\rangle$ in superembedding space. The four-point correlator and its superconformal partial wave expansion are studied in section 5. Our main results are presented in
(\ref{oddscb}), (\ref{N2evenblock}), (\ref{N2blockmodd}), (\ref{N2blockmeven}) and (\ref{N2blockmeven4}). In section 6
we decompose the $\cN=2$ superconformal blocks for long multiplets into $\cN=1$ superconformal blocks, which provide nontrivial consistency checks for our results.
Conclusion and discuss are made in section 7. More details on the superconformal invariants and tensor structures in superembedding space and the conformal integrations will be presented in the Appendices.

\section{$\cN=2$ superconformal algebra representations and selection rules}

\subsection{Representations of $\cN=2$ superconformal algebra}
The $\cN=2$ superconformal algebra $\mathfrak{su}(2,2|2)$ is the supersymmetric extension of the conformal algebra $\mathfrak{so}(4,2)$. Its maximum bosonic subalgebra contains conformal algebra and the R-symmetry algebra $SU(2)_R\times U(1)_r$, with generators $\{{\cR}_i^j, r\}$. In addition, there are eight Poincar\'e supercharges $\{\cQ_\al^i, \bar{\cQ}_{i\ad}\}$ and eight conformal supercharges $\{\cS_\al^i, \bar{\cS}_{i\ad}\}$, where $\al=\pm$, $\ad=\dot{\pm}$ are the Lorentz indices and $i\in\{1, 2\}$ are the $SU(2)_R$ indices.

Representations of $\mathfrak{su}(2,2|2)$ can be constructed from highest weight states, or superconformal primaries which are annihilated by the conformal supercharges $\cS$. The representations are characterized by the quantum numbers $(\Delta, j_1, j_2, R, r)$ of superconformal primaries, where $\Delta$ is the conformal dimension, $j_1, j_2$ are the Lorentz indices and $R, r$ are the $SU(2)_R\times U(1)_r$ Dynkin labels. For general representations they may also have extra quantum numbers corresponding to the flavor symmetries which commute with $\mathfrak{su}(2,2|2)$. While in this work, we focus on the stress-tensor multiplet and operators in the $\cJ\times\cJ$ OPE. These operators are invariant under the flavor symmetries. The superconformal primary and its super-descendants form a supermultiplet. It can be shown from the superconformal algebra that a supermultiplet consists of finite many conformal primaries. In consequence, the superconformal block which captures the contributions on four-point correlator from exchange of a superconformal family can be decomposed into finite many conformal blocks.

Unitarity provides constraints on the quantum numbers of the representations. For a general representation, denoted as  $\cA_{R,r\left( j_1,j_2\right)}^{\Delta}$ following the notation of \cite{Dolan:2002zh},\footnote{See \cite{Dobrev:1985qz, Dobrev:1985qv, Dobrev:1985vh} for early studies on extended superconformal symmetry and their representations.} there is a unitary bound on the conformal dimension
\bea
\Delta\geqslant 2+2j_i+2R-(-1)^ir, \,~~~~~~~~~~~~~~~~~~~~~~~~~~~~~~~~~~~~~~ &j_i\neq0&, \\
\Delta=2j_i+2R-(-1)^ir ~~{\textrm{or}}~~ \Delta\geqslant 2+2j_i+2R-(-1)^ir, ~~&j_i=0&.
\eea
If the conformal dimension saturates the unitary bounds, the superconformal primary can be annihilated by certain combinations of the Poincar\'e supercharges. In another words, part of the superconformal descendants become null and the multiplet shortens. The shortening conditions are classified in \cite{Dolan:2002zh,Kinney:2005ej}. The results are summarized in Table \ref{Reps} (see also (\cite{Beem:2014zpa, Liendo:2015ofa})).
\begin{table}[t]
\centering
\small
\begin{tabular}{|l|l l|l|}
\hline
Shortening & Quantum Number Relations&  & Multiplet
\\ \hline\hline
$\oslash$ & $\Delta \geqslant 2+2j_i+2R - (-1)^ir,$ ~$i=1,2$ &   & $\mathcal{A}^\Delta_{R,r\left( j_1,j_2
\right) }$ \\
\hline
$\mathcal{B}^{1}$ & $\Delta = 2R + r$ & $j_1=0$ & $\mathcal{B}_{R,r\left( 0,j_2
\right) }$ \\ \hline
$\overline{\mathcal{B}}_{2}$ & $\Delta = 2R - r$ & $j_2=0$ & $\bar{\mathcal{B}}_{R,r\left( j_1,0\right)
} $
\\ \hline
$\mathcal{B}^{1}\cap \mathcal{B}^{2}$ & $\Delta = r$ & $R=0$  & $\mathcal{E}_{r\left( 0,
j_2\right) }$
\\ \hline
$\overline{\mathcal{B}}_{1}\cap \overline{\mathcal{B}}_{2}$ & $\Delta = -r$ & $R=0$ & $\bar{\mathcal{E}}%
_{r\left( j_1,0\right) }$
\\ \hline
$\mathcal{B}^{1}\cap \overline{\mathcal{B}}_{2}$ & $\Delta = 2R$ & $j_i=r=0 ~~~~~~~~~~$ & $\hat{\mathcal{B}}_{R}$
\\
\hline\hline
$\mathcal{C}^{1}$ & $\Delta =2+2j_1+2R+r$  & & $\mathcal{C}_{R,r\left( j_1,%
j_2\right) }$
\\ \hline
$\overline{\mathcal{C}}_{2}$ & $\Delta =2+2j_2+2R-r$ &
& $\bar{\mathcal{C}}_{R,r\left( j_1,j_2\right) }$
\\ \hline
$\mathcal{C}^{1}\cap \mathcal{C}^{2}$ & $\Delta =2+2j_1+r$ & $R=0$ & $%
\mathcal{C}_{0,r\left( j_1,j_2\right) }$
\\ \hline
$\overline{\mathcal{C}}_{1}\cap \overline{\mathcal{C}}_{2}$ & $\Delta =2+2%
j_2-r$ & $R=0$ & $\bar{\mathcal{C}}_{0,r\left( j_1,j_2%
\right) }$
\\ \hline
$\mathcal{C}^{1}\cap \overline{\mathcal{C}}_{2}$ & $\Delta =2+2R+j_1+j_2$ & $r=j_2-j_1$ & $\hat{\mathcal{C}}_{R\left( j_1,j_2\right) }$
\\ \hline\hline
$\mathcal{B}^{1}\cap \overline{\mathcal{C}}_{2}$ & $\Delta =1+j_2%
+2R$ & $r=j_2+1$ & $\mathcal{D}_{R\left( 0,j_2\right) }$
\\ \hline
$\overline{\mathcal{B}}_{2}\cap \mathcal{C}^{1}$ & $\Delta =1+j_1+2R$ &
$-r=j_1+1$ & $\bar{%
\mathcal{D}}_{R\left( j_1,0\right) }$
\\ \hline
$\mathcal{B}^{1}\cap \mathcal{B}^{2}\cap \overline{\mathcal{C}}_{2}$ & $%
\Delta =r=1+j_2$ & $R=0$ & $\mathcal{D}_{0\left( 0,%
j_2\right) }$
\\ \hline
$\mathcal{C}^{1}\cap \overline{\mathcal{B}}_{1}\cap \overline{\mathcal{B}}%
_{2}$ & $\Delta =-r=1+j_1$ & $R=0$ & $\bar{\mathcal{D}}_{0\left(
j_1,0\right) }$
\\ \hline
\end{tabular}
\caption{Classification of the unitary irreducible representations of the $\cN=2$ superconformal algebra. \label{Reps}
}
\end{table}
There are two types of shortening conditions corresponding to saturating different unitary bounds given above, namely the $\cB$ type and $\cC$ type:
\begin{eqnarray}
\cB^{i} &:&\cQ_{\alpha }^{i} \cO  =0, \\
\overline{\cB}_{i} &:&\bar{\cQ}_{i\dot{\alpha}}\cO =0, \\
\cC^{i} &:&\left\{
\begin{array}{c}
\varepsilon ^{\alpha \beta }\cQ_{\alpha }^{i} \cO_{\beta }=0,~~~\quad j_1\neq 0, \\
\varepsilon ^{\alpha \beta }\cQ_{\alpha }^{i}\cQ_{\beta }^{i} \cO =0,\quad j_1=0,%
\end{array}%
\right. \\
\overline{\cC}_{i} &:&\left\{
\begin{array}{c}
\varepsilon ^{\dot{\alpha}\dot{\beta}}\bar{\cQ}_{i\dot{\alpha}}
\cO _{\dot{\beta}}=0, ~~~~~~ j_2\neq 0, \\
\varepsilon ^{\dot{\alpha}\dot{\beta}}\bar{\cQ}_{i\dot{\alpha}}\bar{\cQ%
}_{i\dot{\beta}} \cO =0,\quad j_2=0.%
\end{array}%
\right.
\end{eqnarray}
In \cite{Dolan:2002zh} the representations satisfying type $\cB$ ($\cC$) conditions are called short (semi-short) multiplets, as the semi-short superconformal primaries are annihilated by half number of supercharges comparing with the short multiplets.
In Table \ref{Reps}, there are three special classes of shortened multiplets $(\cE_{r}, \hat{\cB}_R, \hat{\cC}_{0(j_1,j_2)})$ which satisfy the maximum number of short or semi-shorten conditions. The multiplets $\cE_{r}$ ($\hat{\cB}_R$) obey two $\cB$ type conditions with the same (opposite) chirality. They are also called $\frac{1}{2}$-BPS multiplets since they are annihilated by half of the Poincar\'e supercharges. In $\cN=2$ theories, the multiplets $\cE_{r}$ correspond to the Coulomb branch physics while the multiplets $\hat{\cB}_R$ have connections with Higgs branch physics. The four-point correlators of the chiral (anti-chiral) superconformal primaries $\cE_{r}$ and the moment map $\hat{\cB}_1$ have been studied using conformal bootstrap in \cite{Beem:2014zpa, Lemos:2015awa}, which lead to strong constraints on the CFT data in Coulomb and Higgs branches.

The multiplet $\hat{\cC}_{0(j_1,j_2)}$ is a special class $(R=0)$ of the shortened multiplet $\hat{\cC}_{R(j_1,j_2)}$, and it obeys an enhanced semi-short conditions $\cC^1\cap \cC^2\cap \bar{\cC}_1\cap\bar{\cC}_2$. Its conformal dimension $\Delta=2+j_1+j_2$ saturates unitary bound for general conformal operators and actually it satisfies a generalized conservation equation. The conserved higher spin operators $(j_1+j_2>0)$ are not allowed in an interacting CFT \cite{Maldacena:2011jn, Alba:2013yda}. However, the conserved multiplet $\hat{\cC}_{0(0,0)}$ contains a conserved spin two currents so it is expected to be the stress-tensor multiplet. Any local $\cN=2$ SCFT that cannot be factorized as a product of two local theories contains a unique $\hat{\cC}_{0(0,0)}$ multiplet. Its superconformal primary is a scalar invariant under R-symmetry transformations.
Moreover, the multiplet includes spin one superconformal descendants which give the conserved currents for R-symmetry $SU(2)_R\times U(1)_r$.

\subsection{$\cJ\times\cJ$ selection rules from $\cN=2$ superspace analysis}
The $\cN=2$ stress-tensor multiplet correlators have been studied in \cite{Kuzenko:1999pi, Liendo:2015ofa}. In these work the three-point functions are solved in terms of the superconformal covariant variables in superspace developed in \cite{Osborn:1998qu,Park:1999pd}. Here we briefly review the analysis in superspace. The results will be reproduced in superembedding space later.

Following the notation in \cite{Kuzenko:1999pi, Liendo:2015ofa}, the stress-tensor multiplet is denoted as a superfield $\cJ$
\be
\cJ(x,\te,\teb) = J(x) + J_{\al \ad\, j}^{i}\te^{\al}_{i} \teb^{j\, \ad} + \ldots \, ,
\ee
 which satisfies the reality condition $\cJ=\bar\cJ$ and the conservation equations
\be
D^{\alpha i}D_{\al }^{j}\cJ=0\, ,\qquad \bar{D}_{%
\dot{\alpha}}^{i}\bar{D}^{j\dot{\alpha}}\cJ=0\, ,
\ee
where $D^{\alpha i}$ and $\bar{D}^{i\dot{\alpha}}$ are covariant derivatives.
The superconformal primary of stress-tensor multiplet $J$ has scaling dimension $2$ and is invariant under the R-symmetry $SU(2)_R\times U(1)_r$,
which is crucial for us to study its correlators in superembedding space.
The $\cN=2$ superconformal two and three-point correlators are constructed based on the superconformal covariant coordinates in superspace $z^I=(\tx^{\ad\al},\te_i,\teb^i)$. With two points $(z_1,z_2)$ one can construct variables which transform as a product of two tensors at $z_i$ under superconformal transformation:
\begin{align}
& \tx^{\ad \al}_{\bar{1}2}   = \tx^{\ad \al}_{1-} -  \tx^{\ad \al}_{2+} -4\mathrm{i}\, \theta^{\al}_{2\,i} \bar{\theta}^{\ad i}_1=  (\tx_{12})^{\ad \al}_-\, , \label{xxbsuperspace}
\\
& z_{12}^I = z_{1}^I - z_{2}^I\, ,
\end{align}
where we have employed the convention $\mx=x^\mu(\sigma_\mu)_{\al\ad}=\mx_{\al\ad},~ \tx=x^\mu(\tilde{\sigma}_\mu)^{\ad\al}=\tx^{\ad\al}$ and the chiral combinations $\tx^{\ad \al}_{\pm} = \tx^{\ad \al} \mp 2{\rm i}\te_{i}^\al \teb^{\ad\, i}$.

With three points $(z_1,z_2,z_3)$ one can construct the superconformal covariant coordinates  $\mathbf{Z}_1 = (\Xbf_1, \Theta_1, \bar{\Theta}_1)$:
\begin{align}
&
\mathbf{X}_{1}  = \tx_{1\bar{2}}^{-1}\tx_{\bar{2}3}\tx_{3\bar{1}}^{-1}\, ,
&
&
\bar{\mathbf{X}}_{1} = \mathbf{X}^{\dagger}_{1}
= -\tx_{1\bar{3}}^{-1}\tx_{\bar{3}2}\tx_{2\bar{1}}^{-1}\, ,
&
\\
&
\tilde{\Theta}^{i}_{1} = \mathrm{i} \left(\tx_{\bar{2}1}^{-1}\bar{\theta}^{i}_{12}- \tx_{\bar{3}1}^{-1}\bar{\theta}^{i}_{13} \right)\, ,
&
&
\tilde{\bar{\Theta}}_{1\, i} = \mathrm{i} \left(\theta_{12\, i}\tx_{\bar{1}2}^{-1}
- \theta_{13\, i}\tx_{\bar{1}3}^{-1} \right)\, \label{Z},
&
\end{align}
where ${\tx}^{-1}=-\frac{\mx}{x^2}$ following the convention $x^2 \equiv x^\mu x_\mu=-\frac{1}{2} \mathrm{tr}({\tx}\, \mx)$.
$\mathbf{Z}_1$ transforms as a ``tangent" vector at $z_1$.
It also satisfies the ``chiral" condition
\be
\bar{\mathbf{X}}_{1\,\al\ad} = \mathbf{X}_{1\,\al\ad}-4 \mathrm{i}\, \Theta^{i}_{1\,\al} \bar{\Theta}_{1\,\ad\, i}\, .\label{Xchiral}
\ee
Moreover, we have following relations
\be
\Xbf_1^2 =\frac{x_{\bar{2}3}^2}{x_{\bar{2}1}^2x_{\bar{1}3}^2}, ~~~~~~~~~~\bar{\Xbf}_1^2 =\frac{x_{\bar{3}2}^2}{x_{\bar{3}1}^2x_{\bar{1}2}^2}\, . \label{X2}
\ee
Variables $\mathbf{Z}_2$ and $\mathbf{Z}_3$ can be constructed similarly by cyclically permuting $(z_1,\,z_2,\,z_3)$ and it is easy to show
\be
\frac{\Xbf_1^2}{\bar{\Xbf}_1^2} =\frac{\Xbf_2^2}{\bar{\Xbf}_2^2}=\frac{\Xbf_3^2}{\bar{\Xbf}_3^2}\equiv\ub . \label{ub}
\ee
The variable $\ub$ is a superconformal invariant which appears both in $\cN=1$ and $\cN=2$ theories \cite{Osborn:1998qu,Park:1999pd}. Another variable that is also invariant under continuous superconformal transformation can be obtained by contracting Lorentz indices between $\Xbf$ and $\bar{\Xbf} $ \footnote{Combining with $\ub+\ub^{-1}$, $\wb^\prime$ is also invariant under superinversion \cite{Park:1999pd}.} \cite{Park:1999pd}
\be
\frac{\Xbf_1\cdot \bar{\Xbf}_1}{\sqrt{\Xbf_1^2} \, \bar{\Xbf}_1^2\;} =-\frac{\mathrm{tr}(\mx_{\bar{2}1}\tx_{\bar{2}3}\mx_{\bar{1}3}\tx_{\bar{1}2}\mx_{\bar{3}2}\tx_{\bar{3}1})}
{2\sqrt{x_{\bar{2}1}^2 x_{\bar{2}3}^2 x_{\bar{1}3}^2 x_{\bar{1}2}^2 x_{\bar{3}2}^2 x_{\bar{3}1}^2}}\equiv-\wb^\prime, \label{wb}
\ee
which also admits a cyclical permutation symmetry as in (\ref{ub}).
In $\cN=1$ superspace, there is only one independent superconformal invariant $\ub$ that can be obtained from three points, while $\wb^\prime$ is actually corresponding to $\ub$ through an identity of $\cN=1$ Grassmann variables. In $\cN=2$ superspace, $\ub$ and $\wb^\prime$ are independent with each other from which two nilpotent superconformal invariants can be constructed. Both of the two nilpotent invariants are necessary to construct general $\cN=2$ superconformal correlators. In the next section, we will show that the two superconformal invariants can be nicely constructed in superembedding space.

In $\cN=2$ superspace, the three-point correlator $\langle \cJ\cJ\cO\rangle$ with general multiplet $\cO$ reads
\be
\langle \cJ(z_1) \cJ(z_2) \cO(z_3) \rangle = \frac{1}{(x_{\bar{1}3})^2 (x_{\bar{3}1})^2 (x_{\bar{2}3})^2 (x_{\bar{3}2})^2} \cH(\mathbf{Z}_3)\, \label{JJO}.
\ee
The Lorentz indices in $\cH(\mathbf{Z}_3)$ and the scaling conditions are fixed by superconformal symmetry.
The three-point function in (\ref{JJO}) is further restricted by the conservation equation of $\cJ$
\be
\frac{\pd^2}{\pd \Theta^{i}_{3\,\al} \pd \Theta^{\al\, j}_{3}} \cH(\mathbf{Z}_3) = 0\, , ~~~
\frac{\pd^2}{\pd \bar{\Theta}^{\ad}_{3\,i} \pd \bar{\Theta}_{3\,\ad j}} \cH(\mathbf{Z}_3) = 0\, . \label{conseq}
\ee
The Grassmann coordinates $\Theta_3, \bar{\Theta}_3$ are antisymmetric, so for two Grassmann coordinates with the same chirality, either the Lorentz $SU(2)$ indices
or the R-symmetry $SU(2)_R$ indices get contracted. Now the conservation equations (\ref{conseq}) suggest that
only the second choice is possible for the correlators $\langle \cJ\cJ\cO\rangle$.
Moreover,
the three-point correlator is invariant under the exchange of $z_1$ and $z_2$, which requires the RHS in (\ref{JJO})
invariant under the reflection $(\Xbf_3, \Theta_3, \bar{\Theta}_3) \leftrightarrow (-\bar{\Xbf}_3, -\Theta_3, -\bar{\Theta}_3)$.

The three-point function $\langle\cJ\cJ\cJ\rangle$ is of special importance for the stress-tensor bootstrap. It has been solved in  \cite{Kuzenko:1999pi} and the corresponding function $\cH(\mathbf{Z}_3)$ in (\ref{JJO}) reads
\be
\cH(\Xbf_3, \Theta_3, \bar{\Theta}_3)=\la_\cJ^{(1)}\left(\frac{1}{\Xbf_3^2}+\frac{1}{\bar{\Xbf}_3^2}\right)+\la_\cJ^{(2)}
\frac{\Theta^{\alpha \beta}_3 \Xbf_{3\,\alpha \dot{\alpha}}\Xbf_{3\,\beta \dot{\beta}}\bar{\Theta}_3^{\dot{\alpha} \dot{\beta}}}{(\Xbf_3^2)^{3}}, \label{JJJ}
\ee
in which the Grassmann variables with symmetric Lorentz indices are defined by
\be
\Theta^{\alpha \beta}_3= \Theta_3^{\alpha i}\Theta_3^{\beta j}\epsilon_{ij}=\Theta^{(\alpha \beta)}_3,~~~~~~~
\bar\Theta^{\ad \bd}_3= \bar\Theta_{3i}^{\ad }\Theta_{3j}^{\bd }\epsilon^{ij}=\bar\Theta^{(\ad \bd)}_3,
\ee
and they are invariant under R-symmetry $SU(2)_R$.
The OPE coefficients $\la_\cJ^{(i)}$ depend on the conformal anomaly coefficients:
\be
\la_\cJ^{(1)}=\frac{3}{64\pi^6}(4a-3c),~~~~~~~~~\la_\cJ^{(2)}=\frac{1}{8\pi^6}(4a-5c).
\ee
The second term in (\ref{JJJ}) is nilpotent and invariant under the reflection transformation due to the following identity
\be
\frac{\Theta^{\alpha \beta} \Xbf_{\alpha \dot{\alpha}}\Xbf_{\beta \dot{\beta}}\bar{\Theta}^{\dot{\alpha} \dot{\beta}}}{(\Xbf^2)^{3}}  = \frac{\Theta^{\alpha \beta} \bar{\Xbf}_{\alpha \dot{\alpha}}\bar{\Xbf}_{\beta \dot{\beta}}\bar{\Theta}^{\dot{\alpha} \dot{\beta}}}{(\bar{\Xbf}^2)^{3}} \, . \label{KuzenkoIdentity}
\ee
For the scalar multiplet with $\Delta>2$, the second term will be modified according to the scaling condition and does not satisfy the reflection symmetry. Therefore there is only one independent OPE coefficient for the scalar multiplets except the supercurrent $\cJ$. For non-supersymmetric theories the conformal anomaly coefficient $a$ appears in the correlation function of stress-tensor and conserved currents, accompanied by complicated tensor structures \cite{Erdmenger:1996yc}.
While in $\cN=2$ theories, these tensor structures are packaged in a unique nilpotent tensor structure because of the supersymmetry. Coefficients of these tensor structures, or the conformal anomaly coefficients are subject to the constraints from causality \cite{Hofman:2008ar, Camanho:2014apa}.

A comprehensive study on the three-point functions $\langle\cJ\cJ\cO\rangle$ with general $\cO$ have been provided in \cite{Liendo:2015ofa}, which gives
all the multiplets $\cO$ that can appear in the $\cJ\times\cJ$ OPE. The three-point functions $\langle\cJ\cJ\cO\rangle$ have been fixed up to
certain free parameters that are corresponding to the dynamics of the theories.
Specifically, it leads to following selection rules:
\begin{align}
\begin{split}
\hat{\cC}_{0(0,0)} \times \hat{\cC}_{0(0,0)} & \sim \cI+ \hat{\cC}_{0(\frac{\ell}{2},\frac{\ell}{2})}
 + \hat{\cC}_{1(\frac{\ell}{2},\frac{\ell}{2})}  +\cC_{\frac{1}{2},\frac{3}{2}(\frac{\ell}{2},\frac{\ell+1}{2})}
+\cC_{0,3(\frac{\ell}{2},\frac{\ell+2}{2})}+\cC_{0,0(\frac{\ell+2}{2},\frac{\ell}{2})} \\
&
+\cC_{0,0(\frac{\ell+4}{2},\frac{\ell}{2})}
+\cA^{\Delta}_{0,0(\frac{\ell}{2},\frac{\ell}{2})}
+\cA^{\Delta}_{0,0(\frac{\ell+2}{2},\frac{\ell}{2})}
+\cA^{\Delta}_{0,0(\frac{\ell+4}{2},\frac{\ell}{2})}
+\textrm{c.c.}\, .  \label{JJOPE}
\end{split}
\end{align}
The selection rules determine the multiplets that contribute to the superconformal partial wave expansions of the four-point correlator $\langle JJJJ\rangle$.
The multiplets $\hat{\cC}_{0(\frac{\ell}{2},\frac{\ell}{2})}$ contain conserved higher spin currents for $\ell\geq1$ so they will not appear in interacting theories except for $\ell=0$. To obtain the general formula for the superconformal partial waves, the most challenge parts are that of the long multiplets $\cA^{\Delta}_{0,0(\frac{\ell}{2},\frac{\ell}{2})}$ and $\cA^{\Delta}_{0,0(\frac{\ell+2}{2},\frac{\ell}{2})}$, which involve rather complex structures. The multiplets $\hat{\cC}_{1(\frac{\ell}{2},\frac{\ell}{2})}$ arise from the long multiplet $\cA^{\Delta}_{0,0(\frac{\ell}{2},\frac{\ell}{2})}$ when $\Delta$ saturates the unitary bound. For the $\cA^{\Delta}_{0,0(\frac{\ell+4}{2},\frac{\ell}{2})}$ and extra $\cC$ multiplets, their superconformal blocks contain only one conformal block with undetermined overall coefficient.

\section{$\cN=2$ superconformal invariants and tensor structures in superembedding space}

We aim to solve the three-point function (\ref{JJO}) in superembedding space, in which the coordinates transform linearly under the superconformal transformations.
In superembedding space, the superconformal invariants can be simply constructed by the scalar products of variables with correct homogeneity.
The local operators can be uplifted to superembedding space with manifest superconformal transformations, and
the superconformal correlators have compact forms. The superembedding formalism, however, is not a complete approach for superconformal theories with extended symmetry that it cannot describe
multiplets carrying nonabelian R-symmetry charges \cite{Fitzpatrick:2014oza}.
Luckily the stress-tensor multiplet in $\cN=2$ theories and the most relevant multiplets in the OPE (\ref{JJOPE}) are invariant under $SU(2)_R$.
In this work, we follow the notation and conventions in \cite{Fitzpatrick:2014oza, Khandker:2014mpa}.

In superembedding formalism, the fundamental elements are the (dual) supertwistors $Z_A (\bar{Z}^A)\in \mathbb{C}^{4|\cN}$:
\be
Z_A =
\begin{pmatrix}
Z_\alpha\\
Z^{\dot\alpha}\\
Z_i
\end{pmatrix}
~~~~~~
\bar Z^A =
\begin{pmatrix}
\bar Z^\alpha &~
\bar Z_{\dot\alpha} &~
\bar Z^i
\end{pmatrix},
\ee
which contain four bosonic components and $\cN$ fermionic components. The (dual) supertwistors $Z_A$ ($\bar Z^A$)
transform as (anti) fundamental of superconformal group $SU(2,2|\cN)$. Superspace is equivalent to the space spanned by a pair of supertwistors $Z_A^a$, $a\in\{1,2\}$ and a pair of dual supertwistors $\bar Z^{\dot{a}A}$, $\dot{a}\in\{1,2\}$ with constraint
\be
\bar Z^{\dot{a}A}Z_A^a=0, ~~~~a, \dot{a}\in\{1,2\}. \label{ZZb}
\ee
Besides, there is a gauge redundancy on the two-planes in supertwistor space which corresponds to a change of basis
\be
Z_A^a\sim Z_A^b M_b^a,  ~~~ M\in\mathrm{GL}(2,\mathbb{C}), \label{redun}
\ee
and similarly for the two-planes spanned by the dual supertwistors. By gauge fixing the redundancy group
$\mathrm{GL}(2,\mathbb{C})\times \mathrm{GL}(2,\mathbb{C})$, the space spanned by the (dual) supertwistors reduces to the ``Poincar\'e section"
\be
Z_A^a = \begin{pmatrix}
\de_\alpha{}^a\\
i\tx_+^{\dot\alpha a}\\
2\theta_i^a
\end{pmatrix},
~~~~~~~
\bar Z^{\dot a A} = \begin{pmatrix}
-i\tx_-^{\dot a \alpha} &~ \de^{\dot a}_{\dot \alpha} &~ 2\bar\theta^{\dot a\,i} \label{gaugefixing}
\end{pmatrix}.
\ee
The constraint (\ref{ZZb}) now gives exactly the chiral condition and the (dual) supertwistor $Z_A$ ($\bar Z^A$) is equivalent to the chiral (anti-chiral) coordinates in superspace. Here the ``Poincar\'e section" gauge fixing breaks the covariance of supertwistor and goes back to the classical superspace, so it will not be applied until the final steps in the computations. Instead, the $\mathrm{GL}(2,\mathbb{C})\times \mathrm{GL}(2,\mathbb{C})$ is partially fixed in another way that keeps the covariance of the formalism, the bi-supertwistors:
\be
\cX_{AB} \equiv Z^a_A Z^b_B\epsilon_{ab},\qquad \bar \cX^{AB}\equiv \bar Z^{\dot a A}\bar Z^{\dot b B}\epsilon_{\dot a \dot b}.
\ee
Apparently the bi-supertwistors are invariant under the $\mathrm{SL}(2,\mathbb{C})\times \mathrm{SL}(2,\mathbb{C})$, which fix the gauge redundancies (\ref{redun}) up to rescaling.
The bi-supertwistor satisfies the null condition
\be
\bar \cX^{AB}\cX_{BC} =0,
\ee
and graded antisymmetry
\be
\cX_{AB} =-(-1)^{p_Ap_B}\cX_{BA},
\ee
where $p_A=1$ for fermion components while vanishes for bosonic components.
More properties of the bi-supertwistors are given in \cite{Fitzpatrick:2014oza}. The superembedding space refers to the space
described by $(\cX,\bar\cX)$. However, sometimes it is quite helpful to go back to the supertwistor, as will be shown later.

By constructions, the scalar products of supertwistors are invariant under $SU(2,2|\cN)$ transformations up to an overall scaling. Therefore the superconformal invariants can be simply obtained by taking care about the scaling weights of each variables. However, when there are several superembedding coordinates contracted consecutively, which is a product of several matrixes in superembedding space, the results turn out to be quite obscure. In the next part, we develop a systematical approach to expand the products of bi-supertwistors in terms of the superconformal covariant variables in superspace.

\subsection{Correspondence between supertraces in superembedding space and superspace}
Superconformal invariants in superembedding space can be obtained from supertraces of products of $\cX$s and $\bar\cX$s.\footnote{Another approach is to contract the superembedding space indices with auxiliary fields. This will be employed later to construct the tensor structures.} However, when there are several $\cX$s and $\bar\cX$s involved in the supertrace, the results are quite obscure to be understood in terms of superspace variables even taking the ``Poincar\'e section" gauge fixing.\footnote{With two points the results can be obtained from two matrices product. With three points, the problem can be partially simplified by taking a special frame with $z_1\rightarrow0, ~z_2\rightarrow\infty$ \cite{Goldberger:2012xb, Li:pr}} In this section we
provide a method that translates the supertraces of bi-supertwistors into superconformal covariant variables in superspace. Interestingly, the problem has a rather simple and clear solution based on supertwistor formalism.

We show that the invariants in superembedding space can be expanded  in terms of superconformal covariant variables in superspace through following identity\footnote{Following the notation in \cite{Fitzpatrick:2014oza, Khandker:2014mpa}, hereinafter the bi-supertwistors $\cX_i$s and $\bar\cX_j$s are denoted by $i$ and $\bar j$ for simplicity.}:

\begin{large}
\be
\mathbf{\langle\bar{i}\,j\,\bar{k}\dots\bar{s}\,t\rangle=\mathrm{tr}(\tx_{\bar{i}j}\,
x_{\bar{k}j}\dots\tx_{\bar{s}t}\,x_{\bar{i}t})}. \label{lemma1}
\ee
\end{large}

The LHS in (\ref{lemma1}) gives the supertrace of bi-supertwistors, while the RHS is the trace in superspace. Superconformal variables $\mx_{\bar{i}j}$ and $\tx_{\bar{i}j}$ are defined in (\ref{xxbsuperspace}).

The $n$-point supertrace in (\ref{lemma1}) is hard to evaluate directly. To prove this identity, we go back to the supertwistors. The trick is that instead of contracting the  indices $(a, b)$, $(\dot{a}, \dot{b})$ to obtain (dual) bi-supertwistors,
the problem is drastically simplified by contracting the $SU(2,2|\cN)$ indices!
Specifically we use the identity
\be
\bar Z_{k}^{\dot{a}A}Z_{lA}^a=-\mathrm{i}\,\tx_{\bar k\,l}^{\dot{a}a}, \label{ZZbcontraction}
\ee
which is obtained by taking the ``Poincar\'e section" (\ref{gaugefixing}) of the supertwistors. With $k=l$, we are back to the chiral condition
$\tx_{\bar k\,k}^{\dot{\al}\al}=\tx^{\ad \al}_{k-} -  \tx^{\ad \al}_{k+} -4\mathrm{i}\, \theta^{\al}_{k\,i} \bar{\theta}^{\ad i}_k=0$, as expected.

Now let us consider the supertrace with two bi-supertwistors. From the identity (\ref{ZZbcontraction}), the supertrace turns into
\be
(\bar{i}\,j)^{A}_D=-\mathrm{i}\,\bar{Z}_i^{\dot{a}A}\epsilon_{\dot{a}\dot{b}}\,\tx_{\bar{i}j}^{\dot{b}b}\,\epsilon_{bc}\,
\,Z_{j\,D}^{c}, \label{ij}
\ee
which can be simply proved by rewriting coordinates $(\bar{i}, j)$ in terms of the supertwistors and recombining them through contractions of the $SU(2,2|\cN)$ indices (\ref{ZZbcontraction}). The supertrace of (\ref{ij}) gives the two-point invariant product\footnote{Note the convention used here differs from that used in embedding space (\ref{Embconvention}) by a rescaling factor $\cX\rightarrow\frac{1}{2}\cX$. This factor has no effect on the definition of the superconformal invariants. }
\be
\langle \bar{i}\,j\rangle=-2 x_{\bar{i}j}^2. \label{sij}
\ee
It is straightforward to generalize this formula to products with more bi-supertwistors. For instance, the three-point product reads
\be
(\bar{i}\,j\,\bar{k})^{AD}=(-\mathrm{i})^2\,\bar{Z}_i^{\dot{a}A}\epsilon_{\dot{a}\dot{b}}\,\tx_{\bar{i}j}^{\dot{b}b}\, \mx_{j\bar{k}\,b\dot{c}}
\,\bar{Z}_k^{\dot{c}D}. \label{ijk}
\ee
Given $k=i$, the formula (\ref{ijk}) becomes
\be
(\bar{i}\,j\,\bar{i})^{AD}=-x_{\bar{i}j}^2\,\bar i^{AD},
\ee
which agrees with the result in \cite{Fitzpatrick:2014oza}.

The formulas (\ref{ij},\ref{ijk}) suggest that the $n$ point products in superembedding space are essentially the contractions of superconformal covariant variables $x_{\bar{i}j}$.  The identity (\ref{lemma1}) follows this conclusion and is a trivial generalization of the form (\ref{sij}).

\subsection{Superconformal invariants in superembedding space}
Superconformal invariants are constructed from the supertraces of products with n points in superembedding space.
The superconformal invariant $\ub$ (\ref{ub}) consists of 2-traces from three points $(1,\,2,\,0)$. A nilpotent superconformal invariant is obtained from $\ub$ by getting rid of its constant part \cite{Goldberger:2012xb, Khandker:2014mpa}
\be
\zb=\frac{\langle\bar{1}2\,\rangle\langle\bar{2}0\rangle\langle\bar{0}1\rangle
-\langle\bar{2}1\rangle\langle\bar{1}0\rangle\langle\bar{0}2\rangle}
{\langle\bar{1}2\rangle\langle\bar{2}0\rangle\langle\bar{0}1\rangle
+\langle\bar{2}1\rangle\langle\bar{1}0\rangle\langle\bar{0}2\rangle}. \label{zb}
\ee
Note that $\zb$ is antisymmetric under exchange $1\leftrightarrow2$, $\zb^3=0$ for $\cN=1$ theories
and $\zb^5=0$ for $\cN=2$ theories.
Besides, we can also construct the superconformal invariants with supertraces involving more points.
It can be shown that the next supertrace with new independent structure is $\langle \bar{1}2\bar{0}1\bar{2}0\rangle$.
All other contractions from the three points either vanish or degenerate to the 2-traces. With this new 6-traces we can build another superconformal invariant
\be
\wb^\prime=\frac{4~\langle \bar{1}2\,\bar{0}1\bar{2}\,0\rangle}
{\left(\langle \bar{1}2\rangle\langle \bar{2}1\rangle \langle \bar{1}0\rangle\langle \bar{0}1\rangle
\langle \bar{0}2\rangle \langle \bar{2}0\rangle\right)^{\frac{1}{2}}},
\ee
which is invariant under the exchange $1\leftrightarrow2$. Using the identity (\ref{lemma1}) it is clear that above formula gives the same result
as in (\ref{wb}). In a non-supersymmetric theory with $\te_i=\teb^i=0$,  we have $\wb^\prime=-1$. Then the new nilpotent superconformal
invariant can be obtained after removing this constant part
\be
\wb=\wb^\prime+1.  \label{wbr}
\ee
As discussed previously, in $\cN=1$ theories, the superconformal invariants $\wb$ is proportional to $\zb^2$ so it does not give a new independent superconformal invariant. This is expected since we know there is only one independent superconformal invariant can be constructed from three points in $\cN=1$ superconformal theories. While for $\cN=2$ theories, there are two independent superconformal invariants from three points \cite{Park:1999pd}. In superembedding space they are given by $\zb$ and $\wb$. In \cite{Park:1999pd} it also shows that there are only two independent superconformal invariants from three points for any superconformal theories $\cN\geq2$.
Although in principle one can write down the irreducible supertraces of products with more points $(n>6)$ in superembedding space. They are just polynomials of invariants $\zb$/$\wb$ instead of independent variables.

\subsection{Superconformal tensor structures in superembedding space}
Like in embedding space, the auxiliary twistors $\cS_A, ~\bar{\cS}^A$ are introduced to absorb the spacetime indices of fields in superembedding space. The fields in superembedding space are subject to gauge redundancies. It is convenient to choose the gauge in which the auxiliary twistors are transverse and null
\be
\bar\cX\cS=\bar\cS\cX=\bar\cS\cS=0.
\ee
The tensor structures of three-point functions are written in terms of the auxiliary twistors.

From two points $(1,~2)$ one can construct two tensor structures \cite{Fitzpatrick:2014oza, Khandker:2014mpa}
\be
S\equiv\frac{\bar{\cS}1\bar{2}\cS}{\langle 1\bar{2}\rangle },~~~~~~ S_*=S|_{1\leftrightarrow2}\equiv\frac{\bar{\cS}2\bar{1}\cS}{\langle 2\bar{1}\rangle},
\ee
and
\be
S_{\pm}=\frac{1}{2}(S\pm S_{*}).
\ee
The symmetric structure $S_+$ is nilpotent and vanishes when $\te_i=0$.
In $\cN=1$ superembedding space, $S_+^2=0$, so the tensor structures terminate at order $S_+$; while in $\cN=2$ superembedding space, $S_+^3=0$ and we have tensor structures up to order $S_+^2$:
\bea
S_{-}^{\ell} & \sim & \frac{1}{2}\left(S^{\ell}+(-1)^{\ell}S_*^\ell\right),\label{Sp} \\
S_{+}\, S_{-}^{\ell-1} & = & \frac{1}{2\ell}\left(S^{\ell}-(-1)^{\ell}S_*^\ell\right) \label{Sm}, \\
S_{+}^2\, S_{-}^{\ell-2} & = & \frac{1}{4(\ell-1)}\left(S^{\ell}+S^{\ell-1}S_*
+(-1)^{\ell}S S_*^{\ell-1}+(-1)^{\ell}S_*^\ell\right). \label{Sp2}
\eea
In (\ref{Sp}) it is not an exact identity. Actually the tensor structure $S_{+}^2\, S_{-}^{\ell-2}$ has the same ``parity" as $S_{-}^{\ell}$ and it also appears in the expansion of RHS of (\ref{Sp}). In the
 following part we denote the RHS of (\ref{Sp}) by the tensor structure $S_{-}^{\ell}$, indicating the parity $(-1)^\ell$ under $1\leftrightarrow2$ exchange. The term proportional to $S_{+}^2\, S_{-}^{\ell-2}$ is assumed implicitly.

 Similar to the superconformal invariant, new tensor structures can be constructed with more points in superembedding space. For instance, we have the following tensor structures from three points $(1,2,0)$
 \be
 H\equiv\frac{\cS\bar{1}2\bar{0}1\bar{2}\cS}{\langle1\bar{2}\rangle\langle2\bar{1}\rangle},~~~~ ~~~~~~
  \bar{H}\equiv\frac{\bar\cS 2\bar{1}0\bar{2}1\bar\cS}{\langle1\bar{2}\rangle\langle2\bar{1}\rangle}, \label{HHb}
 \ee
which contain only chiral or anti-chiral auxiliary supertwistors.
Apparently above tensor structures vanish when $\te_i=0$ so it only appears in supersymmetric theories.
Different from the superconformal invariant $\wb$, the tensor structures $H,
\bar H$ are odd under exchange $1\leftrightarrow2$ as they contain odd number of bi-supertwistors.
They are necessary ingredients to solve the three-point correlators of multiplets with mixed symmetry in the OPE selection rules (\ref{JJOPE}), such as $(\frac{\ell+2}{2},\frac{\ell}{2})$ and $(\frac{\ell+4}{2},\frac{\ell}{2})$.

Following the correspondence in (\ref{lemma1}), we can rewrite the tensor structures built in superembedding space in terms of the superconformal covariant variables in superspace $\mathbf{Z}_0 = (\Xbf, \Theta, \bar{\Theta})$.\footnote{Henceforth we will omit the subindex $0$ in the superspace variables for simplicity.}

To reproduce the tensor structures from the invariants in superembedding space, we remove the auxiliary supertwistors through the action $\left(\bar{\cX}\overrightarrow{\partial_{\bar{\cS}}}\right)^{\dot{\al}}$
or $\left(\overleftarrow{\partial_\cS} \cX\right)_{\alpha}$. Then we obtain products of pure bi-supertwistors, which can be expanded in terms of the superspace variables $Z_3$ according to the rule in (\ref{ijk}). Take the tensor structure $S$ for example:
\bea
S\equiv\frac{\bar{\cS}1\bar{2}\cS}{\langle 1\bar{2}\rangle }
&\rightarrow& \frac{1}{\langle 1\bar{2}\rangle } (\bar{0}1\bar{2}0)^{\ad}_\al
\propto
\frac{1}{x_{1\bar{2}}^2} \bar{Z}_0^{\dot{a}\ad}\left(\epsilon \tx_{\bar{0}1}\, \mx_{1\bar{2}}\tx_{\bar{2}0}\epsilon\right)_{\dot{a}a} Z_{0\,\al}^a  \nonumber \\
& \propto &\frac{{\bar\Xbf}^{\ad}_{\al}}{\bar\Xbf^2},  \label{SinX}
\eea
where we have used the formula (\ref{X2}) and the Poincar\'e section of supertwistors
\be
Z_{0\,\al}^a=\delta_{\al}^a, ~~~~~~~~~~\bar{Z}_{0\,\ad}^{\dot{a}}=\delta_{\ad}^{\dot{a}}.
\ee
By exchanging $1\leftrightarrow2$, the LHS of (\ref{SinX}) gives tensor structure $S_*$ and superconformal
variables in the
RHS become chiral.

The tensor structures $H$ can  be expanded in superspace as follows
\bea
 H\equiv\frac{\cS\bar{1}2\bar{0}1\bar{2}\cS}{\langle1\bar{2}\rangle\langle2\bar{1}\rangle}
 &\rightarrow&
 \frac{1}{\langle1\bar{2}\rangle\langle2\bar{1}\rangle}(0\bar{1}2\bar{0}1\bar{2}0)_{(\al\,\b)}
 \propto  \frac{\Xbf_{\left(\al\right.\ad}\bar{\Xbf}^{\ad}_{\left.\b\right)}}{(\Xbf^2\bar{\Xbf}^2)^{\frac{1}{2}}}  \nonumber \\
 &\propto&  \frac{\Theta_{\left(\al\right.}^i\Xbf_{\left.\b\right)\,\ad}\bar{\Theta}_i^{\ad}}{(\Xbf^2\bar{\Xbf}^2)^{\frac{1}{2}}},
\eea
where the two chiral indices ($\al,\b$) are symmetrized. Besides, we have applied the chiral condition (\ref{Xchiral}) and also the fact that $(X\cdot X)_{(\al\b)}$ vanishes after symmetrizing the two indices.  The tensor structure $\bar H$ can be expanded similarly with two anti-chiral indices symmetrized.

\section{Three-point functions in superembedding space}

In this section, we solve the three-point function $\langle \cJ\cJ\cO\rangle$ in superembedding space. Here the multiplets $\cO$ are restricted to be invariant under the R-symmetry $SU(2)_R$. According to the OPE selection rules (\ref{JJOPE}),
this includes the long multiplets $\cA^{\Delta}_{0,0(\frac{\ell}{2},\frac{\ell}{2})}$, $\cA^{\Delta}_{0,0(\frac{\ell+2}{2},\frac{\ell}{2})}$ and $\cA^{\Delta}_{0,0(\frac{\ell+4}{2},\frac{\ell}{2})}$, as well as several $\cC$ type short multiplets. However, the $\cC$ type short multiplets are either disappear in interacting theories or only contain unique conformal block, except $\hat{\cC}_{0(\frac{\ell}{2},\frac{\ell}{2})}$, whose solutions also closely relate to these of long multiplets. We focus on the correlators involving in long multiplets.

In \cite{Liendo:2015ofa}, the three-point functions have been solved based on the procedure that, firstly write down the most general ansatz  that are consistent with superconformal symmetry and also the conservation equations (\ref{conseq}), then
applying the reflection symmetry $(z_1\leftrightarrow z_2)$ to fix the coefficients of each term appears in the ansatz.
While in superembedding space, these constraints are fulfilled in a more straightforward way.

The procedure to solve the three-point function $\langle\cJ\cJ\cO\rangle$ in superembedding space includes two steps:
\begin{itemize}
\item Write the most general ansatz consistent with the homogeneity and reflection symmetry.
\item Solve the coefficients by imposing the conservation equations (\ref{conseq}).
\end{itemize}
In $\cN=2$ superembedding space, a general superfield $\Phi$ with quantum numbers $(\Delta, j_1, j_2, R=0, r)$ satisfies the homogeneity
\be
\Phi(\la \cX,\bar \la \cX) = \la^{-q-j_1}\bar\la^{-\bar q-j_2}\Phi(\cX,\bar \cX),
\ee
where $q$ and $\bar q$ are the superconformal weights given by
\begin{equation}
q \equiv \frac{1}{2} \left(\Delta + \frac{3}{2}r\right), ~~~~~~ \bar{q} \equiv \frac{1}{2} \left(\Delta - \frac{3}{2}r\right).
\end{equation}
The correlators in superembedding space are consisted of superconformal invariants and subject to the homogeneity of the constituent fields. The superconformal invariants and tensor structures are constructed following the rules we discussed before, which can be decomposed into odd or even parts under the reflection transformation $(z_1\leftrightarrow z_2)$. The reflection symmetry is realized directly when writing down the general ansatz. Then we will apply the correspondences between the variables in superembedding space and superspace to impose the constraints from conservation equations (\ref{conseq}). From this procedure the coefficients in the ansatz can be fixed accordingly.

The fundamental elements to build three-point functions in superembedding space are the superconformal invariants $\zb^n$ with $n\in\{1,2,3,4\}$, $\wb$ and the tensor structures $S_{-}^{\ell}$, $S_{+}\, S_{-}^{\ell-1}$, $S_{+}^2\, S_{-}^{\ell-2}, H, \bar H$.
For the invariant $\wb$, due to the algebra of supercharges, it has several constraints
\be
2\wb \zb=\zb^3, ~~~2\wb \zb^2=\zb^4, ~~~4\wb^2=5\zb^4, ~~~ \wb^3=0.
\ee
There are more restrictions involving the invariants and tensor structures, which significantly reduce the possible terms in the general ansatz.

\subsection{$\cA^{\Delta}_{0,0(\frac{\ell}{2},\frac{\ell}{2})}$}
The general ansatz on the correlators are different for odd/even spin multiplets due to the constraint from reflection symmetry. We start for the multiplets with odd spin.

For odd $\ell$, the most general three-point functions consistent with superconformal symmetry and also the reflection symmetry are\footnote{In principle, one may also consider to introduce the tensor structures $H$ or $\bar H$ in the three-point functions (\ref{oddspin}) and (\ref{evenspin}). However, for the symmetric multiplets, $H$ and $\bar H$ always appear in pairs, in this case, they will not give new independent tensor structures.}
\be
\langle \cJ(1,\bar 1)\cJ(2,\bar 2)\cO(0,\bar 0)\rangle=\frac{S_{-}^{\ell}(\la_1\zb+\la_3\zb^3)+ S_{+}\, S_{-}^{\ell-1}(\la_0+\la_2\zb^2)}
{(\langle\bar{1}2\rangle\langle\bar{2}1\rangle)^{1-\frac{1}{4}(\Delta+\ell)}  \label{oddspin}
(\langle\bar{0}1\rangle\langle\bar{1}0\rangle\langle\bar{0}2\rangle\langle\bar{2}0\rangle)^{\frac{1}{4}(\Delta+\ell)}}.
\ee
There are extra combinations of invariants and tensor structures consistent with the superconformal symmetry and reflection symmetry, such as $S_{+}^2\, S_{-}^{\ell-2} \,\zb$, however, they actually do not give new independent terms. In particular,
the superconformal invariant $\wb$ does not appear in (\ref{oddspin}). The remaining restrictions on the general ansatz are from the conservation equations (\ref{conseq}), which requires that when expanded in terms of variables $\mathbf{Z}_0$, there are no
terms contain the variables $\Theta^{\al\,i}_{0}\Theta^{j}_{0\,\al} $
or $\bar{\Theta}_{0 \,\ad i} \bar{\Theta}^{\ad}_{0\,j}$. The coefficients $\la_i$ are fixed to
\bea
\vec{\la}_\cO = &\la_{\cO} & \left(1, \, -\frac{\Delta+\ell}{2\,(\Delta-2)},\,
\frac{(\Delta+\ell)(2\ell^2-2\ell-8+6\Delta+\ell\Delta-\Delta^2)}{8\,(\Delta-2)},\right. \nonumber \\
& &~~\left.-\frac{(\Delta+\ell)(5\ell^2-8+16\Delta+2\ell\Delta-3\Delta^2)}{48\,(\Delta-2)}   \right). \label{OddSolution}
\eea
The solutions are reminiscent of the results from $\cN=1$ theories \cite{Fortin:2011nq, Khandker:2014mpa}, where the three-point function with one independent
superconformal invariant is fixed up to an overall constant by conservation equations.
Expanding the three-point function (\ref{oddspin}) with above coefficients in terms of superconformal variables in superspace, we obtain the results consistent with these presented in  \cite{Liendo:2015ofa}. Here we have rescaled the coefficients by multiplying an overall factor $(4+\ell-\Delta)$ comparing with the expressions in \cite{Liendo:2015ofa}. Otherwise the coefficients admit a pole at $\Delta=4+\ell$ above the unitary bound, which is of course unphysical. In (\ref{OddSolution}) there is another pole at $\Delta=2$, nevertheless, it is below the unitary bounds of any multiplets with odd spin.

For even $\ell$, we have the most general ansatz from superconformal symmetry and reflection symmetry
\be
\langle \cJ(1,\bar 1)\cJ(2,\bar 2)\cO(0,\bar 0)\rangle=\frac{S_{-}^{\ell}(\la_0+\la_2\zb^2+\la_3\wb+\la_4\zb^4)+ S_{+}\, S_{-}^{\ell-1}\,\la_1\zb
+S_{+}^2\, S_{-}^{\ell-2} \la_5}
{(\langle\bar{1}2\rangle\langle\bar{2}1\rangle)^{1-\frac{1}{4}(\Delta+\ell)}  \label{evenspin}
(\langle\bar{0}1\rangle\langle\bar{1}0\rangle\langle\bar{0}2\rangle\langle\bar{2}0\rangle)^{\frac{1}{4}(\Delta+\ell)}}.
\ee
Here we have two independent nilpotent superconformal invariants in the three-point function. Because of the extra invariant $\wb$, coefficients in the three-point function cannot be fixed up to an overall constant from conservation equations. Instead, there are two independent solutions
\bea
\vec{\la}_\cO^{(1)} &= &\la_{\cO}^{(1)}  \left(1, \,\frac{(\Delta+\ell)(2-\Delta)}{2},  \, \frac{(\Delta+\ell)^2}{8},\, 0,\right. \nonumber \\
& &\hspace{-3mm}\left.
\frac{(\Delta+\ell)(64-3\ell^3+(\Delta-4)^2\Delta-\ell^2(5\Delta-8)-\ell(\Delta^2-48))}{384}, \,
\frac{(2+\ell-\Delta)(2-\ell-\Delta)}{2}   \right), \nonumber \\
~ \label{EvenSolution2}
\eea
and
\bea
\vec{\la}_\cO^{(2)} &= &\la_{\cO}^{(2)}  \left(0, \, \frac{(\Delta+\ell)(6+\ell-3\Delta)}{2},  \, \frac{(\Delta+\ell)(\ell+3\Delta)}{8},\, -\frac{(\Delta+\ell)(4+\ell-\Delta)}{4},\right. \nonumber \\
& &\left.
\frac{(\Delta+\ell)(3(4+3\Delta)-\ell(1-2\Delta+2\ell(3+\ell+\Delta)))}{96}, \,
\frac{3(\Delta-2)^2-2\ell-\ell^2}{2}  \right). \label{EvenSolution1}
\eea
The solutions given in (\ref{EvenSolution1}) vanish in non-supersymmetric theories. They are from the nilpotent structures with superconformal invariant $\wb$ and correspond to the solutions $c_{\cJ\cJ\cO}^{(1)}$ in \cite{Liendo:2015ofa}.
The solutions given in (\ref{EvenSolution2}) do not contain the term with $\wb$. Expanding the two solutions in terms of the variables in superspace, we can reproduce the results in \cite{Liendo:2015ofa}.\footnote{The solutions in (\ref{EvenSolution2}) are actually the linear superpositions of the two solutions in \cite{Liendo:2015ofa} in order to remove the $\wb$ dependence.} Here we have rescaled the solutions  by factors $(4+\ell-\Delta)$ for (\ref{EvenSolution1}) and $\ell$ for (\ref{EvenSolution2}) to remove the unphysical poles. In \cite{Kuzenko:1999pi} the three-point correlator $\langle\cJ\cJ\cJ\rangle$ has been studied in superspace, which admits two independent solutions with coefficients corresponding to the $a$ and $c$ central charges. The two independent coefficients $\la^{(i)}$ will appear in the superconformal blocks as well, and it is expected that through conformal bootstrap, more constraints on these coefficients will be uncovered.

\subsubsection*{Scalar multiplet $\cA^{\Delta}_{0,0(0,0)}$}
The three-point function $\langle\cJ\cJ\cO\rangle$ with a scalar multiplet $\cA^{\Delta}_{0,0(0,0)}$ is more subtle. For a general scalar multiplet with $\Delta>2$, as shown from the analysis in superspace, the  superconformal nilpotent structure (\ref{KuzenkoIdentity}) can not appear in the three-point function and we have only one independent OPE coefficient. This can also be seen from the two solutions (\ref{EvenSolution1})
and (\ref{EvenSolution1}) in superembedding space. Taking $\ell=0$ we expect they give the solutions for the three-point function with scalar multiplet, in which $\la_1=\la_5=0$ by definition. However, in both solutions, by taking $\ell=0$ the two coefficients do not vanish unless we have $\Delta=2$ as well. In another words, the solutions (\ref{EvenSolution2}) and (\ref{EvenSolution1}) do not work for multiplets with $(\Delta>2,\ell=0)$. Instead the unique solution is given by their linear superposition
\be
\vec{\la}_\cO=\vec{\la}_\cO^{(1)}+\vec{\la}_\cO^{(2)}, \label{scalarcoeff}
\ee
with a constraint on the two OPE coefficients
\be
\la_{\cO}^{(1)}=-3\la_{\cO}^{(2)}\equiv\la_{\cO}.
\ee
Now the coefficients in the three-point function with general scalar multiplet are
\be
\vec{\la}_\cO = \la_{\cO}  \left(1, \, 0,  \, 0,\,\frac{1}{12}\Delta(4-\Delta),\, \frac{1}{384} (\Delta -6) (\Delta -4) \Delta  (\Delta +2),\, 0\right), \label{ZeroSolution1}
\ee
in which the two redundant tensor structures vanish $\la_1=\la_5=0$.

When $\Delta=2$ the multiplet $\cA^{\Delta}_{0,0(0,0)}$ hits the unitary bound and splits in several short multiplets
\be
\cA^{\Delta=2}_{0,0(0,0)} = \hat{\cC}_{0(0,0)} +\cD_{1(0,0)} +\bar{\cD}_{1(0,0)}+\hat{\cB}_2, \label{splitJ}
\ee
while only the semi-short multiplet $\hat{\cC}_{0(0,0)}$ can appear in the $\cJ\times\cJ$ OPE and the corresponding three-point function is given in (\ref{JJJ}) in superspace.
In superembedding space this three-point function is given by (\ref{evenspin}), and the coefficients are given
in (\ref{EvenSolution1})
and (\ref{EvenSolution2}) with $(\Delta=2, \ell=0)$:
\bea
\vec{\la}_\cO^{(1)} &= & \frac{3 }{32 \pi ^6}(4 a-3 c) \left(1, \, 0,  \, \frac{1}{2},\, 0,\,
\frac{3}{8}, \,0 \right),  \nn\\
\vec{\la}_\cO^{(2)} &= &\frac{-1}{32 \pi ^6}(4 a-5 c)  \left(0, \, 0,  \, \frac{3}{2},\, -1, \,
\frac{5}{8}, \,
0  \right). \label{JJJac}
\eea

\subsection{$\cA^{\Delta}_{0,0(\frac{\ell+2}{2},\frac{\ell}{2})}$}

In superembedding space, there are $\ell+2$ auxiliary supertwistors $\cS$ and $\ell$ auxiliary dual supertwistors $\bar\cS$ to absorb the indices of the mixed symmetry operator $\cA^{\Delta}_{0,0(\frac{\ell+2}{2},\frac{\ell}{2})}$. To construct the three-point function from these auxiliary (dual) supertwistors, we need to use the tensor structure $H$ that only includes two $\cS$s, which has been defined in (\ref{HHb}).

For even $\ell$, the most general ansatz for the correlator reads
\be
\langle \cJ(1,\bar 1)\cJ(2,\bar 2)\cO(0,\bar 0)\rangle=H\frac{S_{-}^{\ell}(\la_1\zb+\la_3\zb^3)+ S_{+}\, S_{-}^{\ell-1}\,\la_2}
{(\langle\bar{1}2\rangle\langle\bar{2}1\rangle)^{1-\frac{1}{4}(\Delta+\ell+2)}  \label{mevenspin}
(\langle\bar{0}1\rangle\langle\bar{1}0\rangle\langle\bar{0}2\rangle\langle\bar{2}0\rangle)^{\frac{1}{4}(\Delta+\ell+2)}}.
\ee
From the conservation equations (\ref{conseq}) the coefficients are fixed to
\be
\vec{\la}=\la_{\cO}  \left(1, ~~-\frac{2(\Delta-2)}{\Delta+\ell+2},~~ 0\right).  \label{mEvenSolution}
\ee
Remarkably, above results, both the tensor structures (except $H$) and the coefficients are almost the same (up to a shift $\ell\rightarrow\ell+2$) as the solutions of the three-point correlator $\langle\cJ\cJ\cO\rangle$ in $\cN=1$ theories \cite{Khandker:2014mpa}, where the multiplet $\cJ$ is the conserved current and $\cO$ is a general multiplet with odd spin.
Here the tensor structure $H$ relates to the Grassmann superconformal variables and the constraints from conservation equations on the rest part of the three-point function performs as the $\cN=1$ theories. Nevertheless, the superconformal blocks for the multiplets $\cA^{\Delta}_{0,0(\frac{\ell+2}{2},\frac{\ell}{2})}$ are still different from those in the $\cN=1$ theories, for the reasons will be explained in the next section.

For odd $\ell$, the most general ansatz for the correlator reads
\be
\langle \cJ(1,\bar 1)\cJ(2,\bar 2)\cO(0,\bar 0)\rangle=H\frac{S_{-}^{\ell}(\la_0+\la_2\zb^2)+ S_{+}\, S_{-}^{\ell-1}\,\la_1\zb}
{(\langle\bar{1}2\rangle\langle\bar{2}1\rangle)^{1-\frac{1}{4}(\Delta+\ell+2)}  \label{moddspin}
(\langle\bar{0}1\rangle\langle\bar{1}0\rangle\langle\bar{0}2\rangle\langle\bar{2}0\rangle)^{\frac{1}{4}(\Delta+\ell+2)}}.
\ee
We have solutions of the coefficients
\be
\vec{\la}=\la_{\cO}  \left(1,~~ 0, ~~\frac{1}{8}(6+\ell-\Delta)(2+\ell+\Delta)\right).  \label{mOddSolution}
\ee
Again, the solutions (\ref{mOddSolution}) are similar to the coefficients in $\cN=1$ correlators $\langle\cJ\cJ\cO\rangle$, where $\cO$ is a multiplet with even spin $\ell+2$.

\subsection{$\cA^{\Delta}_{0,0(\frac{\ell+4}{2},\frac{\ell}{2})}$}

For this operator, we have four more auxiliary supertwistors and there are two tensor structures $H$ in the three-point function. Therefore up to quadratic order of Grassmann variables $\Theta^i_\al$ or $\bar\Theta_i^{\ad}$, there is only one possible tensor structure in the three-point function
\be
\langle \cJ(1,\bar 1)\cJ(2,\bar 2)\cO(0,\bar 0)\rangle=\frac{\la_{\cO}\, H^2\,S_{-}^{\ell}+\cdots}
{(\langle\bar{1}2\rangle\langle\bar{2}1\rangle)^{1-\frac{1}{4}(\Delta+\ell+4)}  \label{m4spin}
(\langle\bar{0}1\rangle\langle\bar{1}0\rangle\langle\bar{0}2\rangle\langle\bar{2}0\rangle)^{\frac{1}{4}(\Delta+\ell+4)}}.
\ee
The reflection symmetry enforces any correlators with odd $\ell$  vanishing. There are extra terms depending on superconformal invariants $\zb, \wb$ to cancel higher order terms of $\Theta^i_\al$, $\bar\Theta_i^{\ad}$, however, they are irrelevant to the superconformal block analysis, as will be shown later.

\section{Superconformal partial waves}

In this section, we compute the superconformal partial waves $\cW_\cO$ of the four-point correlator $\langle J(1,\bar 1)J(2,\bar 2)J(3,\bar 3)J(4,\bar 4)\rangle$, in which the external operator $J$ is the superconformal primary of the stress-tensor multiplet $\cJ(x,\te,\teb)$
\be
J(x)=\left.\cJ(x,\te,\teb)\right|_{\te_i={\teb}^i=0}.
\ee
This is equivalent to set the fermionic components of external bi-supertwistors $i, \bar i$ to zero. Following the method developed in \cite{Fitzpatrick:2014oza, Khandker:2014mpa}, we will use the supershadow formalism to compute the superconformal partial wave $\cW_\cO$. The idea is to construct a non-local projector operator $\left|\cO\right|$ based on a multiplet $\cO$ and its shadow $\tl\cO$, which projects  the four-point correlator $\langle JJJJ\rangle$ onto its specific part corresponding to the exchange of the multiplet $\cO$.

Given a general multiplet $\cO$ with superconformal weights $(q, \bar q)$ and spin $(j_1,j_2)$, its supershadow operator is defined through
\be
\tl{\mathcal{O}}(1,\bar{1},\cS,\bar{\cS})\equiv\int D[2,\bar{2}]\frac{{\mathcal{O}^\dag}(2,\bar2,2\bar{\cS},\bar{2}\cS)}{\langle 1\bar{2}\rangle ^{2-\cN-q+j_1}\langle \bar{1}2\rangle ^{2-\cN-\bar{q}+j_2}},
\label{shadow}
\ee
where $D[2,\bar{2}]$ gives the superconformal measure in superembedding space and $\cO^\dagger$ with spin $(j_2,j_1)$ is the Lorentz conjugate of $\cO$. The supershadow operator $\tl\cO$ is from non-local linear transformation of multiplet $\cO^\dagger$, so it is expected that $\tl\cO$ shares the same Lorentz indices of $\cO^\dagger$. One the other hand, according to the definition (\ref{shadow}), its homogeneity in superembedding variables is
\be
\tl{\mathcal{O}}(\la\cX,\bar{\la}\bar{\cX},a\cS,\bar{a}\bar{\cS})=\la^{-(2-\cN-q+j_1)}\bar\la^{-(2-\cN-\bar{q}+j_2)}
a^{2j_1}\bar{a}^{2j_2}
\tl{\mathcal{O}}(\cX,\bar\cX,\cS,\bar{\cS}),  \label{homog}
\ee
 which performs as a superconformal multiplet with superconformal weights $(2-\cN-q,2-\cN-\bar q)$ and spin $(j_1, j_2)$ instead of spin $(j_2, j_1)$! The superconformal correlators with supershadow operators are determined by the homogeneity (\ref{homog}) with quantum numbers $(j_1, j_2, 2-\cN-q,2-\cN-\bar q)$.

A conformal projector can be constructed from a pair of operator $\cO$
and its shadow
\begin{equation}
\left|\mathcal{O}\right|= \left. \frac{1}{(2j_1)!^{2}(2j_2)!^{2}}\int D[0,\bar{0}]~|\mathcal{O}(0,\bar{0},\cS,\bar{\cS})\rangle \left(\overleftarrow{\partial_{\cS}}0\overrightarrow{\partial_{\cT}}\right)^{2j_1}
\left(\overleftarrow{\partial_{\bar{\cS}}}\bar{0}\overrightarrow{\partial_{\bar{\cT}}}\right)^{2j_2}\langle \tl{\mathcal{O}}(0,\bar{0},\cT,\bar{\cT})| \hspace{2mm} \right|_M,
\label{Projector}
\end{equation}
which projects the four-point correlator to the superconformal partial wave corresponding to the multiplet $\cO$
\begin{equation}
\mathcal{W}_{\cO}\propto  \langle \cJ\cJ \left|\cO(0,\bar0)\right| \cJ\cJ \rangle \sim \int D[0,\bar{0}]~\langle\cJ\cJ\cO\rangle \overleftrightarrow{\mathcal{D}}_{j_1,j_2} \langle\tl\cO\cJ\cJ\rangle,
\end{equation}
where
\be
\overleftrightarrow{\mathcal{D}}_{j_1,j_2}\equiv\frac{1}{(2j_1)!^2(2j_2)!^2}(\partial_{\cS}0\partial_{\cT})^{2j_1}
(\partial_{\bar{\cS}}\bar{0}\partial_{\bar{\cT}})^{2j_2}.  \label{DL}
\ee
Above superintegrand contains two three-point functions for the exchanged operator and its supershadow.
In our case, we have solved the three-point functions $\langle\cJ\cJ\cO\rangle$ for long multiplets with general spins.
The correlators with supershadow operators are given by the same formulas with proper quantum numbers.
Since we focus on the lowest components of the external operators, the external Grassmann variables are set to zero. The
tensor structures appearing in the three-point functions, as well as the integrand will be simplified. The superconformal integral has been analyzed in \cite{Fitzpatrick:2014oza} for general $\cN$. The original integrals seem to be involved technical problems, nevertheless, it is shown that a superconformal integral can be decomposed into the conformal integrals, whose general solutions have been provided in \cite{SimmonsDuffin:2012uy}.

To solve the supershadow three-point correlator $\langle\tl\cO(0,\bar0,\cT,\bar\cT)\cJ(3,\bar3)\cJ(4,\bar4)\rangle$, we will use the tensor structures $T, \bar T$ and superconformal invariants $\tzb, \twb$ constructed from coordinates $(0, 3,4)$ and their duals. Their definitions are analogously to those from points $(1,2,0)$:
\be
\{T,\bar T, \tzb, \twb\}=\{S,\bar S, \zb, \wb\}|_{\{1,2,\cS,\bar\cS\}\rightarrow\{3,4,\cT,\bar\cT\}}.
\ee
After setting the external Grassmann variables to be zero, the superconformal invariants $(\zb,\wb,\tzb,\twb)$, as well as the nilpotent tensor structures $(S_+, T_+)$ are actually proportional to the Grassmann variables $(\te_{0i},\teb_0^i)$ of the superspace coordinate $z_0$.

As we have shown before, the three-point functions have different tensor structures for odd and even spins. So the superconformal blocks are solved separately for odd and even spins. While for the multiplets $\cA^{\Delta}_{0,0(\frac{\ell+4}{2},\frac{\ell}{2})}$, only those with even $\ell$ have non-zero three-point function and the superconformal blocks are actually equivalent to non-supersymmetric case.

\subsection{$\cA^{\Delta}_{0,0(\frac{\ell}{2},\frac{\ell}{2})}$ with odd $\ell$}

The superconformal partial waves are obtained by inserting the projector $|\cO|$ in the four-point correlator
\bea
\mathcal{W}_{\cO}&\propto&  \langle \cJ(1,\bar1)\cJ(2,\bar2) \left|\cO(0,\bar0)\right| \cJ(3,\bar3)\cJ(4,\bar4) \rangle \nonumber\\
&\propto& \int D[0,\bar{0}]~\langle\cJ(1,\bar1)\cJ(2,\bar2)\cO(0,\bar0,\cS,\bar\cS)\rangle \overleftrightarrow{\mathcal{D}}_\ell \langle\tl\cO(0,\bar0,\cT,\bar\cT)\cJ(3,\bar3)\cJ(4,\bar4)\rangle. \label{WO1}
\eea
The three-point functions of multiplets $\cA^{\Delta}_{0,0(\frac{\ell}{2},\frac{\ell}{2})}$ with odd spin are given in (\ref{oddspin}). By replacing the scaling dimension $\Delta\rightarrow -\Delta$ in (\ref{oddspin}) we obtain the three-point functions of the supershadow operators with OPE coefficients ${\tl\la}_i$.
Applying the results to (\ref{WO1}) we get
\bea
\mathcal{W}_{\mathcal{O}}&\propto& \frac{1}{(\<1\bar{2}\>\<2\bar{1}\>)^{1-\frac{1}{4}(\ell+\Delta)}(\<3\bar{4}\>\<4\bar{3}\>)^{1-\frac{1}{4}(\ell-\Delta)}}
\times  \nonumber\\
& &  \int D[0,\bar{0}]
\frac{\mathcal{N}_{\ell}^{\mathrm{full}}}
{(\<1\bar{0}\> \<2\bar{0}\>\<0\bar{1}\> \<0\bar{2}\>)^{\frac{1}{4}(\ell+\Delta)} (\<3\bar{0}\> \<4\bar{0}\>\<0\bar{3}\> \<0\bar{4}\>)^{\frac{1}{4}(\ell-\Delta)}},  \label{oddpwave}
\eea
where
\bea
\mathcal{N}_{\ell}^{\mathrm{full}} &=&
\Sa\Dh\Ta(\la_1{\tl\la}_1\zb\tzb+\la_1\tl\la_3 \zb\tzb^3+\la_3\tl\la_1 \zb^3\tzb) \nn\\
&&+\Sa\Dh\Tb(\la_1{\tl\la}_0\zb    +\la_1\tl\la_2 \zb\tzb^2+\la_3\tl\la_0 \zb^3)  \nn\\
&&+\Sb\Dh\Ta(\la_0{\tl\la}_1\tzb   +\la_2\tl\la_1 \zb^2\tzb+\la_0\tl\la_3 \tzb^3)  \nn\\
&&+\Sb\Dh\Tb(\la_0{\tl\la}_0       +\la_0\tl\la_2 \tzb^2 +\la_2\tl\la_0 \zb^2).\label{Nodd}
\eea
Here we have ignored the higher order terms that are vanishing when setting the external Grassmann variables to zero. Note that in (\ref{Nodd}) the cubic and quartic terms of nilpotent variables $(\zb,\tzb,\wb,\twb,S_+,T_+)$ are new for $\cN=2$ theories.
The OPE coefficients $\la_i$ are given in (\ref{OddSolution}). The coefficients $\tl\la_i$ are from the three-point functions with supershadow operator $\tl\cO$, and they are given by the same form in (\ref{OddSolution}) with a replacement $\Delta\rightarrow-\Delta$.

After setting the external Grassmann variables to zero, the external bi-supertwistors $\cX, \bar\cX$ degenerate to the bi-twistors, which are $4\times4$ antisymmetric matrices $X_{\al\b}$, $\bar X^{\al\b}$ with twistor indices $\al,\b\in\{1,2,3,4\}$. The bi-twistors $X_{\al\b}$ are equivalent to vectors $X_m$  of the conformal group $SO(4,2)\simeq SU(2,2)$ :
 \be
 X^{\al\b}=\frac{1}{2}X_m\Gamma^{m\al\b},~~~~~~X_{\al\b}=\frac{1}{2}X_m\tl\Gamma_{m\al\b}. \label{Embconvention}
 \ee
 The supertraces of external superembedding variables turn into traces of bi-twistors
\be
\langle \bar i j\rangle\rightarrow -X_i \cdot X_j=\frac{1}{2}X_{ij}.
\ee
The fermionic components of superembedding coordinates $(\cX_0, \bar\cX_0)$ are still important for our analysis since it is the full-fledged multiplet $\cO$ contributing on the superconformal partial wave.

For the superconformal integration (\ref{oddpwave}), it is convenient to integrate over the fermionic components of $(0,\bar0)$ at first, after which we obtain an integration in the embedding space and the superconformal partial wave becomes
\begin{equation}
\mathcal{\left.W_{O}\right|}_{\theta_{\mathrm{ext}}=0}\propto\frac{1}{X_{12}^{2-\frac{1}{2}(\ell+\Delta)}
X_{34}^{2-\frac{1}{2}(\ell-\Delta)}}\int D^{4}X_{0}\left.\partial_{\bar{0}}^{4}\frac{\cN_{\ell}^{\mathrm{full}}}{D_{\ell}}\right|_{\bar{0}=0},
\label{oddpwave1}
\end{equation}
and $D_{\ell}$ denotes the products of supertraces containing superembedding coordinates $0$ or $\bar0$
\begin{equation}
 D_{\ell}\equiv (X_{1\bar{0}}X_{01}X_{2\bar{0}}X_{02})^{\frac{1}{4}(\ell+\Delta)}
 (X_{3\bar{0}}X_{03}X_{4\bar{0}}X_{04})^{\frac{1}{4}(\ell-\Delta)}. \label{Dl}
\end{equation}
Note in (\ref{oddpwave1}) we have partial derivatives $\partial_{\bar{0}}$ to the order four, which appear from the integration over fermionic components of the (dual) supertwistors $(Z_I^a, \bar Z^{\dot a I})$, with $I\in\{1,2\}$ in $\cN=2$ superembedding space.

The factor $\cN_\ell^{\mathrm{full}}$ in (\ref{oddpwave}) contains the elementary tensor structure from four points
\be
\cN_{\ell}\equiv(\bar{\cS}1\bar{2}\cS)^{\ell}\overleftrightarrow{\mathcal{D}}_{\ell}(\bar{\cT}3\bar{4}\cT)^{\ell}
=\frac{1}{(\ell!)^2}(\pd_{\cS}0\pd_\cT)^\ell(\cS\bar2 1\bar0 3\bar4\cT)^\ell,
\label{Nl}
\ee
and their coordinate exchanges, in terms of which the four-point tensor structures can be expanded as
\begin{eqnarray}
S_{-}^{\ell}\overleftrightarrow{\mathcal{D}_{\ell}}T_{-}^{\ell}&=&\frac{\cN_{\ell}}{4\langle 1\bar{2}\rangle ^{\ell}\langle 3\bar{4}\rangle ^{\ell}}+(-1)^{\ell}(1\leftrightarrow2)+(-1)^{\ell}(3\leftrightarrow4),  \\
S_{-}^{\ell}\overleftrightarrow{\mathcal{D}_{\ell}}T_{+}T_{-}^{\ell-1}&=&\frac{\cN_{\ell}}{4\ell\langle 1\bar{2}\rangle ^{\ell}\langle 3\bar{4}\rangle ^{\ell}}+(-1)^{\ell}(1\leftrightarrow2)-(-1)^{\ell}(3\leftrightarrow4),  \\
S_{+}S_{-}^{\ell-1}\overleftrightarrow{\mathcal{D}_{\ell}}T_{-}^{\ell}&=&\frac{\cN_{\ell}}{4\ell\langle 1\bar{2}\rangle ^{\ell}\langle 3\bar{4}\rangle ^{\ell}}-(-1)^{\ell}(1\leftrightarrow2)+(-1)^{\ell}(3\leftrightarrow4),   \\
S_{+}S_{-}^{\ell-1}\overleftrightarrow{\mathcal{D}_{\ell}}T_{+}T_{-}^{\ell-1}&=&\frac{\cN_{\ell}}{4\ell\langle 1\bar{2}\rangle ^{\ell}\langle 3\bar{4}\rangle ^{\ell}}-(-1)^{\ell}(1\leftrightarrow2)-(-1)^{\ell}(3\leftrightarrow4).
\label{Symm}
\end{eqnarray}
So far the formulas on the tensor structures remain the same as for the $\cN=1$ theories. New tensor structures for $\cN=2$ theories will appear in  $\cN_\ell^{\mathrm{full}}$ for multiplet with even spin.

By fixing all the fermionic variables  to zero the tensor structure $\cN_{\ell}$ becomes
\begin{equation}
\cN_\ell|_{\te_{\mathrm{ext}}=0}\equiv N_{\ell} = (-1)^\ell s_0^{\frac{\ell}{2}} C_{\ell}^{(1)}(t_{0}),
\end{equation}
where $C_{\ell}^{(\la)}(y )$ is the Gegenbauer polynomial and
\bea
t_{0}&\equiv &-\frac{X_{13}X_{20}X_{40}}{2\sqrt{X_{10}X_{20}X_{30}X_{40}X_{12}X_{34}}}-\left(1\leftrightarrow2\right)-\left(3\leftrightarrow4\right), \\
s_{0}&\equiv &\frac{1}{2^{12}}X_{10}X_{20}X_{30}X_{40}X_{12}X_{34}.
\eea
$N_\ell$ has parity $(-1)^\ell$ under the coordinate exchange $1\leftrightarrow2$ or $3\leftrightarrow4$.
More complex tensor structures will appear after taking partial derivatives $\partial_{\bar{0}}^4$ in (\ref{oddpwave1}). Most of them cannot be directly reduced to the Gegenbauer polynomials and their conformal integrations are in general unknown. We will resort to their recursion relations that correspond to the specific forms of the tensor structures.
There is an interesting correspondence between the algebra of the tensor structures in embedding space and the properties of
Hypergeometric functions, in terms of which the $4D$ conformal blocks are expressed analytically.

Details on the partial derivatives  and the conformal integrations of the tensor structures
are provided in the Appendices. After some calculations, we obtain the final results
\bea
\mathcal{\left.W_{O}\right|}_{\theta_{\mathrm{ext}}=0}&\propto&
\frac{1}{X_{12}^2
X_{34}^2}\la_\cO \la_{\tl\cO} \times \nn\\
&&\left(\frac{1}{(\Delta+\ell)(\Delta+\ell+2)} g_{\Delta+1,\ell+1}
+\frac{(\ell+2)^2(\Delta-\ell-2)}{\ell^2(\Delta-\ell)(\Delta+\ell)^2}g_{\Delta+1,\ell-1} \right. \nn\\
&&+\left.\frac{(\Delta-1)^2(\Delta-\ell-2)}{16(\Delta-\ell)(\Delta+1)^2(\Delta+\ell+1)(\Delta+\ell+3)}
 g_{\Delta+3,\ell+1} \right. \label{oddpwave2}\\
&&\left.+\frac{(\ell+2)^2(\Delta-1)^2(\Delta-\ell-2)^2}{16\ell^2(\Delta-\ell+1)(\Delta-\ell-1)
(\Delta+1)^2(\Delta+\ell)(\Delta+\ell+2)}g_{\Delta+3,\ell-1}
\right),
\nn
\eea
in which the functions $g_{\Delta,\ell}\equiv g_{\Delta,\ell}^{0,0}(u,v)$ are the classical conformal blocks for
identical external operators. The coefficient $\la_{\tl\cO}$ for the supershadow operator is proportional to
$\la_\cO$. This can be seen by inserting the supershadow transformation of $\cO$ (\ref{shadow}) in the three-point correlator $\langle\cJ\cJ\tl\cO\rangle\propto\la_{\tl\cO}$,
and the exact ratio can be obtained, in principle, by completing the superconformal integration (\ref{shadow})
for the three-point function $\langle\cJ\cJ\cO\rangle$. Nevertheless, in numerical bootstrap we only care about the positivity condition, the overall coefficient, as long as it is positive, is not important for the analysis. For analytical bootstrap, the overall constant
involves the crossing equation and it can be fixed from the singularities and its expansion in terms of superconformal blocks.

After removing the kinematic factors and the supershadow OPE coefficient, we obtain the $\cN=2$ superconformal blocks
\bea
\cG_{\Delta,\ell,\mathrm{odd}}^{\cN=2|JJ;JJ}&\propto& \la_\cO^2
\left(\frac{1}{(\Delta+\ell)(\Delta+\ell+2)} g_{\Delta+1,\ell+1}
+\frac{(\ell+2)^2(\Delta-\ell-2)}{\ell^2(\Delta-\ell)(\Delta+\ell)^2}g_{\Delta+1,\ell-1} \right. \nn\\
&&+\left.\frac{(\Delta-1)^2(\Delta-\ell-2)}{16(\Delta-\ell)(\Delta+1)^2(\Delta+\ell+1)(\Delta+\ell+3)}
 g_{\Delta+3,\ell+1} \right. \label{oddscb}\\
&&\left.+\frac{(\ell+2)^2(\Delta-1)^2(\Delta-\ell-2)^2}{16\ell^2(\Delta-\ell+1)(\Delta-\ell-1)
(\Delta+1)^2(\Delta+\ell)(\Delta+\ell+2)}g_{\Delta+3,\ell-1}
\right).
\nn
\eea
It would be interesting to compare above $\cN=2$ superconformal blocks with those of $\cN=1$ conserved currents, which have been solved in \cite{Fortin:2011nq, Kuzenko:1999pi}:
\be
\cG_{\Delta,\ell,\mathrm{odd}}^{\cN=1|JJ;JJ}= \la_\cO^2
\left(\frac{1}{(\Delta+\ell)(\Delta+\ell+1)} g_{\Delta+1,\ell+1} \label{n1odd}
+\frac{(\ell+2)^2(\Delta-\ell-2)}{\ell^2(\Delta-\ell-1)(\Delta+\ell)^2}g_{\Delta+1,\ell-1} \right)
\ee
for odd spin and
\be
\cG_{\Delta,\ell,\mathrm{even}}^{\cN=1|JJ;JJ}= \la_\cO^2  \label{n1even}
\left(g_{\Delta,\ell}
+\frac{(\Delta-2)^2(\Delta-\ell-2)(\Delta+\ell)}{16\Delta^2(\Delta+\ell+1)(\Delta-\ell-1)} g_{\Delta+2,\ell} \right)
\ee
for even spin. One can see that in (\ref{oddscb}), the coefficients of conformal blocks $g_{\Delta+1,\ell+1}$ and $g_{\Delta+1,\ell-1}$
are similar to the $\cN=1$ superconformal blocks with odd spin (\ref{n1odd}) while the ratio between
 the coefficients of $g_{\Delta+1,\ell+1}$ and $g_{\Delta+3,\ell+1}$ is close to the $\cN=1$ superconformal blocks with even spin (\ref{n1even}). Slightly differences appear in certain factors but these are expected, since with more supercharges, the decomposition of the $\cN=2$ superconformal multiplet to conformal multiplets will be different from the pure $\cN=1$ multiplets. The $\cN=2$ superconformal blocks in (\ref{oddscb}) are obtained through complicated
 computations. The final results are quite compact and organized in an interesting way. It would be very interesting to see if this property can be explained at algebraic level, and the complex computations presented here may not be essential for the superconformal blocks.

 \subsection{$\cA^{\Delta}_{0,0(\frac{\ell}{2},\frac{\ell}{2})}$ with even $\ell$}

 The superconformal blocks for even spin multiplets are more involved since the corresponding three-point functions
 have two independent solutions and contain more tensor structures.
 By inserting the conformal projector we have the superconformal partial wave of the even spin multiplet $\cA^{\Delta}_{0,0(\frac{\ell}{2},\frac{\ell}{2})}$:
 \bea
\mathcal{W}_{\cO} &\propto& \int D[0,\bar{0}]~\langle\cJ(1,\bar1)\cJ(2,\bar2)\cO(0,\bar0,\cS,\bar\cS)\rangle \overleftrightarrow{\mathcal{D}} \langle\tl\cO(0,\bar0,\cT,\bar\cT)\cJ(3,\bar3)\cJ(4,\bar4)\rangle \nn\\
&=& \int D[0,\bar{0}]~\frac{S_{-}^{\ell}(\la_0+\la_2\zb^2+\la_3\wb+\la_4\zb^4)+ S_{+}\, S_{-}^{\ell-1}\,\la_1\zb
+S_{+}^2\, S_{-}^{\ell-2} \la_5}
{(\langle\bar{1}2\rangle\langle\bar{2}1\rangle)^{1-\frac{1}{4}(\ell+\Delta)}
(\langle\bar{0}1\rangle\langle\bar{1}0\rangle\langle\bar{0}2\rangle\langle\bar{2}0\rangle)^{\frac{1}{4}(\ell+\Delta)}}
~\overleftrightarrow{\mathcal{D}}_\ell  \nn\\
&&~~~~~~~~~~\frac{T_{-}^{\ell}(\tla_0+\tla_2\tzb^2+\tla_3\twb+\tla_4\tzb^4)+ T_{+}\, T_{-}^{\ell-1}\,\tla_1\tzb
+T_{+}^2\, T_{-}^{\ell-2} \tla_5}
{(\langle\bar{3}4\rangle\langle\bar{4}3\rangle)^{1-\frac{1}{4}(\ell-\Delta)}
(\langle\bar{0}3\rangle\langle\bar{3}0\rangle\langle\bar{0}4\rangle\langle\bar{4}0\rangle)^{\frac{1}{4}(\ell-\Delta)}},
\label{Spwevenspin}
\eea
where we have applied the three-point functions (\ref{evenspin}) for the multiplet $\cA^{\Delta}_{0,0(\frac{\ell}{2},\frac{\ell}{2})}$ and its supershadow $\tl\cA^{-\Delta}_{0,0(\frac{\ell}{2},\frac{\ell}{2})}$. The OPE coefficients $\la_i$ and $\tl\la_i$ are the linear combinations of the two solutions in (\ref{EvenSolution1}) and (\ref{EvenSolution2})
\be
\la_i=\vec{\la}_\cO^{(1)}+\vec{\la}_\cO^{(2)}, ~~~~~~~~~~~\tl\la_i=\vec{\la}_{\tl\cO}^{(1)}+\vec{\la}_{\tl\cO}^{(2)}.
\ee
As for the odd spin case, the supershadow OPE coefficients $\tl\la_i$ are proportional to the OPE coefficients $\la_i$. However, both the OPE coefficients $\la_i$ and their supershadows include two independent overall coefficients $(\la_\cO^{(1)}, \la_\cO^{(2)})$ or $(\la_{\tl\cO}^{(1)}, \la_{\tl\cO}^{(2)})$ and the transformation from $\la_i$ to supershadow coefficients $\tl\la_i$ is described by a $2\times2$ matrix
\bea
\left( \begin{array}{c}
\la_{\tl\cO}^{(1)} \\
\la_{\tl\cO}^{(2)}
\end{array} \right)
& = &
\left( \begin{array}{c}
\cM(\Delta,\ell)^i_j
\end{array} \right)_{2\times2}
\cdot\left( \begin{array}{c}
\la_\cO^{(1)} \\
\la_\cO^{(2)}
\end{array} \right) .
\eea
The transformation matrix $\cM(\Delta,\ell)$ can be determined by faithfully computating the superconformal integration of the supershadow three-point function. Alternatively, as shown in \cite{Khandker:2014mpa, Li:2016chh}, it can also be fixed up to an overall constant from unitarity
\bea
&&\cM(\Delta,\ell)^i_j\propto \label{ShadowMatrix}\\
&& \left(
\begin{array}{cc}
(\Delta +\ell ) \left((3-\Delta) \Delta ^2+((\Delta -1) \Delta +2) \ell +4\right) & 4 \Delta  (\ell +2) (\ell +3)  \\
 2 \Delta  (\ell -\Delta ) (\Delta +\ell ) & (\ell -\Delta ) \left((\Delta +3) \Delta ^2+\left(\Delta ^2+\Delta +2\right) \ell +4\right) \\
\end{array}
\right).\nn
\eea
One can show that above supershadow transformation matrix satisfies the constraint
\be
\cM(-\Delta,\ell)  \cdot \cM(\Delta,\ell)\propto (\Delta-\ell-2)(\Delta-2)~ \mathbf{I}_{2\times2}, \label{MM0}
\ee
which is expected since it gives the original coefficients by applying the supershadow transformation twice.
Derivation of the supershadow transformation matrix is provided in Appendix D.
In (\ref{MM0}) we have explicitly shown the factor $(\Delta-\ell-2)(\Delta-2)$, which suggests the supershadow transformation becomes pathological at the unitary bound $\Delta=\ell+2$. In particular, for the stress-tensor multiplet with $\Delta=2$, the factor gives second order zeros. We will discuss the origin of this factor and its effect later.

Setting the fermionic components of external coordinates to zero, and integrating out the Grassmann variables in $\cX_0$ and $\bar\cX_0$, we obtain a form with conformal integration in embedding space, like (\ref{oddpwave1}). The difference is the tensor structures $\cN_\ell^{\mathrm{full}}$. Now it has a more complex form
\bea
\mathcal{N}_{\ell}^{\mathrm{full}} &=&
\Sa\Dh\Ta(\la_0{\tl\la}_0+\la_2{\tl\la}_0\zb^2+\la_0{\tl\la}_2\tzb^2+\la_0\tl\la_3 \twb+\la_3\tl\la_0\wb
+\la_4{\tl\la}_0\zb^4+\la_0{\tl\la}_4\tzb^4 \nn\\
  &&~~~~~~~~~~~+\la_2\tl\la_3 \zb^2\twb+\la_3\tl\la_2\wb\tzb^2
    +\la_2{\tl\la}_2\zb^2\tzb^2+\la_3{\tl\la}_3\wb\twb)    \nn\\
&&+\Sa\Dh\Tb(\la_0{\tl\la}_1\tzb+\la_2\tl\la_1 \zb^2\tzb+\la_3\tl\la_1 \wb\tzb)  \nn\\
&&+\Sb\Dh\Ta(\la_1{\tl\la}_0\zb+\la_1\tl\la_2 \zb\tzb^2+\la_1\tl\la_3 \zb\twb)     \nn\\
&&+\Sb\Dh\Tb\la_1{\tl\la}_1\zb\tzb                                  \nn\\
&&+\Sa\Dh\Tc(\la_0+\la_2\zb^2+\la_3\wb)\tl\la_5                      \nn\\
&&+\Sc\Dh\Ta(\tl\la_0+\tl\la_2\tzb^2+\tl\la_3\twb)\la_5               \nn\\
&&+\Sc\Dh\Tb \tl\la_1\tzb+ \Sb\Dh\Tc \la_1\zb                   \nn\\
&&+\Sc\Dh\Tc \la_5\tl\la_5.                \label{Neven}
\eea
The last five terms in (\ref{Neven}) contain new tensor structures. The expansions of these tensor structures involve more elementary four-point tensor structures besides $\cN_\ell$:
\bea
\cM_{\ell}&\equiv&(\bar{\cS}1\bar{2}\cS)^{\ell-1}\bar{\cS}2\bar{1}\cS\overleftrightarrow{\mathcal{D}}
(\bar{\cT}3\bar{4}\cT)^{\ell}
=\frac{1}{(\ell!)^2}(\pd_{\cS}0\pd_\cT)^\ell(\cS\bar2 1\bar0 3\bar4\cT)^{\ell-1}\cS\bar1 2\bar0 3\bar4\cT, \\
\cL_{\ell}&\equiv&(\bar{\cS}1\bar{2}\cS)^{\ell}\overleftrightarrow{\mathcal{D}}(\bar{\cT}3\bar{4}\cT)^{\ell-1}
\bar{\cT}4\bar{3}\cT
=\frac{1}{(\ell!)^2}(\pd_{\cS}0\pd_\cT)^\ell(\cS\bar2 1\bar0 3\bar4\cT)^{\ell-1}\cS\bar2 1\bar0 4\bar3\cT, \\
\cK_{\ell}&\equiv&(\bar{\cS}1\bar{2}\cS)^{\ell-1}\bar{\cS}2\bar{1}\cS\overleftrightarrow{\mathcal{D}}
(\bar{\cT}3\bar{4}\cT)^{\ell-1}
\bar{\cT}4\bar{3}\cT =\frac{1}{\ell(\ell!)^2} (\pd_{\cS}0\pd_\cT)^\ell  \nn\\
&&((\cS\bar2 1\bar0 3\bar4\cT)^{\ell-1}\cS\bar1 2\bar0 4\bar3\cT
+(\ell-1)(\cS\bar2 1\bar0 3\bar4\cT)^{\ell-2}\cS\bar1 2\bar0 3\bar4\cT\cS\bar2 1\bar0 4\bar3\cT).
\eea
Note that after setting all the fermionic variables, including those from $\cX_0$ and $\bar\cX_0$ to zero, above tensor structures go back to $N_\ell$ up to a minus sign. However, when taking the partial derivatives $\partial_{\bar{0}}^{4}$,
these tensor structures perform differently in a subtle way.

The new tensor structures in (\ref{Neven}) can be expanded in terms of the elementary tensor structures $(\cN_\ell,\cM_\ell,\cL_\ell,\cK_\ell)$:
\begin{eqnarray}
\Sa\Dh\Tc&=&\frac{\cN_{\ell}+\cL_\ell}{8(\ell-1)\langle 1\bar{2}\rangle ^{\ell}\langle 3\bar{4}\rangle ^{\ell}}+(-1)^{\ell}(1\leftrightarrow2)+(-1)^{\ell}(3\leftrightarrow4),  \\
\Sc\Dh\Ta&=&\frac{\cN_{\ell}+\cM_\ell}{8(\ell-1)\langle 1\bar{2}\rangle ^{\ell}\langle 3\bar{4}\rangle ^{\ell}}+(-1)^{\ell}(1\leftrightarrow2)+(-1)^{\ell}(3\leftrightarrow4), \\
\Sb\Dh\Tc&=&\frac{\cN_{\ell}+\cL_\ell}{8\ell(\ell-1)\langle 1\bar{2}\rangle ^{\ell}\langle 3\bar{4}\rangle ^{\ell}}-(-1)^{\ell}(1\leftrightarrow2)+(-1)^{\ell}(3\leftrightarrow4), \\
\Sc\Dh\Tb&=&\frac{\cN_{\ell}+\cM_\ell}{8\ell(\ell-1)\langle 1\bar{2}\rangle ^{\ell}\langle 3\bar{4}\rangle ^{\ell}}+(-1)^{\ell}(1\leftrightarrow2)-(-1)^{\ell}(3\leftrightarrow4), \\
\Sc\Dh\Tc&=&\frac{\cN_{\ell}+\cM_\ell+\cL_\ell+\cK_\ell}{16(\ell-1)^2\langle 1\bar{2}\rangle ^{\ell}\langle 3\bar{4}\rangle ^{\ell}}+(-1)^{\ell}(1\leftrightarrow2)+(-1)^{\ell}(3\leftrightarrow4).
\label{Symm1}
\end{eqnarray}
Above tensor structures vanish in embedding space, which are expected for the nilpotent variables. Their
contributions on the superconformal partial wave turn into nontrivial after taking partial derivatives $\partial_{\bar{0}}$. More details on these computations are provided in the Appendices.

The next steps are the standard computations on the embedding space algebra and the conformal integrations.
The supershadow coefficients $\la_{\tl\cO}^{(i)}$ are rewritten in terms of the OPE coefficients
$\la_{\cO}^{(i)}$ using the supershadow transformation $\cM_{\Delta,\ell}$.
The final results of the superconformal blocks are
\be
\cG_{\Delta,\ell,\mathrm{even}}^{\cN=2|JJ;JJ}=a_0 g_{\Delta,\ell}+a_1 g_{\Delta+2,\ell+2}+a_2 g_{\Delta+2,\ell}+a_3 g_{\Delta+2,\ell-2}+a_4 g_{\Delta+4,\ell}, \label{N2evenblock}
\ee
where
\bea
a_0&=& \left(\la_\cO^{(1)}\right)^2, \\
a_1&=& \frac{\left((\ell+\Delta)\la_\cO^{(1)}+2(1+\ell+\Delta)\la_\cO^{(2)}\right)^2}{16(1+\ell+\Delta)(3+\ell+\Delta)}, \\
a_3&=& \frac{\left(\left(4(1-\Delta)+\ell(3+\ell)(2+\ell-\Delta)\right)\la_\cO^{(1)}+2(2+\ell)(3+\ell)(1+\ell-\Delta)\la_\cO^{(2)}\right)^2}
{16(\ell-1)^2\ell^2(\Delta-\ell+1)(\Delta-\ell-1)}, \\
a_4&=& \frac{\left((\ell+\Delta)(4+(3-\Delta)\Delta^2+\ell(2+(\Delta-1)\Delta))
\la_\cO^{(1)}+4(2+\ell)(3+\ell)\Delta\la_\cO^{(2)}\right)^2}
{256(\Delta-\ell+1)(\Delta-\ell-1)(\Delta+1)^2(\Delta+2)^2(\Delta+\ell+1)(\Delta+\ell+3)},
\eea
and
\bea
{\displaystyle{a_2} }&=&
\frac{1}{4 \ell^2(\Delta +1)^2  (\ell -\Delta ) (\Delta +\ell +2)} \times \nn\\
&&\left( (\la_\cO^{(1)})^2 (\Delta +\ell ) \left(4 (\Delta +2)+\Delta ^2 (\ell +4) \left(\ell ^2-2\right)+\Delta ^3 (-(\ell  (\ell +2)+4)) \right.\right.\nn\\
&&~~~~~~~~~\left.\left.+\Delta  \ell  (7 \ell +12)+\ell  (3 \ell  (\ell +2)+4)\right)
+2 \la_\cO^{(1)}\la_\cO^{(2)} (\ell +3) (\Delta +\ell )\times \right.\label{N2evena2}\\
&&~~~~~\left.\left(-4 (\Delta -1) (\Delta +1) (\Delta +2)+\Delta  (\Delta +5) \ell ^2+\left(\Delta  \left(-\Delta ^2+\Delta +14\right)+4\right) \ell \right) \right. \nn\\
&&\left.+4 (\la_\cO^{(2)})^2 (\ell +3)^2 \left(-(\Delta -1) \Delta  (\Delta +1) (\Delta +2)+\left(\Delta ^2+\Delta +1\right) \ell ^2+2 \left(\Delta ^2+\Delta +1\right) \ell \right)  \right).  \nn
\eea
Above coefficients $a_i$ are guaranteed to be non-negative for $i\in\{0,1,3,4\}$. This agrees with the unitarity, which requires that any independent conformal blocks appearing in the superconformal partial wave
$\cW_\cO$ should be positive. In contrast, the coefficient $a_2$ seems to be
disorganized and it is not clear that the whole expression is always positive. This problem
can be clarified by decomposing the $\cN=2$ multiplet into $\cN=1$ multiplets, from which
we can see that actually the coefficient $a_2$ contains several contributions from different $\cN=1$ multiplets.
Most of these contributions are restricted by the $\cN=1$ supersymmetry and the unitarity is satisfied term by term. More details on the $\cN=1$ decomposition will be given in section 6.

The specific expressions of the coefficients $a_i$ shown above rely on the definitions of the OPE coefficients $\vec\la_\cO^{(i)}$, given by (\ref{EvenSolution2}) and (\ref{EvenSolution1}) in our case. While one can always obtain different solutions from their linear superpositions, and the expressions of the coefficients $a_i$ will be modified accordingly.

For a general scalar long multiplet $\cA^{\Delta}_{0,0(0,0)}$ with $\Delta>2$, there is only one independent OPE coefficient and the superconformal blocks are given by (\ref{N2evenblock}) with an extra constraint on the OPE coefficients (\ref{scalarcoeff}). If the scaling dimension saturates the unitary bound $\Delta=2$, the multiplet degenerates to the stress-tensor multiplet and there are again two independent OPE coefficients.
The superconformal block $\cG_{\cJ}^{\cN=2|JJ;JJ}$ corresponding to the exchange of stress-tensor multiplet $\hat{C}_{0(0,0)}$
involves both $a$ and $c$ anomaly coefficients. The expression (\ref{N2evenblock}) is for general long multiplet.
One may consider to obtain the superconformal block  $\cG_{\cJ}^{\cN=2|JJ;JJ}$ by analytically continuing  (\ref{N2evenblock}) to $\Delta=2, \ell=0$ and also using the OPE coefficients
given in (\ref{JJJac}). However, the results obtained in this way should be treated carefully since the supershadow transformation for $\Delta=2, \ell=0$ is pathological with second order zeros (\ref{MM0}). There are unphysical terms arising from the analytical continuation.\footnote{I would like to thank Madalena Lemos for the discussion on this problem.}
It is not clear if there are unphysical terms in the analytical continuation $\Delta\rightarrow \ell+2$ of (\ref{N2evenblock}) for $\ell>0$. It would be interesting to see how they correspond to the superconformal blocks of semi-short multiplets $\hat{C}_{1(\frac{\ell}{2},\frac{\ell}{2})}$. We leave this problem for future study.

\subsection{$\cA^{\Delta}_{0,0(\frac{\ell+2}{2},\frac{\ell}{2})}$ with odd $\ell$}

The superconformal partial waves from exchange of the multiplets $\cA^{\Delta}_{0,0(\frac{\ell+2}{2},\frac{\ell}{2})}$ with odd $\ell$ read
\bea
{\cW}_{\cO}&=&\int D[0,\bar{0}]~\frac{H\, S_{-}^{\ell}(\la_1+\la_2 \zb^2)}
{(\langle\bar{1}2\rangle\langle\bar{2}1\rangle)^{1-\frac{1}{4}(2+\ell+\Delta)}
(\langle\bar{0}1\rangle\langle\bar{1}0\rangle\langle\bar{0}2\rangle\langle\bar{2}0\rangle)^{\frac{1}{4}(2+\ell+\Delta)}}
~\overleftrightarrow{\mathcal{D}}  \nn\\
&&~~~~~~~~~~\frac{\tl{H}\, T_{-}^{\ell}({\tl{\la}}_1+{\tl{\la}}_2\tzb^2)}
{(\langle\bar{3}4\rangle\langle\bar{4}3\rangle)^{1-\frac{1}{4}(2+\ell-\Delta)} \label{moddpwave}
(\langle\bar{0}3\rangle\langle\bar{3}0\rangle\langle\bar{0}4\rangle\langle\bar{4}0\rangle)^{\frac{1}{4}(2+\ell-\Delta)}},
\eea
where we have used the three-point functions $\langle\cJ\cJ \cO\rangle$ (\ref{moddspin}) for the operator $\cA^{\Delta}_{0,0(\frac{\ell+2}{2},\frac{\ell}{2})}$ and its supershadow ${\tl\cA}^{-\Delta}_{0,0(\frac{\ell}{2},\frac{\ell+2}{2})}$. The tensor structure $\tl H$ is given by
\be
\tl{H}\equiv H|_{\{1,2,\cS\}\rightarrow\{3,4,\cT\}} =\frac{\cT\bar{3}4\bar{0}3\bar{4}\cT}{\langle3\bar{4}\rangle\langle4\bar{3}\rangle}.
\ee
Note that for the supershadow operator ${\tl\cA}^{-\Delta}_{0,0(\frac{\ell}{2},\frac{\ell+2}{2})}$, its homogeneity in superembedding space performs like the operators with superconformal weight $q=\bar{q}=-\frac{\Delta}{2}$ and spin $(\frac{\ell+2}{2},\frac{\ell}{2})$. The partial derivative on the auxiliary supertwistors now becomes
\be
\overleftrightarrow{\mathcal{D}}\equiv\frac{1}{\ell!^2(\ell+2)!^2}(\partial_{\cS}0\partial_{\cT})^{\ell+2}
(\partial_{\bar{\cS}}\bar{0}\partial_{\bar{\cT}})^{\ell}.
\ee

Integrating out the fermionic components in (\ref{moddpwave}) we obtain an expression with conformal integration in embedding space
\begin{equation}
\mathcal{\left.W_{O}\right|}_{\theta_{\mathrm{ext}}=0}\propto\frac{1}{X_{12}^{2-\frac{1}{2}(2+\ell+\Delta)}
X_{34}^{2-\frac{1}{2}(2+\ell-\Delta)}}\int D^{4}X_{0}\left.\partial_{\bar{0}}^{4}\frac{\cN^{\mathrm{full}}}{D_{\ell+2}}\right|_{\bar{0}=0},
\label{moddpwave1}
\end{equation}
where the factor $D_{\ell+2}$ is given in (\ref{DL}) with shifted subindex $\ell\rightarrow\ell+2$.
 Moreover, the tensor structure
$\cN^{\mathrm{full}}$ turns into
\bea
\mathcal{N}_{\ell}^{\mathrm{full}} =
H\Sa\Dh \tl{H}\Ta
(\la_1{\tl\la}_1+\la_1\tl\la_2 \tzb^2+\la_2\tl\la_1 \zb^2),\label{Nmodd}
\eea
in which
\be
H S_{-}^{\ell}\overleftrightarrow{\mathcal{D}}\tl{H}T_{-}^{\ell}=
\frac{1}{(\ell+2)!^2}(\pd_{\cS}0\pd_\cT)^{\ell+2} H
\tl{H}\left(
\frac{(\cS\bar2 1\bar0 3\bar4\cT)^\ell}{4\langle 1\bar{2}\rangle ^{\ell}\langle 3\bar{4}\rangle ^{\ell}}+(-1)^{\ell}(1\leftrightarrow2)+(-1)^{\ell}(3\leftrightarrow4)\right). \label{TSmo}
\ee
The OPE coefficients $\la_i$ have been solved in (\ref{mOddSolution}).
The supershadow OPE coefficients $\tl\la_i$ also follow the solutions (\ref{mOddSolution}) with $\Delta\rightarrow-\Delta$.

The partial derivatives $\partial_{\bar{0}}^{4}$  are simplified for the mixed symmetry multiplets. The tensor structures $H$ and $\tl{H}$ in $\cN^{\mathrm{full}}$ are nilpotent, they vanish in embedding space unless acted by derivative $\partial_{\bar{0}}$. Therefore the conformal integrand in  (\ref{moddpwave1}) only includes two parts:
\be
\left.\partial_{\bar{0}}^{4}\frac{\cN^{\mathrm{full}}}{D_{\ell+2}}\right|_{\bar{0}=0}\propto
\left.\left(\partial_{\bar{0}}^{2}H\tl{H}\right)\left(\partial_{\bar{0}}^{2}\frac{\cN_\ell}{D_{\ell+2}}\right)
\right|_{\bar{0}=0}+
2\left.\left(\partial_{\bar{0}}^{A}\partial_{\bar{0}}^{B}H\tl{H}\right)
\left(\partial_{\bar{0}A}\partial_{\bar{0}B}\frac{\cN_\ell}{D_{\ell+2}}\right)\right|_{\bar{0}=0}. \label{HHf}
\ee
The first term reproduces the results of $\cN=1$ conserved currents with spin $\ell+1$. Specifically, we have
\bea
\left.\partial_{\bar{0}}^{A}H~\partial_{\bar{0}A}\tl{H}\right|_{\bar{0}=0}&=&
\frac{1}{4}S12\tl\Gamma^m 12S\times T34\tl\Gamma_m 34T
=-\frac{1}{4^3}X_{12}X_{34}S1\Gamma^m 2 S\times T\tl\Gamma_m 34T \nn\\
&=&\frac{2}{4^3}X_{12}X_{34}(S1T S234T-S2T S134T) \nn\\
&\rightarrow& -\frac{\ell(\ell+1)}{32}X_{12}X_{34} S21034T,   \label{HHt}
\eea
where the auxiliary twistors $S, T$ are the non-supersymmetric parts of the auxiliary supertwistors $\cS, \cT$.
From the first line to second line, we have applied the contraction of the $6D$ gamma matrices
\be
\tl\Gamma^{mAB}\Gamma_{mCD}=2(\delta^A_C\delta^B_D-\delta^A_D\delta^B_C).
\ee
In the last step, we have employed the fact that the tensor structure $SiT$ always leads to a factor $X_{0i}$ regardless of extra tensor structures.
The whole superconformal partial waves remains different from those of $\cN=1$ conserved currents due to the second term that is new for $\cN=2$ theories.

For the second term in (\ref{HHf}), there is another constraint
\bea
\left.\partial_{\bar{0}}^{A}H~\partial_{\bar{0}A} N_\ell\,\right|_{\bar{0}=0} &\propto& S12{\tl\Gamma}^m12S\times S21{\tl\Gamma}_m 34T \nn\\
&\propto& \epsilon_{ABCD} (S12)^A(S21)^B(S21)^C (T43)^D=0,
\eea
and similarly $\partial_{\bar{0}}^{A}\tl H~\partial_{\bar{0}A} N_\ell=0$. The non-vanishing contractions lead to a new tensor structure
\bea
\left(\partial_{\bar{0}}^{A}\partial_{\bar{0}}^{B}H\tl{H}\right)&&
\left.\left(\partial_{\bar{0}A}\partial_{\bar{0}B}\frac{1}{D_{\ell+2}}\right)\right|_{\bar{0}=0}\rightarrow \nn\\ &&\left(X_{1\bar{0}} T234T+ X_{2\bar{0}} T134T- X_{1\bar{0}} T243T- X_{2\bar{0}} T143T \right) \nn\\
&& \times \left(X_{3\bar{0}} S214S+ X_{4\bar{0}} S213S- X_{3\bar{0}} S124S- X_{4\bar{0}} S123S  \right).
\eea
Note it is different from the tensor structure $SPT\times SRT$ appears in the superconformal integration of symmetric long multiplet. Here we denote it as $SRS\times TPT$. Its recursion relation is provided in Appendix B and its conformal integration is given in Appendix C.

After the conformal integrations in embedding space, we obtain the superconformal blocks
\be
\cG_{\Delta,\ell+2|\ell,\mathrm{odd}}^{\cN=2|JJ;JJ}\propto \la_\cO^2 \left(g_{\Delta+1,\ell+1}+
\frac{(\Delta -2) (\Delta -1) (\Delta -\ell -2) (\Delta +\ell +2)}{16 (\Delta +1) (\Delta +2) (\Delta -\ell -1) (\Delta +\ell +3)}g_{\Delta+3,\ell+1}\right). \label{N2blockmodd}
\ee
 One can see that above superconformal blocks are similar to the $\cN=1$ conserved current superconformal blocks with exchanged even spin operator $(\Delta+1, \ell+1)$ (\ref{n1even}).

\subsection{$\cA^{\Delta}_{0,0(\frac{\ell+2}{2},\frac{\ell}{2})}$ with even $\ell$}

With even $\ell$, the superconformal partial waves corresponding to the multiplets $\cA^{\Delta}_{0,0(\frac{\ell+2}{2},\frac{\ell}{2})}$ become
\bea
{\cW}_{\cO}&=&\int D[0,\bar{0}]~\frac{H\, \left(S_{-}^{\ell}\la_1\zb+\Sb\la_2\right)}
{(\langle\bar{1}2\rangle\langle\bar{2}1\rangle)^{1-\frac{1}{4}(2+\ell+\Delta)}
(\langle\bar{0}1\rangle\langle\bar{1}0\rangle\langle\bar{0}2\rangle\langle\bar{2}0\rangle)^{\frac{1}{4}(2+\ell+\Delta)}}
~\overleftrightarrow{\mathcal{D}}  \nn\\
&&~~~~~~~~~~\frac{\tl{H}\, \left(T_{-}^{\ell}{\tl{\la}}_1\tzb+\Tb\tl\la_2  \right)}
{(\langle\bar{3}4\rangle\langle\bar{4}3\rangle)^{1-\frac{1}{4}(2+\ell-\Delta)} \label{mevenpwave}
(\langle\bar{0}3\rangle\langle\bar{3}0\rangle\langle\bar{0}4\rangle\langle\bar{4}0\rangle)^{\frac{1}{4}(2+\ell-\Delta)}}.
\eea
The OPE coefficients $\la_i$ and their supershadow $\tl\la_i$ have been solved in (\ref{mEvenSolution}).
By integrating out the fermionic variables we obtain the same expression  for the superconformal partial waves
as in (\ref{moddpwave1}). While the tensor structures $\cN^{\mathrm{full}}$ now turn into
\bea
\mathcal{N}_{\ell}^{\mathrm{full}} &=&
H\Sa\Dh \tl{H}\Ta\la_1{\tl\la}_1\zb\tzb + H\Sa\Dh \tl{H}\Tb\la_1{\tl\la}_2\zb  \nn\\
&&+ H\Sb\Dh \tl{H}\Ta\la_2{\tl\la}_1 \tzb+H\Sb\Dh \tl{H}\Tb\la_2{\tl\la}_2.\label{Nmeven}
\eea
The new tensor structures in above formula can be expanded as follows
\bea
H S_{-}^{\ell}\overleftrightarrow{\mathcal{D}_{\ell}}\tl{H}\Tb&=&
\frac{1}{(\ell+2)!^2} (\pd_{\cS}0\pd_\cT)^{\ell+2}\times \nn\\
&&H \tl{H}\left(
\frac{(\cS\bar2 1\bar0 3\bar4\cT)^\ell}{4\ell\langle 1\bar{2}\rangle ^{\ell}\langle 3\bar{4}\rangle ^{\ell}}+(-1)^{\ell}(1\leftrightarrow2)-(-1)^{\ell}(3\leftrightarrow4)\right), \nn\\
H \Sb\overleftrightarrow{\mathcal{D}_{\ell}}\tl{H}\Ta&=&
\frac{1}{(\ell+2)!^2} (\pd_{\cS}0\pd_\cT)^{\ell+2}\times\nn\\
&& H \tl{H}\left(
\frac{(\cS\bar2 1\bar0 3\bar4\cT)^\ell}{4\ell\langle 1\bar{2}\rangle ^{\ell}\langle 3\bar{4}\rangle ^{\ell}}-(-1)^{\ell}(1\leftrightarrow2)+(-1)^{\ell}(3\leftrightarrow4)\right), \nn\\
H \Sb\overleftrightarrow{\mathcal{D}_{\ell}}\tl{H}\Tb&=&
\frac{1}{(\ell+2)!^2} (\pd_{\cS}0\pd_\cT)^{\ell+2}\times\nn\\
&&H \tl{H}\left(
\frac{(\cS\bar2 1\bar0 3\bar4\cT)^\ell}{4\ell^2\langle 1\bar{2}\rangle ^{\ell}\langle 3\bar{4}\rangle ^{\ell}}-(-1)^{\ell}(1\leftrightarrow2)-(-1)^{\ell}(3\leftrightarrow4)\right). \nn\\
\eea

Similar to the superconformal partial waves with odd $\ell$, the conformal integrand can be separated into two
parts, one of which reproduces the results of $\cN=1$ conserved current with spin $\ell+2$ and another one is new for $\cN=2$ theories. In particular we have following new tensor structures
\bea
\left.\partial_{\bar{0}}^{A}H~\partial_{\bar{0}A}\tl\zb\,\right|_{\bar{0}=0}&=&-\frac{X_{12}}{X_{30}X_{40}} S21034S, \\
\left.\partial_{\bar{0}}^{A}\tl H~\partial_{\bar{0}A} \zb\,\right|_{\bar{0}=0}&=& \frac{X_{34}}{X_{10}X_{20}} T21034T.
\eea
The recursion relations of these tensor structures and their conformal integrations are presented in Appendix B and Appendix C, respectively.

After some algebra in embedding space and conformal integrations, we obtain the superconformal blocks
\bea
\cG_{\Delta,\ell+2|\ell,\mathrm{even}}^{\cN=2|JJ;JJ}&\propto& \la_\cO^2
\left(
\frac{1}{(\Delta +\ell +2) (\Delta +\ell +3)}g_{\Delta+2,\ell+2} \right. \nn\\
&&\left.~~~~~~~~~~~ +
\frac{(\ell +3) (\ell +4) (\Delta -\ell -2)}{\ell  (\ell +1) (\Delta -\ell -1) (\Delta +\ell+2)^2}g_{\Delta+2,\ell}\right),  \label{N2blockmeven}
\eea
which are reminiscent to the $\cN=1$ superconformal blocks for conserved currents with exchanged odd spin operator
$(\Delta+1,\ell+1)$.

\subsection{$\cA^{\Delta}_{0,0(\frac{\ell+4}{2},\frac{\ell}{2})}$}

We have shown before that for the three-point correlator with the multiplet $\cA^{\Delta}_{0,0(\frac{\ell+4}{2},\frac{\ell}{2})}$, only these with even $\ell$ can have non-zero three-point functions with two stress-tensor multiplet. Moreover, from the three-point functions we have following superconformal partial waves
\begin{equation}
\mathcal{\left.W_{O}\right|}_{\theta_{\mathrm{ext}}=0}\propto\frac{1}{X_{12}^{2-\frac{1}{2}(4+\ell+\Delta)}
X_{34}^{2-\frac{1}{2}(4+\ell-\Delta)}}\int D^{4}X_{0}\left.\partial_{\bar{0}}^{4}\left(H^2\tl{H}^2\frac{\cN_{\ell}}{D_{\ell+4}}\right)\right|_{\bar{0}=0},
\end{equation}
where we have ignored the higher order nilpotent terms which have no contribution on the superconformal partial waves.
Since both $H$ and $\tl H$ are nilpotent, the four partial derivatives have to act on these nilpotent tensor structures.
Moreover, the contraction between two $\partial_{\bar{0}}H$'s vanish
\bea
\left.\partial_{\bar{0}}^{A}H~\partial_{\bar{0}A}H\right|_{\bar{0}=0}=
-\frac{1}{4^3}X_{12}X_{34}S1\Gamma^m 2 S\times S\tl\Gamma_m 12S=0.
\eea
Therefore the contraction in (\ref{HHt}) is the only way to obtain non-zero contributions from nilpotent tensor structures. The superconformal partial waves now become
\bea
\mathcal{\left.W_{O}\right|}_{\theta_{\mathrm{ext}}=0}
&\propto& \int D^{4}X_{0}\frac{N_{\ell+2}}{(X_{10}X_{20})^{\frac{1}{2}(4+\ell+\Delta)}
(X_{30}X_{40})^{\frac{1}{2}(4+\ell-\Delta)}}   \nn\\
&\propto& g_{\Delta+2,\ell+2}^{0,0}(u,v),  \label{N2blockmeven4}
\eea
where we have used the results of conformal integration in embedding space presented in \cite{SimmonsDuffin:2012uy}.
Actually from the representations of $\mathfrak{su}(2,2|2)$ algebra one can show that contributions of the multiplet $\cA^{\Delta}_{0,0(\frac{\ell+4}{2},\frac{\ell}{2})}$ on the four-point correlator are from the component with quantum numbers $(\Delta+2,\ell+2, R=0, r=0)$. Here we provided another simple explanation on this fact.

\subsection{Comments on the semi-short multiplet $\hat{\cC}_{1(\frac{\ell}{2},\frac{\ell}{2})}$}
The only remaining multiplet in the $\cJ\times\cJ$ OPE (\ref{JJOPE}) with nontrivial superconformal partial wave  is the semi-short multiplet $\hat{\cC}_{1(\frac{\ell}{2},\frac{\ell}{2})}$. It arises from the splitting of long multiplet $\cA^{\Delta}_{0,0(\frac{\ell}{2},\frac{\ell}{2})}$ when $\Delta$ hits the unitary bound
\be
\cA^{2+\ell}_{0,0(\frac{\ell}{2},\frac{\ell}{2})} = \hat{\cC}_{0(\frac{\ell}{2},\frac{\ell}{2})}
+\hat{\cC}_{\frac{1}{2}(\frac{\ell-1}{2},\frac{\ell}{2})}
+\hat{\cC}_{\frac{1}{2}(\frac{\ell}{2},\frac{\ell-1}{2})}
+\hat{\cC}_{1(\frac{\ell-1}{2},\frac{\ell-1}{2})}\, . \label{splitO}
\ee
For $\ell=0$ the splitting goes back to the form (\ref{splitJ}). For spinning multiplet $\ell>0$,
there are two semi-short multiplets $ \hat{\cC}_{0(\frac{\ell}{2},\frac{\ell}{2})}$ and $\hat{\cC}_{1(\frac{\ell}{2},\frac{\ell}{2})}$ satisfying the $\cJ\times\cJ$ selection rules. However,
 the semi-short multiplet  $\hat{\cC}_{0(\frac{\ell}{2},\frac{\ell}{2})}$ contains conserved higher spin operator and should be absent in interactive theories \cite{Maldacena:2011jn,Alba:2013yda}.
The multiplets $\hat{\cC}_{1(\frac{\ell}{2},\frac{\ell}{2})}$ contain the ``Schur operators" \cite{Gadde:2011uv} and play important roles in the $4d/2d$ correspondence. Their contributions on the stress-tensor four-point correlator can be described by the $2d$ chiral algebra \cite{Beem:2013sza}.
Unfortunately, this multiplet carries non-zero charge of the R-symmetry $SU(2)_R$. It is not clear how to uplift it to superembedding space and estimate its superconformal block. The superconformal block of $\hat{\cC}_{1(\frac{\ell}{2},\frac{\ell}{2})}$ can be solved indirectly. In the limitation $\Delta\rightarrow2+\ell$, the superconformal block of $\cA^{\Delta}_{0,0(\frac{\ell}{2},\frac{\ell}{2})}$ can be separated into two parts for $\hat{\cC}_{0(\frac{\ell}{2},\frac{\ell}{2})}$ and $\hat{\cC}_{1(\frac{\ell-1}{2},\frac{\ell-1}{2})}$. The multiplet $\hat{\cC}_{0(\frac{\ell}{2},\frac{\ell}{2})}$  can be uplifted to superembedding space and the three-point function $\langle\cJ\cJ \,\hat{\cC}_{0(\frac{\ell}{2},\frac{\ell}{2})}\rangle$ follows (\ref{evenspin}) by applying its semi-short conditions of on the three-point function. Then its superconformal partial wave can be solved accordingly.

Actually as shown in Appendix D, the shortening condition has already appeared in the superconformal partial waves with the supershadow OPE coefficients, that only the coefficients of conformal blocks $g_{\Delta,\ell}$ and $g_{\Delta+2,\ell+2}$ of the semi-short multiplet $\hat{\cC}_{0(\frac{\ell}{2},\frac{\ell}{2})}$ are non-vanishing. Without extra constraints we cannot solve the supershadow OPE coefficients just from these two terms directly, but they are expected to be relevant to the analytical continuation of the supershadow transformation of general long multiplet.
The superconformal partial waves of the semi-short multiplets $\hat{\cC}_{1(\frac{\ell}{2},\frac{\ell}{2})}$, as well as their roles in the $4d/2d$ correspondence will be studied in a future work \cite{Li:2018fu}.

\section{Decomposition of $\cN=2$ Blocks into $\cN=1$ Blocks}

The $\cN=2$ superconformal blocks can be expanded in terms of $\cN=1$ superconformal blocks following from the decomposition
of the $\cN=2$ multiplets into $\cN=1$ multiplets. This relation has been employed in \cite{Poland:2010wg, Fortin:2011nq, Khandker:2014mpa} for consistency check on the $\cN=1$ superconformal blocks, which agree with the $\cN=1$ decomposition of the superconformal blocks of $\cN=2$ global symmetry conserved current solved in \cite{Dolan:2001tt}.
In this work, we show that the $\cN=1$ decomposition can also provide  consistency checks on the $\cN=2$ superconformal blocks we obtained in the previous section.

For a general $\cN=2$ long multiplet $\cA^{\Delta}_{0,0(\frac{\ell}{2},\frac{\ell}{2})}$, its $\cN=1$ decomposition has been solved in \cite{Liendo:2015ofa}:
\begin{align}
\begin{split}
\cA^{\Delta}_{0,0(\frac{\ell}{2},\frac{\ell}{2})}  \rightarrow &A^{\Delta}_{r'=0(\frac{\ell}{2},\frac{\ell}{2})} + A^{\Delta+1}_{r'=0(\frac{\ell-1}{2},\frac{\ell-1}{2})}
 + A^{\Delta+1}_{r'=0(\frac{\ell+1}{2},\frac{\ell+1}{2})}
\\
&
+ A^{\Delta+2}_{r'=0(\frac{\ell}{2},\frac{\ell}{2})}
+A^{\Delta+1}_{r'=0(\frac{\ell-1}{2},\frac{\ell+1}{2})}
+A^{\Delta+1}_{r'=0(\frac{\ell+1}{2},\frac{\ell-1}{2})}+\cdots\, ,  \label{N2N1}
\end{split}
\end{align}
where the $\cN=1$ mutliplets are denoted as $A$ and the subindex $r'\equiv\frac{2}{3}(2R+r)$ is the $U(1)_{r'}$ charge of $\cN=1$ multiplets $A$. There are extra terms with non-zero $r'$ in the decomposition, while they have no contribution on the correlator so are omitted in (\ref{N2N1}). The $\cN=2$ superconformal blocks then can be decomposed into $\cN=1$ accordingly.

Given odd $\ell$, the last two terms in (\ref{N2N1}) have no contribution on the four-point correlator $\langle JJJJ\rangle$, since the components in the multiplets that could appear in the $J\times J$ OPE have odd spin. Therefore the $\cN=1$ decomposition reads
\be
\cG_{\Delta,\ell,\mathrm{odd}}^{\cN=2|JJ;JJ} = \cG_{\Delta,\ell,\mathrm{odd}}^{\cN=1} + b_1\cG_{\Delta+1,\ell-1,\mathrm{even}}^{\cN=1}+ b_2\cG_{\Delta+1,\ell+1,\mathrm{even}}^{\cN=1}+b_3\cG_{\Delta+2,\ell,\mathrm{odd}}^{\cN=1}, \label{N2N1block}
\ee
where the $\cN=1$ superconformal blocks  $\cG_{\Delta,\ell,\mathrm{odd/even}}^{\cN=1}$ have been given in (\ref{n1odd}) and (\ref{n1even}). Using the expressions of  $\cN=1$ superconformal blocks in (\ref{N2N1block}), we obtain an
expansion of $\cN=2$ superconformal blocks in terms of conformal blocks
\bea
\cG_{\Delta,\ell,\mathrm{odd}}^{\cN=2|JJ;JJ}\propto c_1 g_{\Delta+1,\ell+1}
+c_2 g_{\Delta+1,\ell-1} +c_3 g_{\Delta+3,\ell+1} + c_4 g_{\Delta+3,\ell-1},
\eea
where the $c_i$ are abbreviations of coefficients depend on $(b_i, \Delta,\ell)$.
Remarkably,
the coefficients $c_i$ obtained from $\cN=1$ decompositions (\ref{N2N1block}) are restricted to satisfy the constraint
\bea
c_1\frac{(\Delta -\ell -2)}{(\Delta -\ell -1)}&-&c_2\frac{\ell ^2  (\Delta +\ell )}{(\ell +2)^2  (\Delta +\ell +1)}\nn\\
&-&\frac{16 (\Delta +1)^2}{(\Delta -1)^2}\left(c_3\frac{(\Delta +\ell +3)}{ (\Delta +\ell +2)}-c_4 \frac{\ell ^2 (\Delta -\ell +1)}{(\ell +2)^2(\Delta -\ell )}\right)=0, \label{oddconsistency}
\eea
which provides nontrivial consistency check on our previous results.
Adopting the coefficients $c_i$ from the results in (\ref{oddscb}), we find the LHS of (\ref{oddconsistency}) vanishes.

For the multiplets $\cA^{\Delta}_{0,0(\frac{\ell}{2},\frac{\ell}{2})}$ with even $\ell$, the last two terms
in (\ref{N2N1}) $A^{\Delta+1}_{r'=0(\frac{\ell-1}{2},\frac{\ell+1}{2})}$ and
$A^{\Delta+1}_{r'=0(\frac{\ell+1}{2},\frac{\ell-1}{2})}$ now contain conformal primary components with quantum number $(\Delta,\ell,r'=0)$ and satisfy the
selection rules of $J\times J$ OPE. So they have non-zero contributions on the correlator which is described by non-supersymmetric conformal blocks. The $\cN=1$ decomposition of the superconformal blocks becomes
\be
\cG_{\Delta,\ell,\mathrm{even}}^{\cN=2|JJ;JJ} = \cG_{\Delta,\ell,\mathrm{even}}^{\cN=1} + b_1\cG_{\Delta+1,\ell-1,\mathrm{odd}}^{\cN=1}+ b_2\cG_{\Delta+1,\ell+1,\mathrm{odd}}^{\cN=1}+b_3\cG_{\Delta+2,\ell,\mathrm{even}}^{\cN=1}
+b_4g_{\Delta+2,\ell}. \label{N2N1block2}
\ee
Different from the odd spin multiplet, above decomposition does not lead to any cancellation, instead, each
$\cN=1$ superconformal block in (\ref{N2N1block2}) contains a term proportional to the conformal block $g_{\Delta+2,\ell}$. The final results on
$g_{\Delta+2,\ell}$ actually originate from five different parts. This explains the disorganized behavior of $a_2$, the coefficient of $g_{\Delta+2,\ell}$ shown in (\ref{N2evena2}).

The $\cN=1$ decomposition (\ref{N2N1block2}) now provides the constraint on the $\cN=2$ superconformal blocks (\ref{N2evenblock}) on the specific term $g_{\Delta+2,\ell}$. From the coefficients given below (\ref{N2evenblock}), we can solve the $\cN=1$ coefficients $b_i, i\in\{1,2,3\}$, as well as their contributions on the conformal block $g_{\Delta+2,\ell}$. Then by subtracting these contributions from the overall coefficient $a_2$ (\ref{N2evena2}), we obtain the contributions purely from $A^{\Delta+1}_{r'=0(\frac{\ell-1}{2},\frac{\ell+1}{2})}$ and
$A^{\Delta+1}_{r'=0(\frac{\ell+1}{2},\frac{\ell-1}{2})}$, which should be positive according to the unitarity. Since the conformal block contains two independent OPE coefficients $\la_\cO^{(i)}, i\in\{1,2\}$,
it is highly nontrivial that the three terms $(\la_\cO^{(1)}\la_\cO^{(1)}, \la_\cO^{(1)}\la_\cO^{(2)},\la_\cO^{(2)}\la_\cO^{(2)})$ are organized conspiratorially to be positive.
Indeed, from above results we obtain
\be
b_4=\frac{(\ell +2)(\Delta -1)  \left(\la_\cO^{(1)} (\Delta +\ell )+\la_\cO^{(2)} (\ell +3)\Delta  \right){}^2}{2\ell  (\ell +1)^2 \Delta ^2 (\Delta +1) },
\ee
 which perfectly agrees with the unitarity condition.\footnote{The coefficient $b_4$ includes contributions from both $A^{\Delta+1}_{r'=0(\frac{\ell-1}{2},\frac{\ell+1}{2})}$ and
$A^{\Delta+1}_{r'=0(\frac{\ell+1}{2},\frac{\ell-1}{2})}$, so in general the unitarity condition should be satisfied respectively, i.e., $b_4$ should be a summation of two positive terms. However, in our case $A^{\Delta+1}_{r'=0(\frac{\ell-1}{2},\frac{\ell+1}{2})}$ and
$A^{\Delta+1}_{r'=0(\frac{\ell+1}{2},\frac{\ell-1}{2})}$ are conjugate partners in an $\cN=2$ multiplet, their contributions are expected to be equal.}

\section{Conclusion and Discussion}

We have computed the superconformal partial waves of the four-point correlator $\langle JJJJ\rangle$, where the external operator $J$ is the lowest component of the $4D$ $\cN=2$ stress-tensor multiplet, and the exchanged operator is a singlet of R-symmetry $SU(2)_R$. We solve the three-point correlators $\langle\cJ\cJ\cO\rangle$ in superembedding space.
We find a  method to systematically expand the invariants in superembedding space in terms of
superconformal covariant variables in superspace, from which we are able to construct the $\cN=2$ superconformal invariants and tensor structures in superembedding space. Our results agree with the consistency checks by decomposing into the $\cN=1$ superconformal blocks of conserved currents.
It is straightforward to employ the method applied in this work for the superconformal partial waves of the
other interesting $\cN=2$ four-point correlators, for example $\langle JJ\Phi\Phi^\dagger\rangle$, where $\Phi$ is a chiral operator.

Our results provide necessary ingredients for the $\cN=2$ stress-tensor bootstrap. With the $\cN=2$ superconformal blocks it is straightforward to apply the numerical techniques developed in \cite{Rattazzi:2008pe, Kos:2014bka, Poland:2011ey, Simmons-Duffin:2015qma} to bootstrap the crossing equation of the four-point correlator $\langle JJJJ\rangle$. We expect
that the $\cN=2$ stress-tensor bootstrap will further promote the $4D$ $\cN=2$ superconformal bootstrap project initiated in \cite{Beem:2014zpa}, which aims to provide systematical studies on the extremely fruitful $\cN=2$ SCFTs. Specifically,
it is quite promising to bootstrap the simplest rank one Argyres-Douglas theory $H_0$, which saturates the lower bound on the $c$ central charge of any $\cN=2$ SCFTs. One of the characteristics of this theory is that the semi-short multilet $\hat{\mathcal{C}}_{1\left( \frac{1}{2},\frac{1}{2}\right)}$ decouples \cite{Liendo:2015ofa}. From the bootstrap point of view, this is reminiscent of the phenomena discovered in \cite{ElShowk:2012ht,El-Showk:2014dwa,Kos:2014bka} and \cite{Li:2017kck}. The critical $3D$ Ising model locates at the kink of the bound on the CFT data \cite{ElShowk:2012ht,El-Showk:2014dwa,Kos:2014bka}, and this kink corresponds to decoupling of certain spectra. In \cite{Li:2017kck} the correspondence between kink (as a solution of specific CFT) and decouple of certain spectra has been generalized to various $3D$ CFTs with global symmetry or supersymmetry. In particularly there is a putative new $3D$ $\cN=2$ SCFT that corresponds to the decouple of spectrum in its BPS sector, or the chiral ring relation $\Phi^2=0$ \cite{Bobev:2015vsa, Bobev:2015jxa}.\footnote{This putative SCFT has $4D$ $\cN=1$ analogy with the same chiral ring relation \cite{Poland:2011ey, Poland:2015mta, Li:2017ddj}.}
It is tempting to expect this scenario works for $4D$ $\cN=2$ theories as well.

The $\cN=2$ stress-tensor bootstrap provides a nice approach to study the conformal anomaly coefficient $a$. Due to the $\cN=2$ supersymmetry, the crossing equation is significantly simplified  comparing with those of non-supersymmetric theories that also involve $a$-anomaly coefficient \cite{Dymarsky:2017xzb,Dymarsky:2017yzx}.
The $a$-anomaly coefficient satisfies the conformal collider bound \cite{Hofman:2008ar}. This bound has been proven analytically in \cite{Hofman:2016awc, Faulkner:2016mzt}. Remarkably the conformal collider bound automatically appears
in the numerical conformal bootstrap of the $3D$ conserved current $(J^\mu)$ correlator $\langle JJJJ\rangle$ \cite{Dymarsky:2017xzb} and the stress-tensor $(T^{\mu\nu})$ correlator $\langle TTTT\rangle$ \cite{Dymarsky:2017yzx}. It would be very interesting to reproduce the $\cN=2$ supersymmetric version of the bound from numerical $\cN=2$ stress-tensor bootstrap.

With additional assumption on the operator spectra of CFTs, the conformal collider bound in $4D$ can be further restricted to be
$(a-c)/c\leqslant \Delta_{\mathrm{gap}}^{-2}$, where the parameter $\Delta_{\mathrm{gap}}$ is the dimension of lightest
single trace  operator with higher spin $\ell\geqslant3$ \cite{Camanho:2014apa}. This refined constraint on the conformal anomaly coefficients and the sparse spectra has been derived in \cite{Afkhami-Jeddi:2016ntf,Costa:2017twz,Afkhami-Jeddi:2017rmx}, see \cite{Meltzer:2017rtf} for more results on the generalization of the conformal collider bound.
The $\cN=2$ theories provide an ideal laboratory to study this constraint further. With eight Poincar\'e supercharges, we can get access to the stress-tensor OPE from the scalar component in the stress-tensor multiplet. On the other hand, there remains two independent tensor structures in the stress-tensor three-point function and the refined constraint is highly non-trivial.
 In contrast, for $\cN=4$ SCFTs, $a=c$ and the constraint is satisfied trivially. While for $\cN=1$ SCFTs, the stress-tensor is a component of the spin $1$ supercurrent multiplet, and its OPE is still quite involved \cite{Manenti:2018xns}. It is expected to obtain better understanding on the generalized conformal collider bound from the crossing symmetry and unitarity condition of $\cN=2$ stress-tensor four-point correlator. We leave these problems for future studies.

\acknowledgments
I am grateful to C. Beem, M. Costa, M. Lemos, D. Li, P. Liendo, L. Rastelli, B.C. van Rees for valuable discussions. I would like to thank the Princeton Center for Theoretical Science, ICTP-SAIFR in S˜ao Paulo, Institute of Theoretical Physics (Chinese Academy of Sciences) and the organizers of the ``Project Meeting: Analytical Approaches" workshop at Azores for the hospitality during the course of this work. I additionally thank T. Li and C. Pope for their support during the early stage of this work.
This work was supported by the Simons Foundation grant 488637 (Simons collaboration on the Nonperturbative
bootstrap).
Centro de F\'isica do Porto is partially funded by the Foundation for Science and Technology
of Portugal (FCT).


\appendix
\section{Superconformal invariants and tensor structures  in superembedding space}

In $\cN=2$ superembedding space, we can construct two independent superconformal invariants
$\zb$ and $\wb$ from three points $(0, 1, 2)\sim(\cX_0, \cX_1, \cX_2)$:
\bea
\zb&\equiv &\frac{\langle\bar{1}2\,\rangle\langle\bar{2}0\rangle\langle\bar{0}1\rangle
-\langle\bar{2}1\rangle\langle\bar{1}0\rangle\langle\bar{0}2\rangle}
{\langle\bar{1}2\rangle\langle\bar{2}0\rangle\langle\bar{0}1\rangle
+\langle\bar{2}1\rangle\langle\bar{1}0\rangle\langle\bar{0}2\rangle},  \\
\wb&\equiv &\frac{4~\langle \bar{1}2\,\bar{0}1\bar{2}\,0\rangle}
{\left(\langle \bar{1}2\rangle\langle \bar{2}1\rangle \langle \bar{1}0\rangle\langle \bar{0}1\rangle
\langle \bar{0}2\rangle \langle \bar{2}0\rangle\right)^{\frac{1}{2}}}+1.
\eea
In this work we also need the superconformal invariants $\tl\zb$ and $\tl\wb$ constructed similarly with variable replacements $1\rightarrow3, 2\rightarrow4$.
Both $\zb$ and $\wb$ are nilpotent and vanish when setting external Grassmann variables to zero
\be
\begin{split}
\zb^5&=\wb^3=0, \\
\zb|_{\te_\mathrm{ext}=0}&=\wb|_{\te_\mathrm{ext}=0}=0.
\end{split}
\ee
Therefore they actually have no effect on the superconformal integrations for the superconformal partial waves unless acted by the partial derivative $\partial_{\bar{0}}$. The derivatives on $\zb$ are
\bea
\partial_{\bar{0}}^m\zb&=&(\zb^2-1)\left(\frac{X_1^m}{X_{1\bar{0}}}-\frac{X_2^m}{X_{2\bar{0}}}\right),~~~~~~
\partial_{\bar{0}}^2\zb=2\zb(\zb^2-1) \frac{X_{12}}{X_{1\bar{0}}X_{2\bar{0}}}, \nn \\
\partial_{\bar{0}}^m\partial_{\bar{0}}^n\zb&=&2(\zb^2-1)\left(\zb\left(\frac{X_1^m}{X_{1\bar{0}}}-\frac{X_2^m}{X_{2\bar{0}}}\right)
\left(\frac{X_1^n}{X_{1\bar{0}}}-\frac{X_2^n}{X_{2\bar{0}}}\right)+
\left(\frac{X_1^m X_1^n}{X_{1\bar{0}}X_{1\bar{0}}}-\frac{X_2^m X_2^n}{X_{2\bar{0}}X_{2\bar{0}}}\right)\right),
\eea
which are also anti-symmetric under the coordinate permutation $z_1\leftrightarrow z_2$.
Using above results it is straightforward to get the higher order derivatives on $\zb$, as well as the derivatives on $\zb^2, \zb^3, \zb^4$.

The derivative $\partial_{\bar{0}}^m$ on $\wb$ gives
\be
\partial_{\bar{0}}^m\wb=\frac{16~\langle \bar{1}2\,\bar{\Gamma}^m 1\,\bar{2}\,0\rangle}
{\left(X_{1\bar{0}}X_{0\bar{1}}X_{2\bar{0}}X_{0\bar{2}}X_{2\bar{1}}X_{1\bar{2}}\right)^{\frac{1}{2}}}
+(\wb-1)\left(\frac{X_1^m}{X_{1\bar{0}}}+\frac{X_2^m}{X_{2\bar{0}}}\right), \label{p1wb}
\ee
which is symmetric under the coordinate permutation $1\leftrightarrow2$. Higher order derivatives are given by
\bea
\partial_{\bar{0}}^2\wb &=& -(\wb-1)\frac{X_{12}}{X_{1\bar{0}}X_{2\bar{0}}}, \\
\partial_{\bar{0}}^m\partial_{\bar{0}}^n\wb&=& \partial_{\bar{0}}^m\wb\left(\frac{X_1^n}{X_{1\bar{0}}}+\frac{X_2^n}{X_{2\bar{0}}}\right)+
\partial_{\bar{0}}^n\wb\left(\frac{X_1^m}{X_{1\bar{0}}}+\frac{X_2^m}{X_{2\bar{0}}}\right)\nn\\
&&+
(\wb-1)\left(\frac{X_1^m}{X_{1\bar{0}}}-\frac{X_2^m}{X_{2\bar{0}}}\right)
\left(\frac{X_1^n}{X_{1\bar{0}}}-\frac{X_2^n}{X_{2\bar{0}}}\right),  \\
\partial_{\bar{0}}^m\partial_{\bar{0}}^2\wb&=& -\partial_{\bar{0}}^m\wb \frac{X_{12}}{X_{1\bar{0}}X_{2\bar{0}}}
-2(\wb-1)\frac{X_{12}}{X_{1\bar{0}}X_{2\bar{0}}}\left(\frac{X_1^m}{X_{1\bar{0}}}+\frac{X_2^m}{X_{2\bar{0}}}\right),\\
\partial_{\bar{0}}^2\,\partial_{\bar{0}}^2\wb&=& 9(\wb-1)\left(\frac{X_{12}}{X_{1\bar{0}}X_{2\bar{0}}}\right)^2,
\eea
where we have solved the $SO(4,2)$ vector $\langle \bar{1}2\,\bar{\Gamma}^m 1\,\bar{2}\,0\rangle$ in terms of
$\partial_{\bar{0}}^m\wb$ from (\ref{p1wb}) for simplicity. Setting $\wb=0$ above formulas give the results with vanishing
external Grassmann variables $\te_\mathrm{ext}=0$. In particular, we have
\be
\left. \partial_{\bar{0}}^m\wb\right|_{\te_{\mathrm{ext}}=0}=-\frac{X_{12}}{X_{10}X_{20}}X_0^m. \label{wb0}
\ee
This identity can be obtained by evaluating the trace $\langle \bar{1}2\,\bar{\Gamma}^m 1\,\bar{2}\,0\rangle$ explicitly,
or alternatively, the readers can simply convince themselves by showing that the two sides of (\ref{wb0}) are equal by  multiplying vectors $X_{i\,m}$.

Replacing $1\rightarrow3$ and $2\rightarrow4$ in above identities, we obtain the derivatives on $\tl\zb$ and $\tl\wb$.
Moreover, in the full tensor structures $\mathcal{N}_{\ell}^{\mathrm{full}}$, like in (\ref{Nodd}) and (\ref{Neven}),
we also have mixed terms that are proportional to $\zb\tl\zb$, $\zb\tl\wb$, $\wb\tl\wb$, etc. The corresponding derivatives can be evaluated accordingly based on above results.

The tensor structure $\cN_\ell$ in the superconformal integration are
\be
\cN_{\ell}\equiv(\bar{\cS}1\bar{2}\cS)^{\ell}\overleftrightarrow{\mathcal{D}_{\ell}}(\bar{\cT}3\bar{4}\cT)^{\ell}
=\frac{1}{(\ell!)^2}(\pd_{\cS}0\pd_\cT)^\ell(\cS\bar2 1\bar0 3\bar4\cT)^\ell.
\ee
By permuting the variables in $\bar{\cS}1\bar{2}\cS$ or $\bar{\cT}3\bar{4}\cT$  we can also obtain extra tensor structures appearing in $\mathcal{N}_{\ell}^{\mathrm{full}}$, such as $\cM_\ell,\cL_\ell$ and $\cK_\ell$. These tensor structures are constructed based on the supershadow formalism and the variables $\cX_0, \bar{\cX}_0$ originate from
\be
\overleftrightarrow{\mathcal{D}}_{j_1,j_2}\equiv\frac{1}{(2j_1)!^2(2j_2)!^2}(\partial_{\cS}0\partial_{\cT})^{2j_1}
(\partial_{\bar{\cS}}\bar{0}\partial_{\bar{\cT}})^{2j_2}.
\ee
When $j_1\neq j_2$, it gives the tensor structure for mixed symmetry multiplet $\cA^{\Delta}_{0,0(\frac{\ell+2}{2},\frac{\ell}{2})}$ or $\cA^{\Delta}_{0,0(\frac{\ell+4}{2},\frac{\ell}{2})}$.
The derivative $\partial_{\bar{0}}$ acting on the tensor structures is equivalent to acting on $\overleftrightarrow{\mathcal{D}}$, and there is a null condition:
\be
\partial_{\bar{0}}^2 \overleftrightarrow{\mathcal{D}}_{j_1,j_2}\propto \epsilon^{\al\b\gamma\delta}
\partial_{\bar{S}\al}\partial_{\bar{S}\b}\partial_{\bar{T}\gamma}\partial_{\bar{T}\delta}=0.
\ee
Therefore we have following constraint on the general tensor structures $X_\ell\in\{\cN_\ell,\cM_\ell,\cL_\ell,\cK_\ell\}$
\be
\partial_{\bar{0}}^2 X_\ell=0.
\ee
For $\cN=1$ theories, there is at most one derivative $\partial_{\bar{0}}^m$ applied on the tensor structures, and the $SO(4,2)$ index is contracted with another one from the derivative on superconformal invariants or $\frac{1}{D_\ell}$. While for $\cN=2$ theories, we have terms proportional to
$\partial_{\bar{0}}^m\partial_{\bar{0}}^n X_{\ell} $,
which lead to various kinds of tensor structures in embedding space. The problem to obtain superconformal block now is transferred to the conformal integration of the tensor structures, which are usually cannot be expanded in terms of Gegenbauer polynomials directly. For instance, in \cite{Li:2016chh} the $\cN=1$ superconformal blocks corresponding to exchanging of symmetric multiplets with none-zero $U(1)_R$ charges have been computed. One of the major challenges is to evaluate the conformal integration with the tensor structure
\be
\left(\partial_{S}0\partial_{T}\right)^{\ell}\left(S\bar{2}1\bar{0}3\bar{4}T\right)^{\ell-1}\left( X_{30}S\bar{2}1\bar{4}T + X_{40}S\bar{2}1\bar{3}T - X_{30}S\bar{1}2\bar{4}T - X_{40}S\bar{1}2\bar{3}T\right).
\ee
The tensor structures are more complex for $\cN=2$ theories.

Take the odd spin long multiplet $\cA^{\Delta}_{0,0(\frac{\ell}{2},\frac{\ell}{2})}$ for example. The superconformal integrand contains the following term
\be
  \la_1{\tl\la}_1\zb\tzb\frac{1}{D_\ell} \Sa\Dh\Ta.
\ee
Integrating out the fermionic variables we obtain the conformal integration in embedding space:
\bea
&&\left(\partial _{\bar{0}}^2\partial _{\bar{0}}^2\zb\tl\zb\right)\frac{1}{D_{\ell }} N_{\ell }
+4\left(\partial _{\bar{0}}^n\partial _{\bar{0}}{}^2\zb\tl\zb\right)\left(\partial _{\bar{0}n}\frac{1}{D_{\ell }}\right)N_{\ell }
+2\left(\partial _{\bar{0}}^2\zb\tl\zb\right) \left(\partial _{\bar{0}}{}^2\frac{1}{D_{\ell }}\right)N_{\ell }\nn\\
&&+4\left(\partial _{\bar{0}}^n\partial _{\bar{0}}^m\zb\tl\zb\right)\left(\partial _{\bar{0}n}\partial _{\bar{0}m}\frac{1}{D_{\ell }}\right)N_{\ell }
+4\left(\partial _{\bar{0}}^n\partial _{\bar{0}}^2\zb\tl\zb\right)\frac{1}{D_{\ell }}\partial _{\bar{0}n}N_{\ell }
+8\left(\partial _{\bar{0}}^n\partial _{\bar{0}}^m\zb\tl\zb\right) \partial _{\bar{0}m}\frac{1}{D_{\ell }}\partial _{\bar{0}n}N_{\ell } \nn\\
&&+4\left(\partial _{\bar{0}}^2\zb\tl\zb\right) \partial _{\bar{0}}^m\frac{1}{D_{\ell }}\partial _{\bar{0}m}N_{\ell }
+4\left(\partial _{\bar{0}}^m\partial _{\bar{0}}^n \zb\tl\zb\right)\frac{1}{D_{\ell }}\partial _{\bar{0}m}\partial _{\bar{0}n}N_{\ell } \nn\\
&&+(-1)^\ell\left(1\leftrightarrow2\right)+(-1)^\ell\left(3\leftrightarrow4\right).
\eea
After taking the partial derivatives, we set $\te_{\mathrm{ext}}=0$, and above formula becomes
\bea
&&\hspace{-5mm}2\left(\Omega_-\left(\left((\Delta +\ell)^2+8\right)\frac{X_{12}}{X_{1\bar{0}} X_{2\bar{0}}}+ \left((\Delta -\ell)^2+8\right)\frac{X_{34}}{X_{3\bar{0}} X_{4\bar{0}}} \right)  \right.\nn\\
&&~~~~\left.+(2+\ell-\Delta ) (2+\ell+\Delta) \left(\Omega_A \Omega_B+\Omega_- \Omega_+\right)\right)
\frac{N_\ell}{D_\ell} \nn\\
&&+\frac{\Delta +\ell+2 }{2}\frac{\ell}{D_\ell}\frac{X_{1\bar{2}}}{X_{1\bar{0}} X_{2\bar{0}}}(\Omega_B P1-\Omega_- P3) \nn\\
&&-
\frac{\Delta -\ell-2 }{2}\frac{\ell}{D_\ell} \frac{X_{3\bar{4}}}{X_{3\bar{0}} X_{4\bar{0}}}(\Omega_A R1-\Omega_- R3 )   \nn\\
&&+\frac{1}{8}\frac{\ell(\ell-1)}{D_\ell}\frac{X_{12}}{X_{1\bar{0}} X_{2\bar{0}}}\frac{X_{34}}{X_{3\bar{0}} X_{4\bar{0}}}
(P0\,R3+P1\, R1+P2\, R2+ P3\, R0), \label{Oddts}
\eea
where the variables $\Omega_*$, $Pi$  and $Ri$ are given in (\ref{Omegap}-\ref{R3ts}).
As shown in (\ref{Oddts}), the conformal integrand contains rather complicated tensor structures. More complex tensor structures appear in the conformal integrand for even spin. The whole conformal integrand would take too much length to be presented here.
The tensor structures and their recursion relations will be given in the next Appendices.

\section{Tensor structures and their recursion relations}

In this Appendix we summarize the tensor structures appear in the conformal integrand and their recursion relations, from which
we can evaluate the conformal integration in embedding space.

For the simplest tensor structure $N_\ell$, it satisfies the recursion of Gegenbauer polynomials $C_\ell^{(1)}(t)$ with variable
\be
t\equiv\frac{\langle \bar{2}1\bar{0}3\bar{4}0\rangle }{2\sqrt{s}},~~~~~~~~~~~ s\equiv\frac{1}{2^{6}}\langle \bar{0}1\rangle \langle \bar{2}0\rangle \langle \bar{0}3\rangle \langle \bar{4}0\rangle \langle
\bar{2}1\rangle \langle \bar{4}3\rangle. \label{ts}
\ee
While for extra tensor structures, generally they cannot be expanded in terms of Gegenbauer polynomials directly.

The tensor structures, as well as the invariants in embedding space usually are constructed based on certain combinations which are (anti-)symmetric under coordinate permutations $1\leftrightarrow2$ or $3\leftrightarrow4$, it will be convenient to denote these elementary terms as follows
\bea
\Omega_+&=&\frac{X_{13}}{X_{1\bar{0}} X_{3\bar{0}}}+\frac{X_{23}}{X_{2\bar{0}} X_{3\bar{0}}}
+\frac{X_{14}}{X_{1\bar{0}} X_{4\bar{0}}}+\frac{X_{24}}{X_{2\bar{0}} X_{4\bar{0}}}, \label{Omegap}\\
\Omega_-&=&\frac{X_{13}}{X_{1\bar{0}} X_{3\bar{0}}}-\frac{X_{23}}{X_{2 \bar{0}} X_{3 \bar{0}}}
-\frac{X_{14}}{X_{1 \bar{0}} X_{4 \bar{0}}}+\frac{X_{24}}{X_{2 \bar{0}} X_{4 \bar{0}}}, \\
\Omega_A&=&\frac{X_{13}}{X_{1 \bar{0}} X_{3 \bar{0}}}-\frac{X_{23}}{X_{2 \bar{0}} X_{3\bar{0}}}
+\frac{X_{14}}{X_{1 \bar{0}} X_{4 \bar{0}}}-\frac{X_{24}}{X_{2 \bar{0}} X_{4 \bar{0}}},  \\
\Omega_B&=&\frac{X_{13}}{X_{1\bar{0}} X_{3\bar{0}}}+\frac{X_{23}}{X_{2 \bar{0}} X_{3 \bar{0}}}
-\frac{X_{14}}{X_{1\bar{0}} X_{4 \bar{0}}}-\frac{X_{24}}{X_{2 \bar{0}} X_{4 \bar{0}}},
\eea
and  for the tensor structures
\bea
P0&=&X_{1\bar{0}} S234T+ X_{2\bar{0}} S134T+ X_{1\bar{0}} S243T+ X_{2\bar{0}} S143T, \\
P1&=&X_{1\bar{0}} S234T+ X_{2\bar{0}} S134T- X_{1\bar{0}} S243T- X_{2\bar{0}} S143T, \\
P2&=&X_{1\bar{0}} S234T- X_{2\bar{0}} S134T+ X_{1\bar{0}} S243T- X_{2\bar{0}} S143T, \\
P3&=&X_{1\bar{0}} S234T- X_{2\bar{0}} S134T- X_{1\bar{0}} S243T+ X_{2\bar{0}} S143T, \\
&&\nn\\
R0&=&X_{3\bar{0}} S214T+ X_{4\bar{0}} S213T+ X_{3\bar{0}} S124T+ X_{4\bar{0}} S123T, \\
R1&=&X_{3\bar{0}} S214T+ X_{4\bar{0}} S213T- X_{3\bar{0}} S124T- X_{4\bar{0}} S123T, \\
R2&=&X_{3\bar{0}} S214T- X_{4\bar{0}} S213T+ X_{3\bar{0}} S124T- X_{4\bar{0}} S123T, \\
R3&=&X_{3\bar{0}} S214T- X_{4\bar{0}} S213T- X_{3\bar{0}} S124T+ X_{4\bar{0}} S123T.  \label{R3ts}
\eea
Note that the invariant $\Omega_-$ is proportional to the variable $t_0\equiv t|_{\te_\mathrm{ext}=0}$ in embedding space.
The tensor structures $P2$ and $R2$ vanish, and $P3=R3=8\, S21034T$. $P0$ and $R0$ can be simplified to be $SiT$ which is equivalent to $X_{0i}$ up to a constant. So actually the nontrivial terms are $P1$ and $R1$.
Above terms arise from the partial derivative $\partial_{\bar{0}}^m$ acted on the tensor structures.

For $\cN=2$ theories,
we also have higher order derivatives $\partial_{\bar{0}}^m\partial_{\bar{0}}^n$ acted on the tensor structures, which lead to higher order tensor structures like
\bea
&&(X_{1\bar{0}} S234T)^2 \pm (X_{2\bar{0}} S134T)^2\pm (X_{1\bar{0}} S243T)^2\pm (X_{2\bar{0}} S143T)^2, \\
&&(X_{3\bar{0}} S214T)^2 \pm (X_{4\bar{0}} S213T)^2\pm (X_{3\bar{0}} S124T)^2\pm (X_{4\bar{0}} S123T)^2, \\
&&X_{1\bar{0}}X_{3\bar{0}} S234T  S214T  \pm X_{2\bar{0}} X_{4\bar{0}} S134T S213T  \nn\\
&&~~~~~~~~~~~~~~~~~~~~\pm X_{1\bar{0}}X_{3\bar{0}} S243T  S124T  \pm X_{2\bar{0}} X_{4\bar{0}} S143T S123T.
\eea
Above second order terms can be nicely decomposed in terms of $Pi$ and $Ri$. For instances,
\bea
\begin{split}
(X_{1\bar{0}} S234T)^2 &+ (X_{2\bar{0}} S134T)^2 + (X_{1\bar{0}} S243T)^2+ (X_{2\bar{0}} S143T)^2 \\
&=\frac{1}{4}(P0^2+P1^2+P3^2+P3^2), \label{foursq}
\end{split}
\eea
and
\bea
\begin{split}
X_{1\bar{0}}X_{3\bar{0}} S234T  S214T  &+ X_{2\bar{0}} X_{4\bar{0}} S134T S213T  \\
&+ X_{1\bar{0}}X_{3\bar{0}} S243T  S124T  + X_{2\bar{0}} X_{4\bar{0}} S143T S123T \\
&=\frac{1}{4}(P0 R0+P1 R2+P2 R1+P3 R3).
\end{split}
\eea
The formula (\ref{foursq}) is just the {\it Euler's four-square identity} with four arguments fixed to unit.
Other higher order tensor structures can be expanded similarly.

Actually among the second order tensor structures, only $P1^2, R1^2, P1\times R1$ could give nontrivial results. All the other tensor structures either vanish or degenerate to first order tensor structures.
We use the notation
\be
SPT\equiv P1,~~~~~SRT\equiv R1,
\ee
to trace the auxiliary twistors $S, T$ and the letters $P/R$ refer to the combinations of the coordinates in $P1/R1$. For example, in this notation we have
\be
\partial_S 0\partial_T SPT=\mathrm{tr}(0P),~~~~~~~  SPT \overleftarrow\partial_T 0\partial_S SRT=SP0RT.
\ee
Besides, we also adopt the notation
\be
\Delta_x=X_{10}X_{20}X_{30}X_{40} \Omega_x, ~~~x\in\{+,-,A,B\},
\ee
in terms of which we have
\bea
\mathrm{tr}(0P)&=&\frac{1}{8}\Delta_B, ~~~~~~~~~\mathrm{tr}(0R)=\frac{1}{8}\Delta_A, \\
SR0PT&=&S21034T (X_{30}X_{40}X_{12}+X_{10}X_{20}X_{34}-\Delta_+).
\eea

Now we are ready to write down the recursion relations corresponding to the higher order tensor structures. The recursion relations of $SPT$ and $SRT$ are provided in \cite{Li:2016chh}.

Recursion relation of $SPT^2$:
\bea
&&\hspace{-5mm}\left( \partial_S 0\partial_T\right)^{\ell} S21034T^{\ell -2}SPT^2= \nn\\
&&\ell(\ell -1) (\ell -2) (\ell -3)s_0  \left(\partial_S  0\partial_T\right)^{\ell-2} S21034T^{\ell -4}SPT^2\nn\\
&&+\frac{1}{64}\ell ^2 (\ell -1)^2 (\Delta_B)^2\left(\partial_S  0\partial_T\right)^{\ell-2} S21034T^{\ell -2}\nn\\
&&+\frac{1}{512}\ell^2 (\ell -1) (\ell -2)X_{10}X_{20}X_{34} \Delta_A
\left(\partial_S  0\partial_T\right)^{\ell-2} S21034T^{\ell -3} SPT  \\
&&+\frac{1}{512}\ell^2 (\ell -1) (\ell -2)X_{10}X_{20}X_{34} \Delta_B
\left(\partial_S  0\partial_T\right)^{\ell-2} S21034T^{\ell -3} SRT  \nn\\
&&+ \frac{1}{32}\ell(\ell -1)^2 X_{10}X_{20}X_{34}(X_{30}X_{40}X_{12}+X_{10}X_{20}X_{34}-\Delta_+)\left(\partial_S 0\partial_T\right)^{\ell-2} S21034T^{\ell -2}. \nn
\eea

Recursion relation of $SRT^2$:
\bea
&&\hspace{-5mm}\left( \partial_S 0\partial_T\right)^{\ell} S21034T^{\ell -2}SRT^2= \nn\\
&&\ell(\ell -1) (\ell -2) (\ell -3)s_0  \left(\partial_S  0\partial_T\right)^{\ell-2} S21034T^{\ell -4}SRT^2\nn\\
&&+\frac{1}{64}\ell ^2 (\ell -1)^2 (\Delta_A)^2\left(\partial_S  0\partial_T\right)^{\ell-2} S21034T^{\ell -2}\nn\\
&&+\frac{1}{512}\ell^2 (\ell -1) (\ell -2)X_{30}X_{40}X_{12} \Delta_A
\left(\partial_S  0\partial_T\right)^{\ell-2} S21034T^{\ell -3} SPT  \\
&&+\frac{1}{512}\ell^2 (\ell -1) (\ell -2)X_{30}X_{40}X_{12} \Delta_B
\left(\partial_S  0\partial_T\right)^{\ell-2} S21034T^{\ell -3} SRT  \nn\\
&&+ \frac{1}{32}\ell(\ell -1)^2 X_{30}X_{40}X_{12}(X_{30}X_{40}X_{12}+X_{10}X_{20}X_{34}-\Delta_+)\left(\partial_S 0\partial_T\right)^{\ell-2} S21034T^{\ell -2}. \nn
\eea

Recursion relation of $SPT\times SRT$:
\bea
&&\hspace{-5mm}\left( \partial_S 0\partial_T\right)^{\ell} S21034T^{\ell -2}SPT\times SRT= \nn\\
&&\frac{1}{64} \ell(\ell -2)X_{30}X_{40}X_{12} \left(\partial_S  0\partial_T\right)^{\ell-1} S21034T^{\ell -3}SPT^2\nn\\
&&+\frac{1}{8}\ell(\ell -1) \Delta_A\left(\partial_S  0\partial_T\right)^{\ell-1} S21034T^{\ell -2}SPT\nn\\
&&+\ell(X_{30}X_{40}X_{12}+X_{10}X_{20}X_{34}-\Delta_+) \left(\partial_S 0\partial_T\right)^{\ell-1} S21034T^{\ell -1}. \nn
\eea

\subsection*{Tensor structures for mixed symmetry multiplets}
New tensor structures appear in the conformal integrands of the mixed symmetric multiplet $\cA^{\Delta}_{0,0(\frac{\ell+2}{2},\frac{\ell}{2})}$. They are
\bea
TPT&\equiv&\left(X_{1\bar{0}} T234T+ X_{2\bar{0}} T134T- X_{1\bar{0}} T243T- X_{2\bar{0}} T143T \right),  \\
SRS&\equiv& \left(X_{3\bar{0}} S214S+ X_{4\bar{0}} S213S- X_{3\bar{0}} S124S- X_{4\bar{0}} S123S  \right),
\eea
for odd $\ell$ and $S21034S$, $T21034T$ for even $\ell$. They are involved in the following tensor structures in the conformal integrand
\be
\left( \partial_S 0\partial_T\right)^{\ell+2} S21034S\times T21034T\,S21034T^{\ell}, \label{SSTT}
\ee
or
\be
\left( \partial_S 0\partial_T\right)^{\ell+2} SRS\times TPT\,S21034T^{\ell}. \label{SRSTPT}
\ee

Here we provide the recursion relations of these structures. The tensor structure (\ref{SSTT}) can be decomposed into Gegenbauer polynomials directly
\bea
\left( \partial_S 0\partial_T\right)^{\ell+2} &S21034S& T21034T S21034T^{\ell} = \nn\\
&& \hspace{-30mm}
-\frac{\ell+2}{64}\Delta_- \left( \partial_S 0\partial_T\right)^{\ell+1}S21034T^{\ell+1} +2(\ell+1)(\ell+2)^2s_0 \left( \partial_S 0\partial_T\right)^{\ell}S21034T^{\ell}.~~~ \label{SSTTD}
\eea

The recursion relation of tensor structure (\ref{SRSTPT}) is more difficult to solve
\bea
\left( \partial_S 0\partial_T\right)^{\ell+2} &SRS& TPT S21034T^{\ell}=(\ell+2)^2\left( \partial_S 0\partial_T\right)^{\ell+1}
\left(SR0PT-\frac{\Delta_A}{16} SPT\right) S21034T^{\ell} \nn\\
&&+\frac{\ell(\ell+1)^2}{128}\Delta_A\Delta_B \left( \partial_S 0\partial_T\right)^{\ell} S21034T^{\ell} \nn\\
&&+ \frac{\ell  (\ell +1)^2}{512} X_{30}X_{40}X_{12}\Delta_B \left( \partial_S 0\partial_T\right)^{\ell}(SPT) \, S21034T^{\ell-1} \nn\\
&& +\frac{\ell^2}{128} X_{10}X_{20}X_{34}\left( \partial_S 0\partial_T\right)^{\ell+1}(SRS)\, (TRT) S21034T^{\ell-1}\nn\\
&&+\frac{\ell(\ell-1)(\ell+1)(\ell+2)}{2^{13}}s_0 \left( \partial_S 0\partial_T\right)^{\ell}(SRS)\, (TPT)\,S21034T^{\ell-2},\label{SRSTPTD}
\eea
in which the tensor structure $\left( \partial_S 0\partial_T\right)^{\ell+1}(SRS)\, (TRT) S21034T^{\ell-1}$ can be solved through the recursion relation
\bea
&\left( \partial_S 0\partial_T\right)^{\ell+1} &(SRS) (TRT) S21034T^{\ell-1}=\frac{\ell+1}{8}\Delta_A
\left( \partial_S 0\partial_T\right)^{\ell} (SRT)S21034T^{\ell-1}  \nn\\
&&-\frac{\ell  (\ell +1)^2}{32} X_{30}X_{40}X_{12}(X_{30}X_{40}X_{12}+X_{10}X_{20}X_{34}-\Delta_+)
\left( \partial_S 0\partial_T\right)^{\ell-1} S21034T^{\ell-1}\nn\\
&&-\frac{\ell-1}{64}X_{30}X_{40}X_{12}
\left( \partial_S 0\partial_T\right)^{\ell} (SRS)(TPT)S21034T^{\ell-2}.
\eea
From above recursion relations, we can solve the conformal integrations of the tensor structures corresponding to the mixed symmetry multiplets.

\section{Conformal integrations}

In this part, we provide the formulas used in this work to evaluate the conformal integrations.
The conformal integration related to $N_\ell$, or Gegenbauer polynomial $C_{\ell}^{(1)}(x_{0})$ has been studied in \cite{Dolan:2000ut, SimmonsDuffin:2012uy}. The results are given by
\begin{eqnarray}
\int_M D^{4}X_{0}\frac{(-1)^{\ell} C_{\ell}^{(1)}(t_{0})}{X_{10}^{\frac{\Delta+r}{2}}X_{20}^{\frac{\Delta-r}{2}}X_{30}^{\frac{\tl{\Delta}+\tl r}{2}}X_{40}^{\frac{\tl{\Delta}-\tl r}{2}}}=\xi_{\Delta,\tl{\Delta},\tl r,\ell}\left(\frac{X_{14}}{X_{13}}\right)^{\frac{\tl r}{2}}\left(\frac{X_{24}}{X_{14}}\right)^{\frac{r}{2}}X_{12}^{-\frac{\Delta}{2}}X_{34}^{-\frac{\tl{\Delta}}{2}}g_{\Delta,\ell}^{r,\tl r}(u,v),
\label{Cint}
\end{eqnarray}
in which $r\equiv\Delta_1-\Delta_2$, $\tl r\equiv \Delta_3-\Delta_4$ and
\begin{equation}
\xi_{\Delta,\tl{\Delta},\tl r,\ell}\equiv\frac{\pi^{2}\Gamma(\tl{\Delta}+\ell-1)
\Gamma(\frac{\Delta-\tl r+\ell}{2})\Gamma(\frac{\Delta+\tl r+\ell}{2})}{(2-\Delta)\Gamma(\Delta+\ell)
\Gamma(\frac{\tl{\Delta}-\tl r+\ell}{2})\Gamma(\frac{\tl{\Delta}+\tl r+\ell}{2})}.\label{xi}
\end{equation}
The conformal blocks $g_{\Delta,\ell}^{r,\tl r}(u,v)$ are
\bea
g_{\Delta,\ell}^{r,\tl r}(u,v)&=&\frac{\rho\bar{\rho}}{\rho-\bar{\rho}}\left[k_{\Delta+\ell}(\rho)k_{\Delta-\ell-2}(\bar{\rho})-(\rho\leftrightarrow\bar{\rho})\right]\label{g}, \nonumber\\
k_{\beta}(x)&=&x^{\frac{\beta}{2}}{}_{2}F_{1}\left(\frac{\beta-r}{2},\frac{\beta+\tl r}{2},\beta,x\right),
\eea
where $u,\,v$ are the conformal invariants $u=\rho\bar{\rho},~v=(1-\rho)(1-\bar{\rho})$.
We will focus on the superconformal block of the four-point correlator $\langle JJJJ\rangle$, so the $r=\tl r=0$ and $g_{\Delta,\ell}^{r,\tl r}(u,v)\sim g_{\Delta,\ell}$ in our notation, though conformal blocks with $r,\tl r=\pm1$ also appear in the intermediate steps.

For the conformal integration with extra invariant factor, like $\Omega_x \frac{N_\ell}{D_\ell}$, in principle one can expand the invariant factor and get the results for each part directly from the formula (\ref{Cint}). However, due to the analytical properties of the conformal blocks, usually the results can be organized in a compact way. The conformal integration
of $\Omega_{+/-} \frac{N_\ell}{D_\ell}$ have been provided in \cite{Khandker:2014mpa} with $r=\tl r=0$ and in \cite{Li:2016chh}
for general $r, \tl r$. In this work, we need to evaluate the conformal integration with two extra invariant factors, as shown in (\ref{Oddts}).

Note in our case, the factor $D_\ell$ for symmetric long multiplet is defined by
\begin{equation}
 D_{\ell}\equiv (X_{1\bar{0}}X_{01}X_{2\bar{0}}X_{02})^{\frac{1}{4}(\ell+\Delta)}
 (X_{3\bar{0}}X_{03}X_{4\bar{0}}X_{04})^{\frac{1}{4}(\ell-\Delta)}. \label{D2}
\end{equation}
The conformal integration with two $\Omega_x$ are given as follows
{\small
\bea
\int  &D^{4}X_{0} &\Omega_+^2 \frac{N_{\ell}}{D_{\ell}}
=\frac{2^{-6\ell}\;\xi_{\Delta+2,2-\Delta,2,\ell}}
{X_{12}^{\frac{\Delta-\ell}{2}}X_{34}^{-\frac{\Delta+\ell}{2}}}
\left(\frac{16 ((\Delta -\ell ) (\Delta +\ell )-2 \ell )}{(\Delta +\ell +2) (\Delta -\ell )} g_{\Delta,\ell}
+\frac{2 (\Delta -\ell -2) (\Delta +\ell )}{(\Delta +\ell +2) (\Delta -\ell )} g_{\Delta+2,\ell} \right. \nn\\
&&\left.+\frac{((\Delta -\ell ) (\Delta +\ell )-2 \ell ) (\Delta -\ell -2) (\Delta +\ell )}{16 (\Delta +\ell +3) (\Delta +\ell +1) (\Delta -\ell +1) (\Delta -\ell -1)}g_{\Delta+4,\ell}  \right. \\
&&\left.-\frac{\Delta +\ell }{(\Delta +\ell +3) (\Delta -\ell ) (\Delta +\ell +1)}g_{\Delta+2,\ell+2} -\frac{\Delta -\ell -2}{(\Delta -\ell -1) (\Delta -\ell +1) (\Delta +\ell +2)}g_{\Delta+2,\ell-2}  \right) \nn \\
\nn\\
\int  &D^{4}X_{0} &\Omega_-^2 \frac{N_{\ell}}{D_{\ell}}
=\frac{2^{-6\ell}\;\xi_{\Delta+2,2-\Delta,2,\ell}}
{X_{12}^{\frac{\Delta-\ell}{2}}X_{34}^{-\frac{\Delta+\ell}{2}}}
\left(\frac{2 (\Delta -\ell -2) (\Delta +\ell )}{(\Delta +\ell +2) (\Delta -\ell )} g_{\Delta+2,\ell}
 \right. \nn \\
&&\left.+\frac{(\Delta +\ell ) (\Delta -\ell -1)}{(\Delta +\ell +3) (\Delta -\ell )}g_{\Delta+2,\ell+2}
+\frac{(\Delta -\ell -2) (\Delta +\ell +1)}{(\Delta -\ell +1) (\Delta +\ell +2)}g_{\Delta+2,\ell-2}\right) \\
\nn\\
\int  &D^{4}X_{0} &\Omega_A^2 \frac{N_{\ell}}{D_{\ell}}
=\frac{2^{-6\ell}\;\xi_{\Delta+2,2-\Delta,2,\ell}}
{X_{12}^{\frac{\Delta-\ell}{2}}X_{34}^{-\frac{\Delta+\ell}{2}}}
\left(-\frac{\Delta  (\Delta -\ell -2) (\Delta +\ell )}{8 (\Delta +\ell +3)
(\Delta +\ell +1) (\Delta -\ell +1) (\Delta -\ell -1)} g_{\Delta+4,\ell}
 \right. \nn \\
&&\left.+\frac{\Delta +\ell }{(\Delta +\ell +3) (\Delta -\ell )} g_{\Delta+2,\ell+2}
+\frac{\Delta -\ell -2}{(\Delta -\ell +1) (\Delta +\ell +2)} g_{\Delta+2,\ell-2}\right)  \\
\nn\\
\int  &D^{4}X_{0} &\Omega_B^2 \frac{N_{\ell}}{D_{\ell}}
=\frac{2^{-6\ell}\;\xi_{\Delta+2,2-\Delta,2,\ell}}
{X_{12}^{\frac{\Delta-\ell}{2}}X_{34}^{-\frac{\Delta+\ell}{2}}}
\left(\frac{32 \Delta }{(\Delta +\ell +2) (\Delta -\ell )} g_{\Delta,\ell}
 \right.  \\
&&\left.-\frac{(\Delta -\ell -1) (\Delta +\ell )}{(\Delta +\ell +3) (\Delta -\ell ) (\Delta +\ell +1)} g_{\Delta+2,\ell+2}
-\frac{(\Delta -\ell -2) (\Delta +\ell +1)}{(\Delta -\ell -1) (\Delta -\ell +1) (\Delta +\ell +2)} g_{\Delta+2,\ell-2}\right)\nn     \\
\nn\\
\int  &D^{4}X_{0} &(\Omega_A\Omega_B-\Omega_+\Omega_-) \frac{N_{\ell}}{D_{\ell}}
=\frac{2^{-6\ell}\;\xi_{\Delta+2,2-\Delta,2,\ell}}
{X_{12}^{\frac{\Delta-\ell}{2}}X_{34}^{-\frac{\Delta+\ell}{2}}}
\left(\frac{4 (\Delta +\ell )}{\Delta +\ell +2} g_{\Delta+1,\ell+1}
+4\frac{\Delta -\ell -2}{\Delta-\ell  } g_{\Delta+1,\ell-1} \right. \nn\\
&&\left.+\frac{(\Delta -\ell -2) (\Delta +\ell )^2}{4 (\Delta-\ell  ) (\Delta +\ell +1) (\Delta +\ell +3)} g_{\Delta+3,\ell+1}  \right. \nn\\
&&\left.+\frac{(\Delta -\ell -2)^2 (\Delta +\ell )}{4 (\Delta -\ell +1) (\Delta -\ell -1) (\Delta +\ell +2)}g_{\Delta+3,\ell-1}  \right).
\eea
}

Denote the tensor structures involving $SPT$ and $SRT$ as:
\bea
P_\ell\equiv \frac{1}{{\ell!}^2} \left(\partial_{S}0\partial_{T}\right)^{\ell}\left(S\bar{2}1\bar{0}3\bar{4}T\right)^{\ell-1}\times&&\nonumber \\
 && \hspace{-40mm}  \left( X_{10}S\bar{2}3\bar{4}T + X_{20}S\bar{1}3\bar{4}T - X_{10}S\bar{2}4\bar{3}T - X_{20}S\bar{1}4\bar{3}T\right),  \label{R0}\\
R_\ell\equiv \frac{1}{{\ell!}^2}  \left(\partial_{S}0\partial_{T}\right)^{\ell} \left(S\bar{2}1\bar{0}3\bar{4}T\right)^{\ell-1} \times&& \nonumber\\
&& \hspace{-40mm} \left( X_{30}S\bar{2}1\bar{4}T + X_{40}S\bar{2}1\bar{3}T - X_{30}S\bar{1}2\bar{4}T - X_{40}S\bar{1}2\bar{3}T\right).  \label{P0}
\eea
Above definitions are slightly different from those in \cite{Li:2016chh}.
Their conformal integrations have been solved in \cite{Li:2016chh}. In this work we also need to evaluate their conformal integrations with factors $\Omega_x$:
{\small
\begin{eqnarray}
\int D^{4}X_{0} \frac{X_{12}}{X_{10}X_{20}}\Omega_A\frac{P_\ell}{D_\ell}
&=&\frac{2^{3-6\ell}\,\xi_{\Delta+3,1-\Delta,0,\ell-1}}{X_{12}^{\frac12(\Delta-\ell)}X_{34}^{-\frac12(\Delta+\ell)}}
\left(
\frac{4 (\Delta -\ell +1)}{\Delta  \ell  (\ell -\Delta )}
 g_{\Delta+2,\ell} \right.\nonumber\\
&&
\left.-\frac{4 (\Delta +1) (\ell +1)}{\Delta  \ell  (\Delta +\ell )}
g_{\Delta+2,\ell-2} +\frac{\Delta +\ell +2}{4 (\Delta +\ell +1) (\Delta +\ell +3)}g_{\Delta+4,\ell}\right)  \\
\nonumber\\
\int D^{4}X_{0} \frac{X_{12}}{X_{10}X_{20}}\Omega_B\frac{P_\ell}{D_\ell}
&=&\frac{2^{3-6\ell}\,\xi_{\Delta+3,1-\Delta,0,\ell-1}}{X_{12}^{\frac12(\Delta-\ell)}X_{34}^{-\frac12(\Delta+\ell)}}
\left(
-\frac{1}{\Delta  \ell }
 g_{\Delta+3,\ell-1} \right.\nn\\
&&\left.-\frac{16 (\Delta +1) (\ell +1) (\Delta -\ell +1)}{\Delta  \ell  (\Delta +\ell ) (\Delta -\ell )}
g_{\Delta+1,\ell-1} \right.\nonumber\\
&&
\left. +\frac{(\Delta +\ell +2) (\Delta -\ell +1)}{(\Delta -\ell ) (\Delta +\ell +1) (\Delta +\ell +3)} g_{\Delta+3,\ell+1}\right)  \\
\nonumber\\
\int D^{4}X_{0} \frac{X_{34}}{X_{30}X_{40}}\Omega_A\frac{R_\ell}{D_\ell}
&=&\frac{2^{3-6\ell}\,\xi_{\Delta+1,3-\Delta,0,\ell-1}}{X_{12}^{\frac12(\Delta-\ell)}X_{34}^{-\frac12(\Delta+\ell)}}
\left(
-\frac{\Delta +\ell }{(\Delta +\ell +1) (\Delta -\ell -2)}
 g_{\Delta+1,\ell+1} \right. \\
&&
\left. +\frac{1}{\Delta\ell}
g_{\Delta+1,\ell-1}+\frac{(\Delta -1) (\ell +1) (\Delta +\ell ) (\Delta -\ell )}{16 \Delta  \ell  (\Delta +\ell +1) (\Delta -\ell -1) (\Delta -\ell +1)} g_{\Delta+3,\ell-1}\right) \nn  \\
\nn  \\
\int D^{4}X_{0} \frac{X_{34}}{X_{30}X_{40}}\Omega_B\frac{R_\ell}{D_\ell}
&=&\frac{2^{3-6\ell}\,\xi_{\Delta+1,3-\Delta,0,\ell-1}}{X_{12}^{\frac12(\Delta-\ell)}X_{34}^{-\frac12(\Delta+\ell)}}
\left(
-\frac{4}{\Delta -\ell -2}
 g_{\Delta,\ell} \right. \\
&&
\left.+\frac{\Delta +\ell }{4 \Delta  \ell  (\Delta +\ell +1)}
g_{\Delta+2,\ell} +\frac{(\Delta -1) (\ell +1) (\Delta -\ell )}{4 \Delta  \ell  (\Delta -\ell -1) (\Delta -\ell +1)} g_{\Delta+2,\ell-2}\right).\nn
\end{eqnarray}
}
The conformal integrations of tensor structures $P1^2$ and $R1^2$ are
{\small
\bea
\int &D^{4}X_{0} &\left(\frac{X_{12}}{X_{10}X_{20}}\right)^2 \frac{1}{D_\ell} \frac{\ell(\ell-1)}{(\ell!)^2}
\left(\partial_{S}0\partial_{T}\right)^{\ell}\left(S\bar{2}1\bar{0}3\bar{4}T\right)^{\ell-2}SPT^2 = \nn\\
&&
\frac{2^{-6(\ell-1)}\,\xi_{\Delta+2,2-\Delta,0,\ell-2}}{X_{12}^{\frac12(\Delta-\ell)}X_{34}^{-\frac12(\Delta+\ell)}}
\left(\frac{\ell(\ell +1)}{\Delta +\ell +1}g_{\Delta+2,\ell-2}  \right. \nn\\
&&\left.~~~~~~~~~~~~ ~~~~~~~~~~~~~~~~~~~-\frac{\ell(\ell -1)  \Delta   (\Delta +\ell ) (\Delta +\ell +2)}{16 (\Delta +2) (\Delta +\ell +1)^2 (\Delta +\ell +3)}g_{\Delta+4,\ell}  \right), \\
\nn\\
\int &D^{4}X_{0} &\left(\frac{X_{34}}{X_{30}X_{40}}\right)^2 \frac{1}{D_\ell} \frac{\ell(\ell-1)}{(\ell!)^2}
\left(\partial_{S}0\partial_{T}\right)^{\ell}\left(S\bar{2}1\bar{0}3\bar{4}T\right)^{\ell-2}SRT^2 = \nn\\
&&
\frac{2^{-6(\ell-1)}\,\xi_{\Delta,4-\Delta,0,\ell-2}}{X_{12}^{\frac12(\Delta-\ell)}X_{34}^{-\frac12(\Delta+\ell)}}
\left(\ell(\ell -1)\frac{\Delta +\ell -2}{(\Delta -\ell -2) (\Delta +\ell -1)}g_{\Delta,\ell} \right. \nn\\
 &&~~~~~~~~~~~~~~~~~~~~~~~~~~~~~\left.-\ell(\ell+1) \frac{(\Delta -2) (\Delta -\ell ) (\Delta +\ell -2)}{16 \Delta  \left((\ell -\Delta )^2-1\right) (\Delta +\ell -1)} g_{\Delta+2,\ell-2}  \right).
\eea
}
The conformal integrations of tensor structures with $SPT\times SRT$ vanish.

Above conformal integrations are valid only for spin $\ell\geqslant 1$. For $\ell=0$, the unphysical conformal blocks $g_{\Delta,-2}$ do not vanish automatically and their contributions on the identities should be replaced by those with non-negative second subindex $\ell\geqslant0$. Nevertheless, the final superconformal blocks obtained from the recursion
relations can be correctly continued to $\ell=0$ case.

We have new tensor structures for the mixed symmetry multiplet given in (\ref{SSTT}) and (\ref{SRSTPT}).
The conformal integration of (\ref{SSTT}) can be obtained directly from the relation (\ref{SSTTD}).
The conformal integration of (\ref{SRSTPT}) can be solved from its recursion relation (\ref{SRSTPTD}). The result reads
\bea
\int &D^{4}X_{0}& \frac{1}{((\ell+2)!)^2} \frac{1}{D_{\ell+2}} \left( \partial_S 0\partial_T\right)^{\ell+2} (SRS)(TPT)S21034T^{\ell} =
\frac{2^{-6(\ell+1)}\,\xi_{\Delta+1,3-\Delta,0,\ell+1}}{X_{12}^{\frac12(\Delta-\ell-2)}X_{34}^{-\frac12(\Delta+\ell+2)}} \nn\\
&&\frac{(3+\ell)}{(2+\ell)\Delta}\left(g_{\Delta+1,\ell+1}-
 \frac{(\Delta -1) (\Delta -\ell -2) (\Delta +\ell +2)}{16 (\Delta +1) (\Delta -\ell -1) (\Delta +\ell +3)} g_{\Delta+3,\ell+1}  \right).
\eea

\section{Supershadow transformation of the OPE coefficients}

In this part we show how to solve the supershadow transformation matrix $\cM(\Delta,\ell)$ (\ref{ShadowMatrix}) from the unitarity condition.

By computing the superconformal integration in (\ref{Spwevenspin}) and getting rid of the kinematic factors, we obtain the superconformal  blocks containing supershadow coefficients $\la_{\tl\cO}^{(i)}$
\be
\cG_{\Delta,\ell,\mathrm{even}}^{\cN=2|JJ;JJ}=c_0 g_{\Delta,\ell}+c_1 g_{\Delta+2,\ell+2}+c^\prime_2 g_{\Delta+2,\ell}+c_3 g_{\Delta+2,\ell-2}+c_4 g_{\Delta+4,\ell}, \label{N2evenblockshadow}
\ee
where
\bea
c_0&\propto&  \la_\cO^{(1)}
\left( (\ell -\Delta ) \left(4+(\Delta +3) \Delta ^2+\ell\left(\Delta ^2+\Delta +2\right)  \right)\la_{\tl\cO}^{(1)}
-4  (\ell +2) (\ell +3)\Delta  \la_{\tl\cO}^{(2)} \right)
, \label{SC0}\\
c_1&\propto&\left(\la_\cO^{(1)}  (\Delta +\ell )+2\la_\cO^{(2)}  (\Delta +\ell +1)\right) \left(\la_{\tl\cO}^{(1)} (\ell -\Delta )+2 \la_{\tl\cO}^{(2)} (-\Delta +\ell +1)\right), \label{SC1}\\
c_3&\propto& (\Delta-\ell-2)\times \nn \\
&&\left(\la_\cO^{(1)} (4(1-\Delta) +\ell  (\ell +3) (2+\ell-\Delta))+2 \la_\cO^{(2)} (\ell +2) (\ell +3) (1+\ell-\Delta)\right) ~~~~~~~\nn \\
&&\times \left(\la_{\tl\cO}^{(1)} (4 (1+\Delta)+\ell (\ell +3) (\Delta +\ell +2))+2 \la_{\tl\cO}^{(2)}(\ell +2) (\ell +3) (\Delta +\ell +1)\right), \label{SC3}\\
c_4&\propto& (\Delta-\ell-2)\times \nn \\
&&\left(\la_\cO^{(1)} (\Delta +\ell ) \left((2-\Delta) \Delta ^2+((\Delta -1) \Delta +2) \ell +4\right)+4 \la_\cO^{(2)} \Delta  (\ell +2) (\ell +3)\right)\la_{\tl\cO}^{(1)}, \nn\\ \label{SC4}
\eea
where we have ignored the constant factors of the coefficients $c_i$ that have no effect on the unitarity condition.
For $c^\prime_2$, as we discussed previously, it includes several contributions and the independent parts $c_2$from multiplets $A^{\Delta+1}_{r'=0(\frac{\ell-1}{2},\frac{\ell+1}{2})}$ and
$A^{\Delta+1}_{r'=0(\frac{\ell+1}{2},\frac{\ell-1}{2})}$ are
\be
c_2\propto (\Delta-\ell-2)\left(\la_\cO^{(1)} (\Delta +\ell )+ \la_\cO^{(2)} (\ell +3)\Delta \right) \left(\la_{\tl\cO}^{(1)} (\ell-\Delta )- \la_{\tl\cO}^{(2)}\Delta  (\ell +3) \right). \label{SC2}
\ee
Note that for the coefficients $c_2, c_3$ and $c_4$ (also for $c^\prime_2$) there is a factor $\Delta-\ell-2$, which suggests the corresponding conformal blocks disappear when the scaling dimension of the multiplet saturates the unitary bound. This is consistent with
the multiplet splitting at the unitary bound (\ref{splitO}). The results in (\ref{N2evenblockshadow}) with non-vanishing coefficients $c_1$ and $c_2$ are expected to give the superconformal block of semi-short multiplet $\hat{C}_{0(\frac{\ell}{2},\frac{\ell}{2})}$.

The unitarity condition requires that the coefficients $c_i$ in (\ref{N2evenblockshadow}) are non-negative for general OPE coefficients $\la_{\cO}^{(i)}$. Therefore the supershadow coefficients $\la_{\tl\cO}^{(i)}$ should be the linear superpositions of $\la_{\cO}^{(i)}$ so that $c_i$ are  positive   quadratic polynomials of $\la_{\cO}^{(i)}$.

As a $2\times2$ matrix, the four elements in $\cM(\Delta,\ell)$ can be solved from the unitarity constraints of (\ref{SC0}), (\ref{SC1}) and (\ref{SC4})
\bea
&&\cM(\Delta,\ell)^i_j\propto \label{ShadowMatrix1}\\
&& \left(
\begin{array}{cc}
(\Delta +\ell ) \left((3-\Delta) \Delta ^2+((\Delta -1) \Delta +2) \ell +4\right) & 4 \Delta  (\ell +2) (\ell +3)  \\
 2 \Delta  (\ell -\Delta ) (\Delta +\ell ) & (\ell -\Delta ) \left((\Delta +3) \Delta ^2+\left(\Delta ^2+\Delta +2\right) \ell +4\right) \\
\end{array}
\right).\nn
\eea
While the unitarity conditions from coefficients $c_2$ and $c_3$ provide nontrivial consistency check on the supershadow transformation matrix (\ref{ShadowMatrix1}).

For a general scalar multiplet $\cA^{\Delta}_{0,0(\frac{\ell}{2},\frac{\ell}{2})}$ with $\Delta>2$, in the three-point function $\langle\cJ\cJ\cA^{\Delta}_{0,0(\frac{\ell}{2},\frac{\ell}{2})}\rangle$, there is a constraint on the two OPE coefficients
\be
\la_{\cO}^{(1)}=-3\la_{\cO}^{(2)}\equiv\la_{\cO}.
\ee
We should have the same constraint on the supershadow coefficients $\la_{\tl\cO}^{(i)}$. From the supershadow transformation matrix, indeed we have
\bea
\left( \begin{array}{c}
\la_{\tl\cO}^{(1)} \\
\la_{\tl\cO}^{(2)}
\end{array} \right)
& = &
\left( \begin{array}{c}
\cM(\Delta,\ell)^i_j
\end{array} \right)_{2\times2}
\cdot\la_\cO\left.\left( \begin{array}{c}
1 \\
-\frac{1}{3}
\end{array} \right)\right|_{\ell=0}
\propto \la_\cO \left( \begin{array}{c}
1 \\
-\frac{1}{3}
\end{array} \right).
\eea
Besides, the supershadow transformation matrix also satisfies the constraint
\be
\cM(-\Delta,\ell)  \cdot \cM(\Delta,\ell)\propto (\Delta-\ell-2)(\Delta-2) \mathbf{I}_{2\times2}. \label{MM}
\ee
For general long multiplets with $\Delta>\ell+2$, above identity corresponds to the fact that by taking the supershadow transformation twice, we obtain the original OPE coefficients. However the product becomes null at the unitary bound.
This is expected since we solved the supershadow transformation matrix from the unitarity condition of one of the vanishing coefficients.


\bibliographystyle{JHEP}
\bibliography{bibliography}{}

\providecommand{\href}[2]{#2}\begingroup\raggedright\begin{thebibliography}{10}

\bibitem{Polyakov:1974gs}
A.~M. Polyakov, {\it {Nonhamiltonian approach to conformal quantum field
  theory}},  {\em Zh. Eksp. Teor. Fiz.} {\bf 66} (1974) 23--42.

\bibitem{Ferrara:1973yt}
S.~Ferrara, A.~F. Grillo, and R.~Gatto, {\it {Tensor representations of
  conformal algebra and conformally covariant operator product expansion}},
  {\em Annals Phys.} {\bf 76} (1973) 161--188.

\bibitem{Mack:1975jr}
G.~Mack, {\it {Duality in quantum field theory}},  {\em Nucl. Phys.} {\bf B118} (1977) 445--457.

\bibitem{Rattazzi:2008pe}
R.~Rattazzi, V.~S. Rychkov, E.~Tonni, and A.~Vichi, {\it {Bounding scalar
  operator dimensions in 4D CFT}},  {\em JHEP} {\bf 0812} (2008) 031,
  [\href{http://arxiv.org/abs/0807.0004}{{\tt arXiv:0807.0004}}].

\bibitem{ElShowk:2012ht}
  S.~El-Showk, M.~F.~Paulos, D.~Poland, S.~Rychkov, D.~Simmons-Duffin and A.~Vichi,
  Phys.\ Rev.\ D {\bf 86}, 025022 (2012)
  doi:10.1103/PhysRevD.86.025022
  [arXiv:1203.6064 [hep-th]].

\bibitem{El-Showk:2014dwa}
  S.~El-Showk, M.~F.~Paulos, D.~Poland, S.~Rychkov, D.~Simmons-Duffin and A.~Vichi,
  J.\ Stat.\ Phys.\  {\bf 157}, 869 (2014)
  doi:10.1007/s10955-014-1042-7
  [arXiv:1403.4545 [hep-th]].

\bibitem{Kos:2014bka}
  F.~Kos, D.~Poland and D.~Simmons-Duffin,
  JHEP {\bf 1411}, 109 (2014)
  doi:10.1007/JHEP11(2014)109
  [arXiv:1406.4858 [hep-th]].

\bibitem{Poland:2018epd}
  D.~Poland, S.~Rychkov and A.~Vichi,
  arXiv:1805.04405 [hep-th].


\bibitem{Gaiotto:2009we}
D.~Gaiotto, {\it {N=2 dualities}},  {\em JHEP} {\bf 08} (2012) 034,
  [\href{http://arxiv.org/abs/0904.2715}{{\tt arXiv:0904.2715}}].

\bibitem{Gaiotto:2009hg}
D.~Gaiotto, G.~W. Moore, and A.~Neitzke, {\it {Wall-crossing, Hitchin Systems,
  and the WKB Approximation}},  \href{http://arxiv.org/abs/0907.3987}{{\tt
  arXiv:0907.3987}}.

\bibitem{Dymarsky:2017yzx}
  A.~Dymarsky, F.~Kos, P.~Kravchuk, D.~Poland and D.~Simmons-Duffin,
  JHEP {\bf 1802}, 164 (2018)
  doi:10.1007/JHEP02(2018)164
  [arXiv:1708.05718 [hep-th]].
\bibitem{Hofman:2008ar}
  D.~M.~Hofman and J.~Maldacena,
  JHEP {\bf 0805}, 012 (2008)
  doi:10.1088/1126-6708/2008/05/012
  [arXiv:0803.1467 [hep-th]].
\bibitem{Dymarsky:2017xzb}
  A.~Dymarsky, J.~Penedones, E.~Trevisani and A.~Vichi,
  arXiv:1705.04278 [hep-th].

\bibitem{Beem:2014zpa}
C.~Beem, M.~Lemos, P.~Liendo, L.~Rastelli, and B.~C. van Rees, {\it {The
  ${\mathcal N}=2$ superconformal bootstrap}},
  \href{http://arxiv.org/abs/1412.7541}{{\tt arXiv:1412.7541}}.


\bibitem{Lemos:2015awa}
  M.~Lemos and P.~Liendo,
  JHEP {\bf 1601}, 025 (2016)
  doi:10.1007/JHEP01(2016)025
  [arXiv:1510.03866 [hep-th]].

\bibitem{Cornagliotto:2017snu}
  M.~Cornagliotto, M.~Lemos and P.~Liendo,
  JHEP {\bf 1803}, 033 (2018)
  doi:10.1007/JHEP03(2018)033
  [arXiv:1711.00016 [hep-th]].

\bibitem{Beem:2013qxa}
  C.~Beem, L.~Rastelli and B.~C.~van Rees,
  Phys.\ Rev.\ Lett.\  {\bf 111}, 071601 (2013)
  doi:10.1103/PhysRevLett.111.071601
  [arXiv:1304.1803 [hep-th]].

\bibitem{Beem:2016wfs}
  C.~Beem, L.~Rastelli and B.~C.~van Rees,
  Phys.\ Rev.\ D {\bf 96}, no. 4, 046014 (2017)
  doi:10.1103/PhysRevD.96.046014
  [arXiv:1612.02363 [hep-th]].

\bibitem{Beem:2015aoa}
  C.~Beem, M.~Lemos, L.~Rastelli and B.~C.~van Rees,
  Phys.\ Rev.\ D {\bf 93}, no. 2, 025016 (2016)
  doi:10.1103/PhysRevD.93.025016
  [arXiv:1507.05637 [hep-th]].

\bibitem{Liendo:2015ofa}
  P.~Liendo, I.~Ramirez and J.~Seo,
  JHEP {\bf 1602}, 019 (2016)
  doi:10.1007/JHEP02(2016)019
  [arXiv:1509.00033 [hep-th]].

\bibitem{Ramirez:2016lyk}
  I.~A.~Ramírez,
  JHEP {\bf 1605}, 043 (2016)
  doi:10.1007/JHEP05(2016)043
  [arXiv:1602.07269 [hep-th]].



\bibitem{Beem:2013sza}
C.~Beem, M.~Lemos, P.~Liendo, W.~Peelaers, L.~Rastelli, and B.~C. van Rees,
  {\it {Infinite Chiral Symmetry in Four Dimensions}},  {\em Commun. Math.
  Phys.} {\bf 336} (2015), no.~3 1359--1433,
  [\href{http://arxiv.org/abs/1312.5344}{{\tt arXiv:1312.5344}}].


\bibitem{Eden:2000qp}
  B.~U.~Eden, P.~S.~Howe, A.~Pickering, E.~Sokatchev and P.~C.~West,
  Nucl.\ Phys.\ B {\bf 581}, 523 (2000)
  doi:10.1016/S0550-3213

\bibitem{Dolan:2000ut}
F.~Dolan and H.~Osborn, {\it {Conformal four point functions and the operator
  product expansion}},  {\em Nucl.Phys.} {\bf B599} (2001) 459--496,
  [\href{http://arxiv.org/abs/hep-th/0011040}{{\tt hep-th/0011040}}].

\bibitem{Dolan:2003hv}
F.~Dolan and H.~Osborn, {\it {Conformal partial waves and the operator product
  expansion}},  {\em Nucl.Phys.} {\bf B678} (2004) 491--507,
  [\href{http://arxiv.org/abs/hep-th/0309180}{{\tt hep-th/0309180}}].


\bibitem{Dolan:2011dv}
F.~Dolan and H.~Osborn, {\it {Conformal Partial Waves: Further Mathematical
  Results}},  \href{http://arxiv.org/abs/1108.6194}{{\tt arXiv:1108.6194}}.

\bibitem{Dolan:2001tt}
  F.~A.~Dolan and H.~Osborn,
  Nucl.\ Phys.\ B {\bf 629}, 3 (2002)
  doi:10.1016/S0550-3213(02)00096-2
  [hep-th/0112251].

\bibitem{Dolan:2004mu}
  F.~A.~Dolan, L.~Gallot and E.~Sokatchev,
  JHEP {\bf 0409}, 056 (2004)
  doi:10.1088/1126-6708/2004/09/056
  [hep-th/0405180].


\bibitem{Chester:2014fya}
  S.~M.~Chester, J.~Lee, S.~S.~Pufu and R.~Yacoby,
  JHEP {\bf 1409}, 143 (2014)
  doi:10.1007/JHEP09(2014)143
  [arXiv:1406.4814 [hep-th]].

\bibitem{Chester:2014mea}
  S.~M.~Chester, J.~Lee, S.~S.~Pufu and R.~Yacoby,
  JHEP {\bf 1503}, 130 (2015)
  doi:10.1007/JHEP03(2015)130
  [arXiv:1412.0334 [hep-th]].

\bibitem{Liendo:2015cgi}
  P.~Liendo, C.~Meneghelli and V.~Mitev,
  Commun.\ Math.\ Phys.\  {\bf 350}, no. 1, 387 (2017)
  doi:10.1007/s00220-016-2715-7
  [arXiv:1512.06072 [hep-th]].

\bibitem{Liendo:2016ymz}
  P.~Liendo and C.~Meneghelli,
  JHEP {\bf 1701}, 122 (2017)
  doi:10.1007/JHEP01(2017)122
  [arXiv:1608.05126 [hep-th]].

\bibitem{Lemos:2016xke}
  M.~Lemos, P.~Liendo, C.~Meneghelli and V.~Mitev,
  JHEP {\bf 1704}, 032 (2017)
  doi:10.1007/JHEP04(2017)032
  [arXiv:1612.01536 [hep-th]].

\bibitem{Beem:2016wfs}
  C.~Beem, L.~Rastelli and B.~C.~van Rees,
  Phys.\ Rev.\ D {\bf 96}, no. 4, 046014 (2017)
  doi:10.1103/PhysRevD.96.046014
  [arXiv:1612.02363 [hep-th]].

\bibitem{Poland:2010wg}
  D.~Poland and D.~Simmons-Duffin,
  JHEP {\bf 1105}, 017 (2011)
  doi:10.1007/JHEP05(2011)017
  [arXiv:1009.2087 [hep-th]].

\bibitem{Fortin:2011nq}
J.-F. Fortin, K.~Intriligator, and A.~Stergiou, {\it {Current OPEs in
  Superconformal Theories}},  {\em JHEP} {\bf 09} (2011) 071,
  [\href{http://arxiv.org/abs/1107.1721}{{\tt arXiv:1107.1721}}].

\bibitem{Li:2017ddj}
  D.~Li, D.~Meltzer and A.~Stergiou,
  JHEP {\bf 1707}, 029 (2017)
  doi:10.1007/JHEP07(2017)029
  [arXiv:1702.00404 [hep-th]].

\bibitem{Bobev:2015vsa} 
  N.~Bobev, S.~El-Showk, D.~Mazac and M.~F.~Paulos,
  Phys.\ Rev.\ Lett.\  {\bf 115}, no. 5, 051601 (2015)
  doi:10.1103/PhysRevLett.115.051601
  [arXiv:1502.04124 [hep-th]].

\bibitem{Bobev:2015jxa}
  N.~Bobev, S.~El-Showk, D.~Mazac and M.~F.~Paulos,
  JHEP {\bf 1508}, 142 (2015)
  doi:10.1007/JHEP08(2015)142
  [arXiv:1503.02081 [hep-th]].

\bibitem{Fitzpatrick:2014oza}
  A.~L.~Fitzpatrick, J.~Kaplan, Z.~U.~Khandker, D.~Li, D.~Poland and D.~Simmons-Duffin,
  JHEP {\bf 1408}, 129 (2014)
  doi:10.1007/JHEP08(2014)129
  [arXiv:1402.1167 [hep-th]].



\bibitem{Khandker:2014mpa}
  Z.~U.~Khandker, D.~Li, D.~Poland and D.~Simmons-Duffin,
  JHEP {\bf 1408}, 049 (2014)
  doi:10.1007/JHEP08(2014)049
  [arXiv:1404.5300 [hep-th]].
\bibitem{Li:2016chh}
  Z.~Li and N.~Su,
  JHEP {\bf 1605}, 163 (2016)
  doi:10.1007/JHEP05(2016)163
  [arXiv:1602.07097 [hep-th]].
\bibitem{Bobev:2017jhk}
  N.~Bobev, E.~Lauria and D.~Mazac,
  JHEP {\bf 1707}, 061 (2017)
  doi:10.1007/JHEP07(2017)061
  [arXiv:1705.08594 [hep-th]].



\bibitem{Dirac:1936fq}
  P.~A.~M.~Dirac,
  Annals Math.\  {\bf 37}, 429 (1936).
  doi:10.2307/1968455
\bibitem{Mack:1969rr}
  G.~Mack and A.~Salam,
  Annals Phys.\  {\bf 53}, 174 (1969).
  doi:10.1016/0003-4916(69)90278-4
\bibitem{Ferrara:1973eg}
  S.~Ferrara, R.~Gatto and A.~F.~Grillo,
  Springer Tracts Mod.\ Phys.\  {\bf 67}, 1 (1973).
  doi:10.1007/BFb0111104
\bibitem{Weinberg:2010fx}
  S.~Weinberg,
  Phys.\ Rev.\ D {\bf 82}, 045031 (2010)
  doi:10.1103/PhysRevD.82.045031
  [arXiv:1006.3480 [hep-th]].
\bibitem{Costa:2011mg}
  M.~S.~Costa, J.~Penedones, D.~Poland and S.~Rychkov,
  JHEP {\bf 1111}, 071 (2011)
  doi:10.1007/JHEP11(2011)071
  [arXiv:1107.3554 [hep-th]].
\bibitem{Costa:2011dw}
  M.~S.~Costa, J.~Penedones, D.~Poland and S.~Rychkov,
  JHEP {\bf 1111}, 154 (2011)
  doi:10.1007/JHEP11(2011)154
  [arXiv:1109.6321 [hep-th]].
\bibitem{SimmonsDuffin:2012uy}
  D.~Simmons-Duffin,
  JHEP {\bf 1404}, 146 (2014)
  doi:10.1007/JHEP04(2014)146
  [arXiv:1204.3894 [hep-th]].

\bibitem{Elkhidir:2014woa}
  E.~Elkhidir, D.~Karateev and M.~Serone,
  JHEP {\bf 1501}, 133 (2015)
  doi:10.1007/JHEP01(2015)133
  [arXiv:1412.1796 [hep-th]].
\bibitem{Costa:2014rya}
  M.~S.~Costa and T.~Hansen,
  JHEP {\bf 1502}, 151 (2015)
  doi:10.1007/JHEP02(2015)151
  [arXiv:1411.7351 [hep-th]].

\bibitem{Echeverri:2015rwa}
  A.~C.~Echeverri, E.~Elkhidir, D.~Karateev and M.~Serone,
  JHEP {\bf 1508}, 101 (2015)
  doi:10.1007/JHEP08(2015)101
  [arXiv:1505.03750 [hep-th]].
\bibitem{Rejon-Barrera:2015bpa}
  F.~Rejon-Barrera and D.~Robbins,
  JHEP {\bf 1601}, 139 (2016)
  doi:10.1007/JHEP01(2016)139
  [arXiv:1508.02676 [hep-th]].
\bibitem{Iliesiu:2015akf}
  L.~Iliesiu, F.~Kos, D.~Poland, S.~S.~Pufu, D.~Simmons-Duffin and R.~Yacoby,
  arXiv:1511.01497 [hep-th].
\bibitem{Echeverri:2016dun}
  A.~C.~Echeverri, E.~Elkhidir, D.~Karateev and M.~Serone,
  arXiv:1601.05325 [hep-th].

\bibitem{Costa:2016xah}
  M.~S.~Costa, T.~Hansen, J.~Penedones and E.~Trevisani,
  JHEP {\bf 1607}, 057 (2016)
  doi:10.1007/JHEP07(2016)057
  [arXiv:1603.05552 [hep-th]].

\bibitem{Costa:2016hju}
  M.~S.~Costa, T.~Hansen, J.~Penedones and E.~Trevisani,
  JHEP {\bf 1607}, 018 (2016)
  doi:10.1007/JHEP07(2016)018
  [arXiv:1603.05551 [hep-th]].


\bibitem{Goldberger:2011yp}
  W.~D.~Goldberger, W.~Skiba and M.~Son,
  Phys.\ Rev.\ D {\bf 86}, 025019 (2012)
  doi:10.1103/PhysRevD.86.025019
  [arXiv:1112.0325 [hep-th]].
\bibitem{Goldberger:2012xb}
  W.~D.~Goldberger, Z.~U.~Khandker, D.~Li and W.~Skiba,
  Phys.\ Rev.\ D {\bf 88}, 125010 (2013)
  doi:10.1103/PhysRevD.88.125010
  [arXiv:1211.3713 [hep-th]].
\bibitem{Siegel:2012di}
  W.~Siegel,
  arXiv:1204.5679 [hep-th].
\bibitem{Maio:2012tx}
  M.~Maio,
  Nucl.\ Phys.\ B {\bf 864}, 141 (2012)
  doi:10.1016/j.nuclphysb.2012.06.011
  [arXiv:1205.0389 [hep-th]].
\bibitem{Kuzenko:2012tb}
  S.~M.~Kuzenko,
  JHEP {\bf 1210}, 135 (2012)
  doi:10.1007/JHEP10(2012)135
  [arXiv:1206.3940 [hep-th]].

\bibitem{Erdmenger:1996yc}
  J.~Erdmenger and H.~Osborn,
  Nucl.\ Phys.\ B {\bf 483}, 431 (1997)
  doi:10.1016/S0550-3213(96)00545-7
  [hep-th/9605009].

\bibitem{Park:1997bq}
  J.~H.~Park,
  Int.\ J.\ Mod.\ Phys.\ A {\bf 13}, 1743 (1998)
  doi:10.1142/S0217751X98000755
  [hep-th/9703191].

\bibitem{Osborn:1998qu}
  H.~Osborn,
  Annals Phys.\  {\bf 272}, 243 (1999)
  doi:10.1006/aphy.1998.5893
  [hep-th/9808041].

\bibitem{Park:1999pd}
  J.~H.~Park,
  Nucl.\ Phys.\ B {\bf 559}, 455 (1999)
  doi:10.1016/S0550-3213(99)00432-0
  [hep-th/9903230].

\bibitem{Kuzenko:1999pi}
S.~M. Kuzenko and S.~Theisen, {\it {Correlation functions of conserved currents
  in N=2 superconformal theory}},  {\em Class. Quant. Grav.} {\bf 17} (2000)
  665--696, [\href{http://arxiv.org/abs/hep-th/9907107}{{\tt hep-th/9907107}}].

\bibitem{Arutyunov:2001qw}
G.~Arutyunov, B.~Eden, and E.~Sokatchev, {\it {On nonrenormalization and OPE in
  superconformal field theories}},  {\em Nucl. Phys.} {\bf B619} (2001)
  359--372, [\href{http://arxiv.org/abs/hep-th/0105254}{{\tt hep-th/0105254}}].



\bibitem{Dolan:2002zh}
F.~A. Dolan and H.~Osborn, {\it {On short and semi-short representations for
  four-dimensional superconformal symmetry}},  {\em Annals Phys.} {\bf 307}
  (2003) 41--89, [\href{http://arxiv.org/abs/hep-th/0209056}{{\tt
  hep-th/0209056}}].

\bibitem{Dobrev:1985qz} 
  V.~K.~Dobrev and V.~B.~Petkova,
  Fortsch.\ Phys.\  {\bf 35}, 537 (1987).

\bibitem{Dobrev:1985qv} 
  V.~K.~Dobrev and V.~B.~Petkova,
  Phys.\ Lett.\  {\bf 162B}, 127 (1985).
  doi:10.1016/0370-2693(85)91073-1

\bibitem{Dobrev:1985vh} 
  V.~K.~Dobrev and V.~B.~Petkova,
  Lett.\ Math.\ Phys.\  {\bf 9}, 287 (1985).
  doi:10.1007/BF00397755



\bibitem{Kinney:2005ej}
  J.~Kinney, J.~M.~Maldacena, S.~Minwalla and S.~Raju,
  Commun.\ Math.\ Phys.\  {\bf 275}, 209 (2007)
  doi:10.1007/s00220-007-0258-7
  [hep-th/0510251].
\bibitem{Gadde:2011uv}
  A.~Gadde, L.~Rastelli, S.~S.~Razamat and W.~Yan,
  Commun.\ Math.\ Phys.\  {\bf 319}, 147 (2013)
  doi:10.1007/s00220-012-1607-8
  [arXiv:1110.3740 [hep-th]].

\bibitem{Maldacena:2011jn}
J.~Maldacena and A.~Zhiboedov, {\it {Constraining Conformal Field Theories with
  A Higher Spin Symmetry}},  {\em J. Phys.} {\bf A46} (2013) 214011,
  [\href{http://arxiv.org/abs/1112.1016}{{\tt arXiv:1112.1016}}].

\bibitem{Alba:2013yda}
V.~Alba and K.~Diab, {\it {Constraining conformal field theories with a higher
  spin symmetry in d=4}},  \href{http://arxiv.org/abs/1307.8092}{{\tt
  arXiv:1307.8092}}.

\bibitem{Li:pr}
D. Li.  \emph{Private communication}.

\bibitem{Poland:2011ey}
  D.~Poland, D.~Simmons-Duffin and A.~Vichi,
  JHEP {\bf 1205}, 110 (2012)
  doi:10.1007/JHEP05(2012)110
  [arXiv:1109.5176 [hep-th]].
\bibitem{Simmons-Duffin:2015qma}
  D.~Simmons-Duffin,
  JHEP {\bf 1506}, 174 (2015)
  doi:10.1007/JHEP06(2015)174
  [arXiv:1502.02033 [hep-th]].
\bibitem{Li:2018fu}
Z. Li.  \emph{Work in progress}.

\bibitem{Li:2017kck}
  Z.~Li and N.~Su,
  arXiv:1706.06960 [hep-th].

\bibitem{Poland:2015mta}
  D.~Poland and A.~Stergiou,
  JHEP {\bf 1512}, 121 (2015)
  doi:10.1007/JHEP12(2015)121
  [arXiv:1509.06368 [hep-th]].


\bibitem{Hofman:2016awc}
  D.~M.~Hofman, D.~Li, D.~Meltzer, D.~Poland and F.~Rejon-Barrera,
  JHEP {\bf 1606}, 111 (2016)
  doi:10.1007/JHEP06(2016)111
  [arXiv:1603.03771 [hep-th]].
\bibitem{Faulkner:2016mzt}
  T.~Faulkner, R.~G.~Leigh, O.~Parrikar and H.~Wang,
  JHEP {\bf 1609}, 038 (2016)
  doi:10.1007/JHEP09(2016)038
  [arXiv:1605.08072 [hep-th]].

\bibitem{Camanho:2014apa}
  X.~O.~Camanho, J.~D.~Edelstein, J.~Maldacena and A.~Zhiboedov,
  JHEP {\bf 1602}, 020 (2016)
  doi:10.1007/JHEP02(2016)020
  [arXiv:1407.5597 [hep-th]].
\bibitem{Afkhami-Jeddi:2016ntf}
  N.~Afkhami-Jeddi, T.~Hartman, S.~Kundu and A.~Tajdini,
  JHEP {\bf 1712}, 049 (2017)
  doi:10.1007/JHEP12(2017)049
  [arXiv:1610.09378 [hep-th]].
\bibitem{Costa:2017twz}
  M.~S.~Costa, T.~Hansen and J.~Penedones,
  JHEP {\bf 1710}, 197 (2017)
  doi:10.1007/JHEP10(2017)197
  [arXiv:1707.07689 [hep-th]].
\bibitem{Afkhami-Jeddi:2017rmx}
  N.~Afkhami-Jeddi, T.~Hartman, S.~Kundu and A.~Tajdini,
  arXiv:1709.03597 [hep-th].
\bibitem{Meltzer:2017rtf}
  D.~Meltzer and E.~Perlmutter,
  arXiv:1712.04861 [hep-th].
\bibitem{Manenti:2018xns}
  A.~Manenti, A.~Stergiou and A.~Vichi,
  arXiv:1804.09717 [hep-th].



\end{thebibliography}\endgroup

\end{document}